\documentclass[10pt]{article}

\usepackage[margin=1in]{geometry}

\usepackage{graphicx} 

\usepackage[unicode=true,pdfusetitle,
bookmarks=true,bookmarksnumbered=false,bookmarksopen=false, pdfborder={0 0 0},pdfborderstyle={},backref=false,colorlinks=true]{hyperref}
 
\hypersetup{citecolor=blue} 
\usepackage{appendix,microtype,url}
\usepackage[normalem]{ulem} 
\usepackage{enumitem}
\usepackage{comment}

\usepackage[utf8]{inputenc} 
\usepackage[T1]{fontenc}    
\usepackage{tgtermes}

\usepackage{url}            
\usepackage{booktabs}       
\usepackage{amsfonts}       
\usepackage{nicefrac}       
\usepackage{microtype}      
\usepackage{amsmath}
\usepackage{enumitem}
\usepackage{amssymb}
\usepackage{graphicx}
\usepackage{bm}
\usepackage{theorem}

\usepackage{xcolor}
\usepackage{todonotes} %
\usepackage[showdeletions]{color-edits} %

\usepackage{mathtools}

\usepackage{comment}
\usepackage{dsfont}
\usepackage{lipsum}
\usepackage{algorithm}
\usepackage{algorithmic}

\usepackage[size=small,labelfont=bf]{caption}
\usepackage{harpoon}
\usepackage[normalem]{ulem}

\addauthor[Luke]{ld}{orange} %
\addauthor[Qiaomin]{qx}{red} %
\addauthor[Yudong]{yc}{purple} %
\addauthor[JY]{jy}{blue}
\newcommand{\citep}{\cite}

\theoremstyle{plain}
\newtheorem{theorem}{Theorem}[section]

\newtheorem{lemma}[theorem]{Lemma}
\newtheorem{corollary}[theorem]{Corollary}

\newtheorem{assumption}{Assumption}

\newtheorem{definition}{Definition}

\newtheorem{remark}{Remark}

\usepackage{dsfont,subcaption,float}
\usepackage{mathtools}
\usepackage{lipsum}

\newcommand{\RN}[1]{%
\textup{\uppercase\expandafter{\romannumeral#1}}%
}

\newcommand{\TV}{\texttt{TV}}

\newcommand{\Var}{\mathbb{V}}
\newcommand{\Tr}{\operatorname*{Tr}}
\newcommand{\op}{\mathrm{op}}

\newcommand{\indic}[1]{\mathds{1}\left\{ #1 \right\}} 

\newcommand{\PP}{\mathds{P}}    
\newcommand{\vdot}[2]{\langle #1, #2 \rangle}

\newcommand{\Exs}{\mathbb{E}}

\allowdisplaybreaks

\newif\ifdraft
\newif\ifarxiv
\drafttrue  

\newcommand{\R}{\mathbb{R}}
\arxivtrue

\title{Two-Timescale Linear Stochastic Approximation: \\ Constant Stepsizes Go a Long Way}

\usepackage{authblk}
\author{Jeongyeol Kwon}
\author{Luke Dotson}
\author{Yudong Chen}
\author{Qiaomin Xie}
\affil{University of Wisconsin-Madison}

\begin{document}

\maketitle


\begin{abstract}
  Previous studies on two-timescale stochastic approximation (SA) mainly focused on bounding mean-squared errors under diminishing stepsize schemes. In this work, we investigate {\it constant} stpesize schemes through the lens of Markov processes, proving that the iterates of both timescales converge to a unique joint stationary distribution in Wasserstein metric. We derive explicit geometric and non-asymptotic convergence rates, as well as the variance and bias introduced by constant stepsizes in the presence of Markovian noise. Specifically, with two constant stepsizes $\alpha < \beta$, we show that the biases scale linearly with both stepsizes as  $\Theta(\alpha)+\Theta(\beta)$ up to higher-order terms, while the variance of the slower iterate (resp., faster iterate) scales only with its own stepsize as $O(\alpha)$ (resp., $O(\beta)$). Unlike previous work, our results require no additional assumptions such as $\beta^2 \ll \alpha$ nor extra dependence on dimensions. These fine-grained characterizations allow tail-averaging and extrapolation techniques to reduce variance and bias, improving mean-squared error bound to $O(\beta^4 + \frac{1}{t})$ for both iterates. 
\end{abstract}

\section{Introduction}

Stochastic Approximation (SA) is an iterative procedure to find the root of unknown operators from their noisy samples \citep{robbins1951stochastic}. There has been a long line of work understanding the convergence behavior of SA both asymptotically \citep{borkar2000ode, harold1997stochastic} and in a finite-time \citep{srikant2019finite}, with a wide range of applications including stochastic optimization \citep{harold1997stochastic, moulines2011non} and reinforcement learning \citep{sutton2018reinforcement, lakshminarayanan2018linear, srikant2019finite}.

Two-Timescale Stochastic Approximation (TTSA) is a variant of the SA algorithm, designed to find the root of two intertwined operators \citep{borkar1997stochastic}. In particular, given two operators $F$ and $G$, we aim to find the solution $(x^*, y^*)$ satisfying the fixed-point equations 
\begin{align*}
    \begin{cases}
        F(x^*,y^*) = 0, \\ 
        G(x^*,y^*) = 0.
    \end{cases}
\end{align*}
This work considers linear TTSA with constant stepsizes driven by Markovian data as the following:  
\begin{equation}
    \begin{aligned}
    x_{t+1} &= x_t - \alpha_t (F(x_t, y_t) + w^x(x_t,y_t;\xi_t)), \\
    y_{t+1} &= y_t - \beta_t (G(x_t, y_t) + w^y(x_t,y_t;\xi_t)),  
    \end{aligned}
    \quad t\ge 0,
    \label{eq:basic_ttsa_equation}
\end{equation}
where $\alpha_t \equiv \alpha$, $\beta_t \equiv \beta > 0$ are constant stepsizes for slower and faster iterates respectively, $F$ and $G$ are linear operators, and $w^x$ and $w^y$ are linear Markovian noises driven by {\it exogenous} Markovian states $\xi_t$ (see Section \ref{section:prelim} for precise formulation). 

The updates in \eqref{eq:basic_ttsa_equation} arise in many applications: examples include popular reinforcement learning algorithms such as actor-critic \citep{konda1999actor, haarnoja2018soft} and gradient temporal-difference (GTD) methods \citep{maei2009convergent, szepesvari2022algorithms}, and iterative algorithms for stochastic Bilevel optimization \citep{colson2007overview, ghadimi2018approximation, hong2023two, kwon2023fully}. The core idea of TTSA is the use of different stepsizes for two iterations, which establishes a hierarchy between the two fixed-point equations. For example, in actor-critic algorithms \citep{haarnoja2018soft}, the $y$-variable minimizes the temporal-difference (TD) error, while the $x$-variable represents policy parameters to maximize long-term rewards. To ensure that the policy parameters are updated based on accurate value estimates, we set $\beta \gg \alpha$, meaning that $y$ converges faster, staying close to the minimizer of the TD-error given the current policy parameter $x$.

Classical results have established asymptotic convergence of TTSA with diminishing step sizes, $\alpha_t, \beta_t \rightarrow 0$, under the requirement of order-wise different timescales, {\it i.e.,} $\frac{\alpha_t}{\beta_t} \rightarrow 0$ \citep{borkar1997stochastic, konda2004convergence, mokkadem2006convergence}. With the recent advances in large-scale optimization, several papers have focused on analyzing the finite-time convergence of TTSA under similar vanishing step-size conditions. Earlier analyses reported suboptimal convergence rates of $O(t^{-2/3})$ \citep{dalal2018finite, doan2022nonlinear}, which have been improved to the best possible rate of $O(t^{-1})$ in more recent studies as long as $\beta_t^2 \lesssim \alpha_t$ \citep{kaledin2020finite, dalal2020tale, haque2023tight, doan2024fast, han2024finite, hu2024central}. The key to recent improvements lies in eliminating the need for diminishing stepsize ratios, achieved through a more refined analysis of the cross-correlations between the two intertwined iterations \citep{kaledin2020finite, haque2023tight, han2024finite}.

More recently, SA with constant stepsizes has attracted attention due to its simplicity, fast convergence, and good empirical performance, both for single- and two-timescale cases (see Section \ref{subsec:related_work} for details). 
However, existing results for TTSA are often limited to only providing upper bounds for $\Exs[\|x_t - x^*\|_2^2]$ and $\Exs[\|y_t - y^*(x_t)\|_2^2]$, {\it i.e.,} mean-squared errors (MSE) from the fixed point of operators, leaving the non-asymptotic behavior of TTSA iterations with constant stepsizes unexplored. Through the lens of the Markov process on TTSA iterations, we break down the sources of MSE and demonstrate the advantages of a finer understanding, particularly when employing techniques like tail-averaging and extrapolation.


\paragraph{Our Contributions.} We study the behaviors of Markovian TTSA iterations \eqref{eq:basic_ttsa_equation} with constant stepsizes. We focus on {\it linear} TTSA when the two operators $F,G$ and Markovian noise fields $w^x, w^y$ are linear in the iterates. Our contributions are summarized as follows:
\begin{itemize}
    \item While the iterates do not converge pointwise with constant stepsizes, under the standard assumptions for TTSA, we show that the joint process $(x_t, y_t, \xi_t)$ of iterates and  Markovian noises converges to a unique {\it biased} stationary distribution. 
    \item For the stationary distribution of slower iterates $x_{\infty}$,  
    we show that its bias has a dominating term growing linearly with $\alpha$ and $\beta$, while its variance is $O(\alpha)$. Therefore, the asymptotic MSE of order $O(\alpha)$ for slower iterates reported in prior work (which requires the assumption $\beta^2 \le \alpha$) in fact admits the following bias-variance decomposition:
    \begin{align*}
        \texttt{x-MSE} \asymp \underbrace{\|\Exs[x_\infty] - x^*\|_2^2}_{\text{squared-bias:} \ O(\alpha+\beta)^2}  + \underbrace{\Exs[\|x_\infty - \Exs[x_{\infty}]\|_2^2]}_{\text{variance:} \ O(\alpha)}.
    \end{align*}
    \item Based on our distributional convergence results, we show the benefits of simple Polyak-Ruppert averaging \citep{polyak1992acceleration} and Richardson-Romberg Extrapolation \citep{stoer2013introduction} along with the use of constant stepsizes in TTSA iterations. Specifically, through combining the above techniques, we can achieve (1) exponentially-fast decaying optimization error, (2) variance decaying at $O(1/t)$ rate, and (3) order-wise improvement of asymptotic biases:
    \ifarxiv
    \begin{align*}
    &\Exs[\|\tilde{x}_t-x^*\|_2^2] \;\; \asymp \;\; \underbrace{\Exs[\|\tilde{x}_t - \tilde{x}_{\infty}\|_2^2]}_{\mathclap{\qquad \qquad\text{optimization error:} \, \exp(-\Theta(t))}} + \underbrace{\Exs[\|\tilde{x}_\infty - \Exs[\tilde{x}_{\infty}]\|_2^2]}_{\text{variance:} \, O(1/t)} + \underbrace{\|\Exs[\tilde{x}_\infty] - x^*\|_2^2}_{\mathclap{\text{reduced-bias:} \, O(\beta^4)}}.
    \end{align*}
    \else
    \begin{align*}
        &\Exs[\|\tilde{x}_t-x^*\|_2^2] \;\; \asymp \;\; \underbrace{\Exs[\|\tilde{x}_t - \tilde{x}_{\infty}\|_2^2]}_{\mathclap{\qquad \qquad\text{optimization error:} \, \exp(-\Theta(t))}} \\
        &\ + \underbrace{\Exs[\|\tilde{x}_\infty - \Exs[\tilde{x}_{\infty}]\|_2^2]}_{\text{variance:} \, O(1/t)} + \underbrace{\|\Exs[\tilde{x}_\infty] - x^*\|_2^2}_{\mathclap{\text{reduced-bias:} \, O(\beta^4)}}.
    \end{align*}
    \fi
\end{itemize}

We emphasize that our convergence results do not impose the restriction $\beta^2 \le \alpha$, or involve additional dimension-dependent constants as prior work in \cite{kaledin2020finite, haque2023tight, han2024finite}. 

\subsection{Related Work}
\label{subsec:related_work}

 The literature on (two-timescale) SA is vast. Here we discuss prior work most relevant to us.

\paragraph{Weak Convergence of Constant Stepsize SA.} Recent studies have shown that under regularity conditions, SA iterates with constant stepsizes weakly converge to a stationary distribution \citep{bhandari2018finite, dieuleveut2020bridging, durmus2024finite, chen2024lyapunov, lauand2023curse, allmeier2024computing}. In particular, a line of work has developed an approach based on the Wasserstein distance measure when operators are global contraction mapping \citep{dieuleveut2020bridging, durmus2021tight, huo2023bias, zhang2024constant, lauand2024revisiting}. For cases where operators possess only local contraction or star-convexity properties, other studies have shown convergence in total variation distance under additional assumptions on the noise distribution's support \citep{yu2021analysis, vlatakis2024stochastic}. Our result adopts the approach based on Wasserstein metrics, providing a more explicit convergence rates without requiring assumptions on the noise support, even when the overall iterates $(x_t, y_t)$ do not exhibit global contraction.

\paragraph{Existing Results for TTSA.} 
TTSA arises as a popular iterative solution in various domain; from the classical iterate-averaging schemes \citep{polyak1992acceleration} and off-policy reinforcement learning algorithms \citep{sutton2018reinforcement} to gradient descent-ascent algorithms for saddle-point problems \citep{jin2020local} and single-loop algorithms for Bilevel optimization \citep{hong2023two}. Asymptotic convergence and central limit theorems for TTSA with diminishing step sizes were initially established for linear cases with i.i.d.\ noise \citep{konda2004convergence}, followed by extensions to non-linear and Markovian noise settings \citep{mokkadem2006convergence, hu2024central}.

More recent work has shifted focus to non-asymptotic results, deriving finite-time convergence rates for both linear \citep{dalal2018finite, dalal2020tale} and nonlinear cases \citep{kaledin2020finite, han2024finite, haque2023tight}. However, these studies primarily address MSE bounds with diminishing stepsizes. In contrast, we investigate distributional convergence under constant stepsizes, providing explicit decoupling of biases and variances. Additionally, we establish new results for tail-averaging and extrapolation in TTSA schemes.

\section{Problem Setup}
\label{section:prelim}
Let $F: \mathbb{R}^{d_x}\times\mathbb{R}^{d_y} \rightarrow \mathbb{R}^{d_x}$ and $G: \mathbb{R}^{d_x}\times\mathbb{R}^{d_y} \rightarrow \mathbb{R}^{d_y}$ be linear mean-field operators in the following form:
\begin{align*}
    F(x,y) = J_{11} x + J_{12} y + b_1, \\
    G(x,y) = J_{21} x + J_{22} y + b_2,
\end{align*}
where $J_{11},\ldots,J_{22}$ (resp., $b_1,b_2$) are fixed matrices (resp., vectors), and linear Markovian noises defined as the following:
\begin{align*}
    w^x(x,y;\xi) = W_{11}(\xi) x + W_{12} (\xi) y + u_1(\xi), \\
    w^y(x,y;\xi) = W_{21}(\xi) x + W_{22} (\xi) y + u_2(\xi).
\end{align*} 
Let $J_{\max} := \max_{i,j\in\{1,2\}} \|J_{ij}\|_{\op}$ be the smoothness parameter of the system. The first assumption is on the mean-field operators being Hurwitz:
\begin{assumption}
    \label{assumption:linear_Hurwitz}
    The matrices $-J_{22}$ and $-\Delta := -J_{11} + J_{12} J_{22}^{-1} J_{21}$ are Hurwitz, that is, all real parts of the eigenvalues of $J_{22}$ and $\Delta$ are strictly positive.
\end{assumption}
Therefore, a fixed point in the slower timescale is uniquely defined $y^*(x) = -J_{22}^{-1} (J_{21} x + b_2)$ for every $x$, and the target joint fixed point $(x^*,y^*)$ is given as:
\begin{align*}
    x^* &= -\Delta^{-1} (b_1 - J_{12}J_{22}^{-1} b_2) \\
    y^* &= -J_{22}^{-1} (J_{21} x^* + b_2).
\end{align*}
Assumption \ref{assumption:linear_Hurwitz} is standard in the study of TTSA to ensure the stability of the system \citep{gupta2019finite, doan2022nonlinear}. The main difference from single timescale SA is the star-type stability of slower iterations, {\it i.e.,} we only assume that $-H(x) := -F(x,y^*(x))$ is Hurwitz, while the entire operation $-\begin{bmatrix} F(x,y) \\ G(x,y) \end{bmatrix}$ may not. Therefore, existing results for single-timescale SA cannot be directly applied. 

Next, we assume that the noise fields are controlled by a geometrically mixing exogenous ({\it i.e.,} state evolves independent of TTSA iterations) Markov chain $\{\xi_t\}_{t\ge 0}$:
\begin{assumption}
    \label{assumption:noise_field}
    Let $\{\xi_t\}_{t\ge0}$ be an exogenous Markovian chain on a countable state-space $\Xi$ with a transition kernel $\mathcal{P}$ and a unique stationary distribution $\pi$. Furthermore, $\{\xi_t\}_{t\ge 0}$ is geometrically mixing:
    \begin{align*}
        \|\mathcal{P}^n \pi_0 - \pi\|_{1} \le c_\rho \rho^n,
    \end{align*}
    for some absolute constant $c_\rho > 0$, $\rho \in [0,1)$ and any initial distribution $\xi_0 \sim \pi_0$ for all $n \ge 1$. 
\end{assumption}
We also assume that the noise fields are bounded and unbiased at the stationary limit:
\begin{assumption}
    \label{assumption:noise_norm_bound}
    For all $j \in \{1,2\}$ and $\xi \in \Xi$, we have
    \begin{align*}
        \Exs_{\xi \sim \pi}[W_{ij}(x,y;\xi)] &=0, \qquad \forall i,j \in \{1,2\}, \\
        \Exs_{\xi \sim \pi} [u_i(\xi)] &= 0, \qquad \forall i \in \{1,2\}.
    \end{align*}
    Furthermore, for all $\xi \in \Xi$, the following holds:
    \begin{align*}
        \|W_{ij}(\xi)\|_{\op} \le W_{\max}, \ \|u_i(\xi)\|_{2} \le u_{\max}, \forall i,j \in \{1,2\}.
    \end{align*}
    for some constants $W_{\max}, u_{\max} \ge 0$. For simplicity, we further assume that $W_{\max} \le J_{\max}$.
\end{assumption}
The above two assumptions are common in  the analysis of SA schemes with Markovian noises \citep{dalal2020tale, huo2023bias}. We introduce the notion of noise variances in our setting:
\begin{align}   
    \label{eq:sigma_define}
    \sigma_x^2 &:= \max_{\xi\in\Xi} \|W_{11}(\xi) x^* + W_{12}(\xi) y^* + u_1(\xi) \|_2^2, \nonumber \\
    \sigma_y^2 &:= \max_{\xi\in\Xi} \|W_{21}(\xi) x^* + W_{22}(\xi) y^* + u_2(\xi)\|_2^2,
\end{align}
which reflect the mean-squared fluctuation of the stochastic update around the fixed point. 

We study the convergence of TTSA iterations \eqref{eq:basic_ttsa_equation} via $L^2$-Wasserstein distance \citep{villani2009optimal}. Let $\mathcal{P}_2(\mathbb{R}^d)$ denote the space of square-integrable distributions on $\mathbb{R}^d$ where $d := d_x + d_y$.
Note that $L_2$-Wasserstein distance between two distributions $\mu$ and $\nu$ in $\mathcal{P}_2(\mathbb{R}^d)$ is defined as the following:
\begin{align*}
    \mathcal{W}_2(\mu, \nu) &:= \left(\inf_{\Gamma \in \Pi(\mu, \nu)} \int_{\mathbb{R}^d \times \mathbb{R}^d} \|u - v\|_2^2 \, \mathrm{d}\Gamma(u,v) \right)^{1/2}, 
\end{align*}
where $\Pi(\mu,\nu)$ is a set of all possible couplings with marginal distributions $\mu$ and $\nu$. To study the distribution convergence of the joint iterate-data sequence $(x_t, y_t, \xi_t)_{t\ge 0}$, we slightly extend the definition above to add hamming distance in $\Xi$.  Let $\mathcal{P}_2(\R^d\times \Xi)$ be the set of distributions $\bar{\mu}$ on $\R^d \times \Xi$ with
the property that the marginal of $\bar{\mu}$ on $\R^d$ is square-integrable.

\begin{definition}
    \label{def:joint_wasserstein}
    For any two probability measures $\mu, \nu$ in $\mathcal{P}_2(\R^{d_x+d_y}\times \Xi)$ over $(x,y,\xi)$, we define the distance between $\mu$ and $\nu$ as
    \ifarxiv
    \begin{align}
    &\bar{\mathcal{W}}_2 (\mu, \nu) := \inf_{\Gamma \in \Pi(\mu, \nu)} \Big\{ \Exs_{((x_t,y_t,\xi_t), (x'_t,y'_t,\xi'_t)) \sim \Gamma} \left[\indic{\xi_t \neq \xi'_t}+ \|x_t -x'_t\|_{2}^2 + \|y_t-y'_t \|_{2}^2 \right]^{1/2} \Big\},
    \end{align}
    \else
    \begin{align}
        &\bar{\mathcal{W}}_2 (\mu, \nu) := \inf_{\Gamma \in \Pi(\mu, \nu)} \Big\{ \Exs_{((x_t,y_t,\xi_t), (x'_t,y'_t,\xi'_t)) \sim \Gamma} \nonumber \\
        &\left[\indic{\xi_t \neq \xi'_t}+ \|x_t -x'_t\|_{2}^2 + \|y_t-y'_t \|_{2}^2 \right]^{1/2} \Big\},
    \end{align}
    \fi
    where $\Pi(\mu,\nu)$ is a set of all possible couplings with marginal distributions $\mu,\nu$.
\end{definition}

To establish the finite-time convergence of TTSA iterations \eqref{eq:basic_ttsa_equation}, we define a few error metrics. Let $Q_x, Q_{y} \succ 0$ be the unique solutions of the Lyapunov equations
\begin{align*}
    Q_x \Delta + \Delta^\top Q_x = I, \\
    Q_{y} J_{22} + J_{22}^\top Q_{y} = I.
\end{align*}
The solutions $Q_x, Q_y$, which are guaranteed to exist since $-\Delta, -J_{22}$ are Hurwitz under Assumption~\ref{assumption:linear_Hurwitz} \citep{chen1984linear},  are used for constructing the drift of potentials in our analysis. For the slower and faster iterates, we use $\|\cdot\|_{Q_x}$ and $\|\cdot\|_{Q_y}$ norms respectively, and define $\mu_x := \|Q_x\|_{\mathrm{op}}^{-1}$ and $\mu_y := \|Q_y\|_{\mathrm{op}}^{-1}$. Note that $\sigma_{\min}(\Delta) \ge \mu_x/2$ and $\sigma_{\min}(Q_x) \ge \|\Delta\|_{\op}^{-1}/2$, and similarly for $Q_y$ and $\mu_y$. Consequently, we let the condition number of two iterations as $\kappa_x := \frac{ \kappa_y J_{\max}}{\mu_x}$ and $\kappa_y := \frac{J_{\max}}{\mu_y}$. 

\paragraph{Notation.} For a positive definite matrix $Q \succ 0$ let $\|x\|_Q := \sqrt{x^\top Q x}$ for a vector $x$. With a general real-valued matrix $A$, we define $\|A\|_Q := \max_{\|x\|_Q=1} \|Ax\|_Q$. Let $\vdot{a}{b}_Q := a^\top Q b$ for two vectors $a,b$. For two real-valued matrices $A,B$, we denote $\vdot{A}{B} = \Tr(A^\top B)$. We define 1-Schatten norm $\|A\|_1 := \sum_{i} |\sigma_i(A)|$ as the absolute sum of singular values (sometimes we call it $S^1$-norm), and $\infty$-Schatten norm $\|A\|_{\infty} := \max_i |\sigma_i(A)|$ be the maximum singular value, which is equivalent to matrix operator norm $\|A\|_{\op}$. For a positive semi-definite matrix $Q \succeq 0$, $\|Q\|_1 = \Tr(Q) = \sum_i Q_{ii}$ is the sum of diagonal elements. For a random vector $x$, we denote the covariance $\mathbb{V}(x) := \Exs[(x - \Exs[x])(x - \Exs[x])^\top]$. We often use shorthands $w_t^x := w^x(x_t,y_t;\xi_t)$ and $w_t^y := w^y(x_t,y_t;\xi_t)$. We denote the fixed point of the faster iterates given $x$ as $y^*(x)$, such that $G(x, y^*(x)) = 0$. 
If we just write $y^*$, then it means $y^*(x^*)$. For two probability distributions $p,q$, we denote $\|p-q\|_1$ as the total-variation distance between $p$ and $q$. We use the notation $O(\cdot)$ to hide absolute constants, and $O_{\texttt{P}}(\cdot)$ to omit up to polynomial factors in instance-dependent constants (smoothness, minimum eigenvalues, and noise variances) and up to logarithmic factors in stepsizes. 

\section{Main Results}
\label{section:main_result}

We start with two conditions for stepsizes to ensure the stability of TTSA iterations:
\begin{assumption}
    \label{assumption:stepsize}
    We assume that the stepsizes $(\alpha,\beta)$ satisfy the following:
    \begin{align}
        \label{eq:step_size_conditions}
        \beta \tau_{\alpha} \le \frac{c_1}{ J_{\max} \kappa_y^2\kappa_x^2}, 
        \quad
        \frac{\alpha}{\beta} \le \frac{c_2}{\kappa_y^3 \kappa_x}.
    \end{align}
    where $\tau_{\alpha} := \frac{\log(\alpha \mu_x/c_\rho)}{\log \rho}$ with some sufficiently small absolute constants $c_1, c_2 > 0$.
\end{assumption}
The first condition in \eqref{eq:step_size_conditions} ensures $\beta$ less than the inverse smoothness of operators, and the second condition bounds the ratio between two-timescale iterations. We mention that the dependence on the condition numbers is not fully optimized. In the sequel, we start with a fine-grained convergence in MSE in Section \ref{subsec:conv_mse}. We then show the convergence in distribution and characterize the biases and variances of the limit distribution in \ref{subsec:conv_dist}, which is followed by our final result on tail-averaging and extrapolation in Section \ref{subsec:conv_avg_extra}.

\subsection{Convergence in MSE}
\label{subsec:conv_mse}
We analyze the MSE convergence of linear TTSA in terms of the centered iterates $\bar{x}_t := x_t - x^*$, $\bar{y}_t := y_t - y^*(x_t)$. To this end, we first rewrite the stochastic recursion as the following:
\begin{lemma}
    \label{lemma:TTSA_diff_rewrite}
    Let $\bar{x}_t = x_t - x^*$, $\bar{y}_t = y_t - y^*(x_t)$. Then equation \eqref{eq:basic_ttsa_equation} can be rewritten as:
    \ifarxiv
    \begin{align}
        \bar{x}_{t+1} &= (I - \alpha \Delta)\bar{x}_t - \alpha J_{12} \bar{y}_{t} - \alpha w^x(x_t,y_t;\xi_t) \nonumber \\
        \bar{y}_{t+1} &= (I - \beta J_{22}) \bar{y}_t - \beta w^y(x_t,y_t;\xi_t) - \alpha J_{22}^{-1} J_{21} (J_{12} \bar{y}_t + \Delta \bar{x}_t + w^x(x_t,y_t;\xi_t)) \label{eq:new_recursion}
    \end{align}
    \else
    \begin{align}
        \bar{x}_{t+1} &= (I - \alpha \Delta)\bar{x}_t - \alpha J_{12} \bar{y}_{t} - \alpha w^x(x_t,y_t;\xi_t) \nonumber \\
        \bar{y}_{t+1} &= (I - \beta J_{22}) \bar{y}_t - \beta w^y(x_t,y_t;\xi_t) \nonumber \\
        &- \alpha J_{22}^{-1} J_{21} (J_{12} \bar{y}_t + \Delta \bar{x}_t + w^x(x_t,y_t;\xi_t)) \label{eq:new_recursion}
    \end{align}
    \fi
\end{lemma}
Note that the slower iterates view the error in faster iterates as an additional noise. We are now ready to state our first main convergence theorem with constant step-sizes.
\begin{theorem}
    \label{theorem:MSE_convergence}
    Suppose Assumptions \ref{assumption:linear_Hurwitz}-\ref{assumption:noise_norm_bound} hold, and the step sizes $\alpha,\beta$ satisfy Assumption \ref{assumption:stepsize}.
    Then, for all $t \ge 0$ following the TTSA recursion \eqref{eq:basic_ttsa_equation}, we have
    \ifarxiv
    \begin{align*}
        \Exs[\|\bar{x}_t\|_{Q_x}^2] &\le \exp(-\alpha \mu_x t/4) V_0 + O_{\texttt{P}}(\alpha \sigma_x^2 + (\alpha + \beta^2) \sigma_y^2), \\
        \Exs[\|\bar{y}_t\|_{Q_y}^2] &\le \exp(-\beta \mu_y t/2) U_0 +  O_{\texttt{P}}(\beta) \exp(-\alpha \mu_x t/4) V_0 + O_{\texttt{P}} ((\alpha/\beta + \beta) \alpha \sigma_x^2 + \beta \sigma_y^2),
    \end{align*}
    \else
    \begin{align*}
        \Exs[\|\bar{x}_t\|_{Q_x}^2] &\le \exp(-\alpha \mu_x t/4) V_0 + O_{\texttt{P}}(\alpha \sigma_x^2 + (\alpha + \beta^2) \sigma_y^2), \\
        \Exs[\|\bar{y}_t\|_{Q_y}^2] &\le \exp(-\beta \mu_y t/2) U_0 +  O_{\texttt{P}}(\beta) \exp(-\alpha \mu_x t/4) V_0 \\
        &\ + O_{\texttt{P}} ((\alpha/\beta + \beta) \alpha \sigma_x^2 + \beta \sigma_y^2),
    \end{align*}
    \fi
    where we define potential functions as
    \ifarxiv
    \begin{align*}
        U_0 &:= \Exs[\|\bar{y}_0\|_{Q_y}^2] + O_{\texttt{P}}( \frac{\alpha}{\beta}) \|Q_{y}^{1/2} \Exs[\bar{y}_0 \bar{x}_0^\top]\|_1, \\
        V_0 &:= \Exs[\|\bar{x}_0\|_{Q_x}^2] + O_{\texttt{P}} ( \frac{\alpha^2}{\beta^2} ) \Exs[\|\bar{y}_0\|_{Q_y}^2] + O_{\texttt{P}} (\frac{\alpha}{\beta}) \|Q_{y}^{1/2} \Exs[\bar{y}_0 \bar{x}_0^\top]\|_1.
    \end{align*}
    \else
    $U_0 := \Exs[\|\bar{y}_0\|_{Q_y}^2] + O_{\texttt{P}}( \frac{\alpha}{\beta}) \|Q_{y}^{1/2} \Exs[\bar{y}_0 \bar{x}_0^\top]\|_1$, and $V_0 := \Exs[\|\bar{x}_0\|_{Q_x}^2] + O_{\texttt{P}} ( \frac{\alpha^2}{\beta^2} ) \Exs[\|\bar{y}_0\|_{Q_y}^2] + O_{\texttt{P}} (\frac{\alpha}{\beta}) \|Q_{y}^{1/2} \Exs[\bar{y}_0 \bar{x}_0^\top]\|_1$.
    \fi
\end{theorem}
The theorem states that after sufficiently large iterations $t \gg \alpha^{-1}$, the convergence of TTSA in MSE can be characterized as the following:
\begin{enumerate}
    \item $\lim_{t\rightarrow \infty} \Exs[\|\bar{x}_t\|_{Q_x}^2] = O_{\texttt{P}} (\alpha)\sigma_x^2 + O_{\texttt{P}}(\alpha + \beta^2) \sigma_y^2$.
        \item $\lim_{t\rightarrow \infty} \Exs[\|\bar{y}_t\|_{Q_y}^2] = O_{\texttt{P}} \left(\frac{\alpha^2}{\beta} + \alpha \beta\right) \sigma_x^2 + O_{\texttt{P}}(\beta) \sigma_y^2$.
\end{enumerate}
To our best knowledge, this is the first result that explicitly characterizes the fine-grained scaling of MSE w.r.t.\ the stepsizes and noise variances of each iteration. The work in \cite{dalal2018finite, srikant2019finite} only obtained an $O(\beta^2/\alpha)$ asymptotic bound for the slower iterate. More recent work in  \cite{kaledin2020finite, haque2023tight} obtained an $O(\alpha)$ asymptotic bound but required $\beta^2 \le \alpha$, hence not strong enough to reveal the dependence on $\beta$. Our result shows that noises from slower iterates only change $x_t$ by $O(\alpha)$, while noises from faster iterates influence $x_t$ by $O(\alpha+\beta^2)$, without requiring $\beta^2 \le \alpha$. 


\subsection{Convergence to a Limit Distribution}
\label{subsec:conv_dist}
Now we state the distributional convergence of the process $(x_t,y_t,\xi_t)$ in Wasserstein distance as defined in Definition \ref{def:joint_wasserstein}. We require a mild assumption on the fourth-order moments of initial distributions:
\begin{assumption}
    \label{assumption:bounded_fourth_order}
    We assume that the fourth-order moments of the initial distribution are bounded, i.e., 
    $
        \Exs[\|\bar{x}_0\|_2^4 + \|\bar{y}_0\|_2^4] < \infty.
    $
\end{assumption} 
Our main theorem establishes the linear convergence of the Markovian process $(x_t,y_t, \xi_t)_{t \ge 0}$ in $\bar{\mathcal{W}}_2$-distance to a unique stationary distribution:
\begin{theorem}
\label{theorem:distribution_conv}
    Suppose Assumptions \ref{assumption:linear_Hurwitz}-\ref{assumption:noise_norm_bound} hold, and step sizes $\alpha,\beta$ satisfy Assumption \ref{assumption:stepsize}. If we start from an arbitrary initial distribution $(x_0, y_0, \xi_0) \sim \mu_0$ satisfying Assumption \ref{assumption:bounded_fourth_order}, then there exists a unique stationary distribution $\mu$ such that the process $(x_t, y_t, \xi_t) \sim \mu_t$ linearly converges in $\bar{\mathcal{W}}_2$-distance:
    \begin{align*}
        \bar{\mathcal{W}}_2^2 (\mu_t, \mu) \le O_{\texttt{P}} (1) \cdot \exp(-\alpha \mu_x t / 8).
    \end{align*}
    Furthermore, there exists vectors $\bar{b}_{i}^x, \bar{b}_{i}^y$ independent of $\alpha,\beta$ with $\|\bar{b}_i^x\|_2, \|\bar{b}_i^y\|_2 = O_{\texttt{P}}(1)$ for $i\in\{1,2\}$, such that for $(x_{\infty}, y_{\infty}, \xi_{\infty}) \sim \mu$, 
    \begin{equation}
    \label{eq:bias}
    \begin{aligned}
        \Exs[x_{\infty} - x^*] &= \alpha \bar{b}_1^x + \beta \bar{b}_{2}^x + O_{\texttt{P}}(\beta^2), \\
        \Exs[y_{\infty} - y^*(x_{\infty})] &= \alpha \bar{b}_1^y + \beta \bar{b}_2^y + O_{\texttt{P}}(\beta^2),
    \end{aligned}
    \end{equation}
    and variances of $x_{\infty}$ and $y_{\infty}$ are bounded by
    \begin{equation}
    \label{eq:variance}
    \begin{aligned}
        \Tr(\Var(x_{\infty})) = O_{\texttt{P}} (\alpha), \ \Tr(\Var(y_{\infty})) = O_{\texttt{P}} (\beta). 
    \end{aligned}
    \end{equation}
\end{theorem}
A few remarks follow below. First, the theorem states that any sequence following TTSA \eqref{eq:basic_ttsa_equation} converges to some unique stationary distribution depending on problem instances and step sizes. Given the existence of the unique stationary distribution $\mu$, henceforth, we can define random variables from the limit distribution $(x_{\infty}, y_{\infty}, \xi_{\infty}) \sim \mu$.

Second, the limit distribution has a bias, whose dominating term grows linearly with the stepsizes. The $\beta$-wise growth in the bias of faster iterates $y_t$ is not surprising in light of known results for Markovian single-timescale SA \citep{lauand2023curse,huo2023bias}. More interesting is the bias of the slower iterates $x_t$, which also grows linearly with $\beta$, even though the size of the update is only $O(\alpha)$ in each slow iteration. This is a unique phenomenon of two-timescale SA: the slower iterate effectively views the error from faster iterates, $y_t - y^*(x_t)$, as additional ``biased'' noise. 

Finally, the theorem shows that the limit distribution of slower iterates has an interesting property: the bias in $x$ (slower iterates) is dominated by the faster step-size $\beta$, while its variance only scales with the slower step-size $\alpha$. This is another key property of two-timescale SA that has been overlooked in prior work. In particular, we can deduce that the asymptotic MSE of slower iterates is resulted from two factors:
\begin{align*}
    \Exs[\|x_{\infty}-x^*\|_{2}^2] \;\; \asymp \;\; \underbrace{\alpha (\sigma_x^2+\sigma_y^2)}_{\text{variance}} \;\; + \!\! \underbrace{\beta^2 \sigma_y^2}_{\text{squared bias}}
\end{align*}
Focusing separately on the two iterates, we have the following more fine-grained convergence results:
\begin{corollary}    \label{theorem:fine_grained_distributional_convergence_rate}
    Suppose Assumptions \ref{assumption:linear_Hurwitz}-\ref{assumption:bounded_fourth_order} hold. Define
    $
       U_0 := \Exs[\|x_0 - \Exs[x_{\infty}]\|_{2}^2] + \Exs[\|\bar{y}_0 - \Exs[\bar{y}_{\infty}]\|_{2}^2] + O_{\texttt{P}} (\beta), 
    $ 
    and 
    $
        V_0 := \Exs[\|x_0 - \Exs[x_{\infty}]\|_{2}^2] + O_{\texttt{P}} (\alpha/\beta) U_0.
    $
    Then for all $t \ge 0$, we have the bounds
    \ifarxiv
    \begin{equation*}
    \begin{aligned}
        \mathcal{W}_2^2 (\mu_t(\bar{y}_t), \mu(\bar{y}_{\infty})) &\le  O_{\texttt{P}} (\beta) \exp(-\alpha \mu_x t / 8) V_0 + O_{\texttt{P}}(1) \exp(-\beta \mu_y t / 8) U_0,  \\
        \mathcal{W}_2^2 (\mu_t(x_t), \mu(x_{\infty})) &\le  O_{\texttt{P}} (1) \exp(-\alpha \mu_x t / 8) V_0. 
    \end{aligned}
    \end{equation*}
    \else
    \begin{equation*}
    \begin{aligned}
        \mathcal{W}_2^2 (\mu_t(\bar{y}_t), \mu(\bar{y}_{\infty})) &\le  O_{\texttt{P}} (\beta) \exp(-\alpha \mu_x t / 8) V_0 \\
        &\quad + O_{\texttt{P}}(1) \exp(-\beta \mu_y t / 8) U_0,  \\
        \mathcal{W}_2^2 (\mu_t(x_t), \mu(x_{\infty})) &\le  O_{\texttt{P}} (1) \exp(-\alpha \mu_x t / 8) V_0. 
    \end{aligned}
    \end{equation*}
    \fi
\end{corollary}
This corollary explicitly states how the optimization error decays from arbitrary initial points, and will be used in showing the convergence of tail-averaging next. 


\subsection{Tail-Averaging and Extrapolation}
\label{subsec:conv_avg_extra}
Using the explicit characterization of bias and variance in Theorem \ref{theorem:distribution_conv}, we derive improved convergence rates for tail-averaging and extrapolation.

\subsubsection{Averaging}
We first consider the tail-averaging variant of Polyak-Ruppert averaging \citep{jain2018parallelizing}:
\begin{align}
    \tilde{x}_{t} := \frac{1}{t - t_0} \sum_{t'=t_0}^t x_{t'},  \ \tilde{y}_{t} := \frac{1}{t - t_0} \sum_{t'=t_0}^t y_{t'},
    \label{eq:tail_averaging}
\end{align}
where $t_0 \gtrsim \alpha^{-1}$ is the length of the warm-up period. With the result from Theorem~\ref{theorem:MSE_convergence}, we can analyze the MSE of tail-averaged sequence: 
\begin{theorem}
\label{theorem:tail_averaging}
    Suppose Assumptions \ref{assumption:linear_Hurwitz}-\ref{assumption:bounded_fourth_order} hold and $t_0 > C(\alpha\mu_x)^{-1}$ for some sufficiently large absolute constant $C > 0$. Then for all $t > t_0$
    \begin{align*}
        \Exs[\|\tilde{x}_t - x^*\|_{2}^2] &= O_{\texttt{P}}(\beta^2) + \frac{O_{\texttt{P}}(1)}{t-t_0}, \\
        \Exs[\|\tilde{y}_t - y^*\|_{2}^2] &= O_{\texttt{P}}(\beta^2) + \frac{O_{\texttt{P}} \left( 1 + \sqrt{\beta^2/\alpha} \right)}{t-t_0}. 
    \end{align*}
\end{theorem}
In the above result, we omitted an additional optimization error $\exp(-c \alpha\mu_x t_0)$ since it is dominated by other terms with $t_0 \gg 1/(\alpha\mu_x)$. 
As we can observe, $O(\beta^2)$ is attribted to the squared-bias, and $O(1/t)$ convergence is the variance decaying effect of tail-averaging. We also observe that the faster iterates has extra $O(\frac{1}{t}\sqrt{\beta^2/\alpha})$-term. In part, this is because we measure the MSE of $\tilde{y}_t$ from $y^* = y^*(x^*)$, not from $y^*(\tilde{x}_t)$. However, we are not fully aware whether this is an artifact of an analysis, or can be removed, and we leave the question as an open problem. Note that when $\beta^2 \le \alpha$, both iterates enjoy the same $O(1/t)$-decaying rate of variances as if the two iterates are decoupled.

\begin{figure}
\centering
\begin{subfigure}{.45\textwidth}
  \centering
  \includegraphics[width=.9\linewidth,height=0.55\linewidth]{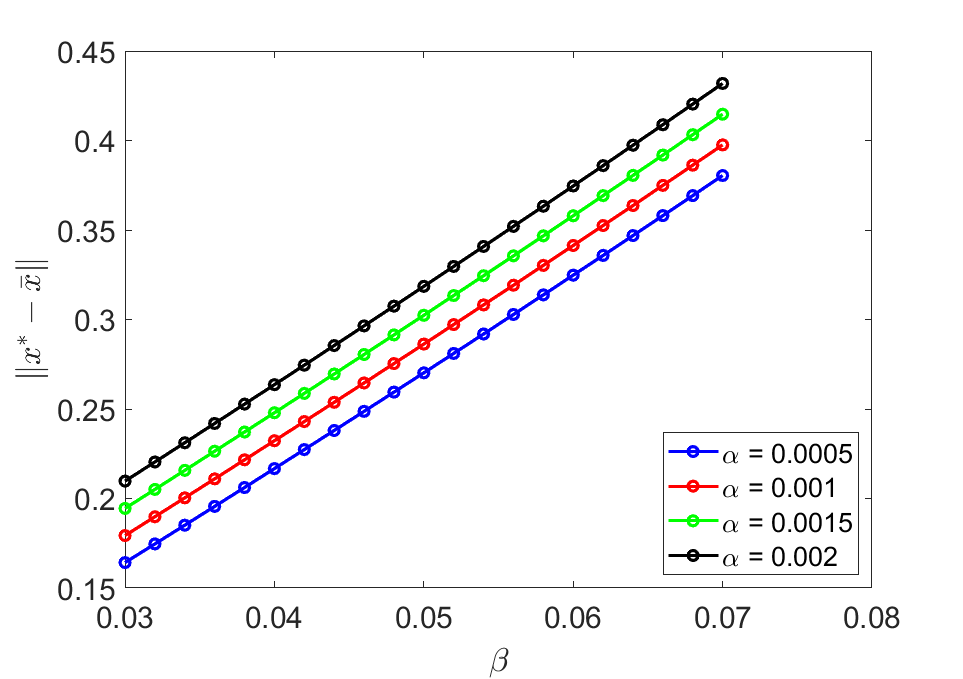}
  \label{fig1:sfig1}
\end{subfigure}
\begin{subfigure}{.45\textwidth}
  \centering
  \includegraphics[width=.9\linewidth,height=0.55\linewidth]{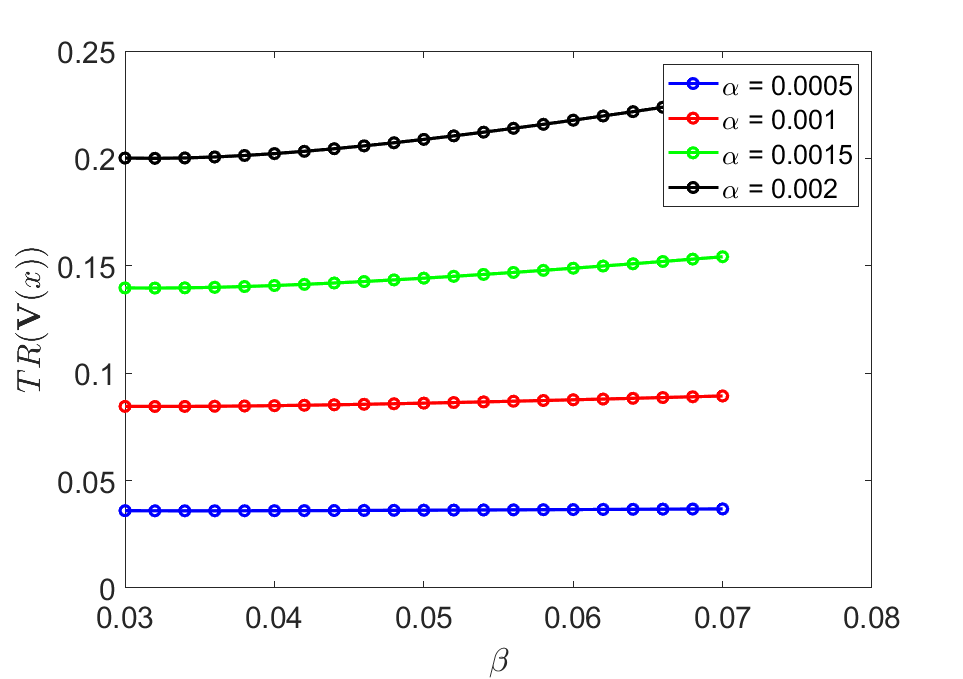}
  \label{fig1:sfig2}
\end{subfigure}
\caption{Bias (top) and variance (bottom) versus $\beta$ at different $\alpha$ for the slower iterate $x_t$.}
\label{fig:fig1}
\end{figure}

\begin{figure}
\centering
\begin{subfigure}{.45\textwidth}
  \centering
  \includegraphics[width=.9\linewidth,height=0.55\linewidth]{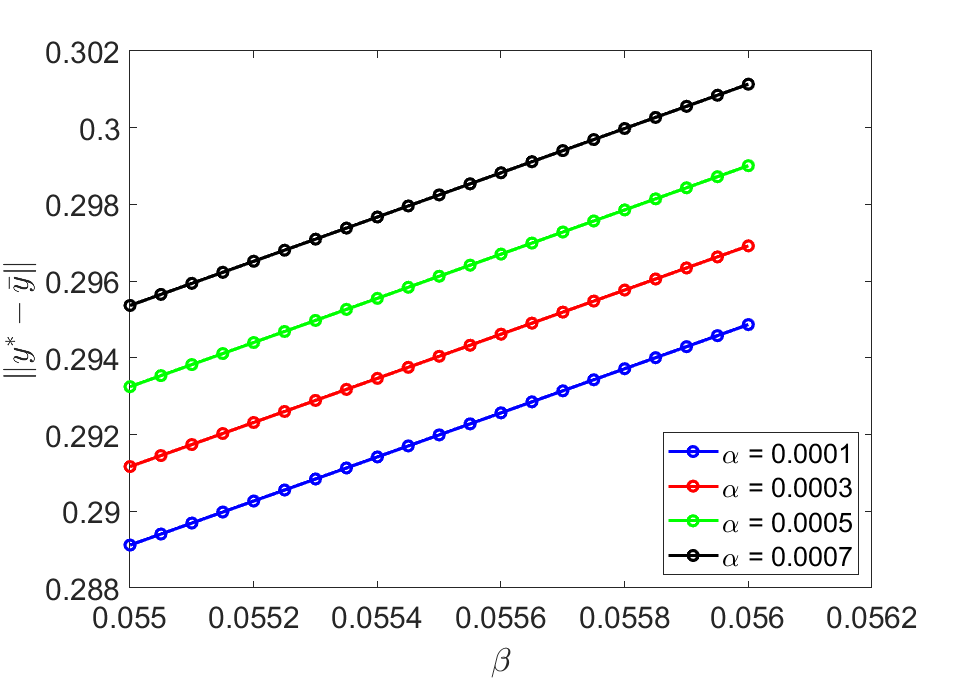}
  \label{fig2:sfig1}
\end{subfigure}
\begin{subfigure}{.45\textwidth}
  \centering
  \includegraphics[width=.9\linewidth,height=0.55\linewidth]{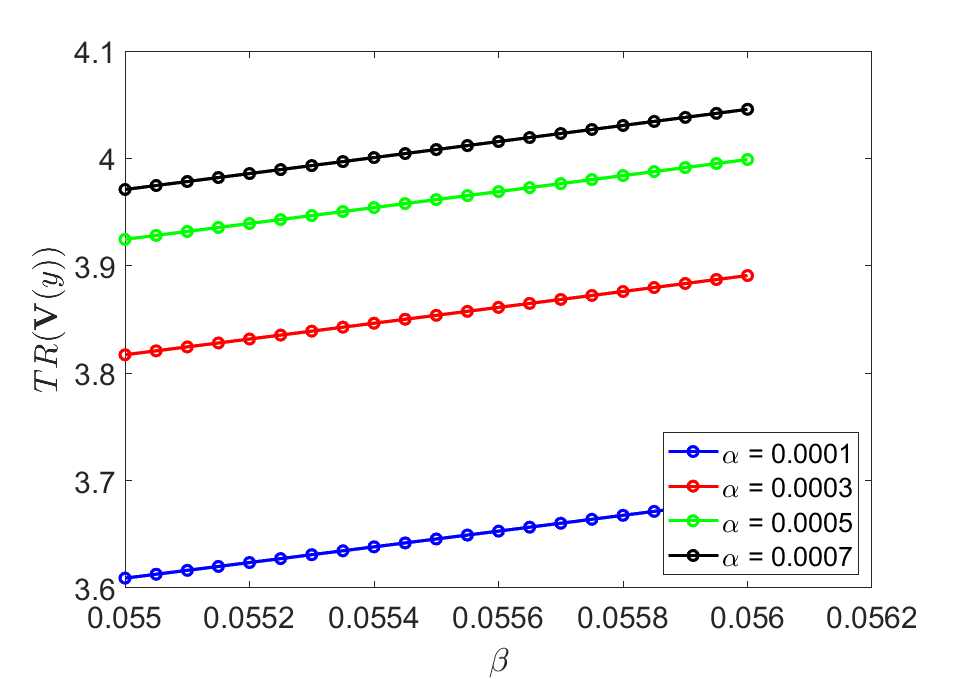}
  \label{fig2:sfig2}
\end{subfigure}
\caption{Bias (top) and variance (bottom) versus $\beta$ at different $\alpha$ for the faster iterate $y_t$.}
\label{fig:fig2}
\end{figure}

\subsubsection{Extrapolation}
When tail-averaging can reduce the variance, extrapolation can reduce the biases of each iterate. As one example, using the fact that biases of iterates grow linearly with step sizes, we can extrapolate two sequences, $(x_t^{\alpha,\beta}, y_{t}^{\alpha,\beta})$ and $(x_t^{2\alpha,2\beta}, y_{t}^{2\alpha,2\beta})$ with pairing stepsizes $(\alpha,\beta)$ and $(2\alpha,2\beta)$. The extrapolated iterates are computed as
\begin{align*}
    \zeta_t^x := 2 \tilde{x}_t^{\alpha,\beta} - \tilde{x}_t^{2\alpha, 2\beta}, \ \zeta_t^y := 2 \tilde{y}_t^{\alpha,\beta} - \tilde{y}_t^{2\alpha, 2\beta}.
\end{align*}

As a corollary of our main theorems, we have the following result characterizing the MSE of the extrapolated sequences. Extrapolation achieves reduced biases by canceling out the leading $\alpha$ and $\beta$ terms in the asymptotic biases~\eqref{eq:bias}, improving the MSE bounds of both iterates from $\beta^2$ to $\beta^4$. 
\begin{corollary}
    \label{corollary:extrapolation}
    Suppose Assumptions \ref{assumption:linear_Hurwitz}-\ref{assumption:bounded_fourth_order} hold and $t_0 > C(\alpha\mu_x)^{-1}$ for some sufficiently large absolute constant $C > 0$. Then for all $t > t_0$,
    \begin{align*}
        \Exs[\|\zeta_t^x - x^*\|_{2}^2] &= O_{\texttt{P}} (\beta^4) + \frac{O_{\texttt{P}} (1)}{t-t_0}, \\
        \Exs[\|\zeta_t^y - y^*\|_{2}^2] &= O_{\texttt{P}}(\beta^4) + \frac{O_{\texttt{P}} \left(1 + \sqrt{\beta^2/\alpha}\right)}{t-t_0}.
    \end{align*}
\end{corollary}


\begin{remark}
    If one uses pairing stepsizes $(\alpha,\beta)$ and $(\alpha,2\beta)$, then only the leading $\beta$ terms in the asymptotic biases~\eqref{eq:bias} are cancelled.
\end{remark}
\begin{remark}
It is possible to further reduce the order of bias via higher-order extrapolation using more than two sets of stepsizes as in \cite{huo2023bias,huo2024inference}, though it comes at the price of potentially slower convergence and higher variance due to using additional stepsizes \citep{durmus2021tight, srikant2019finite}.
\end{remark}

\section{Experiments}

\begin{figure*}
\centering
\begin{subfigure}{.46\textwidth}
  \centering
  \includegraphics[width=\linewidth]{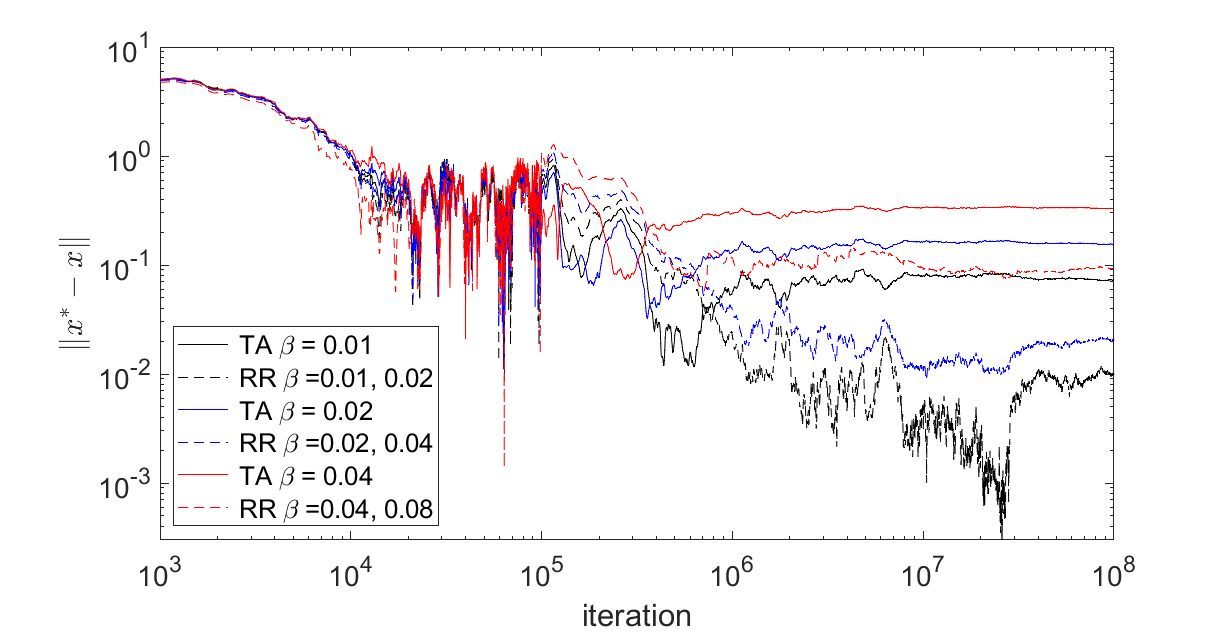}
  \caption{Absolute error in the slower timescale.}
  \label{fig5:sfig1}
\end{subfigure}
\begin{subfigure}{.46\textwidth}
  \centering
  \includegraphics[width=\linewidth]{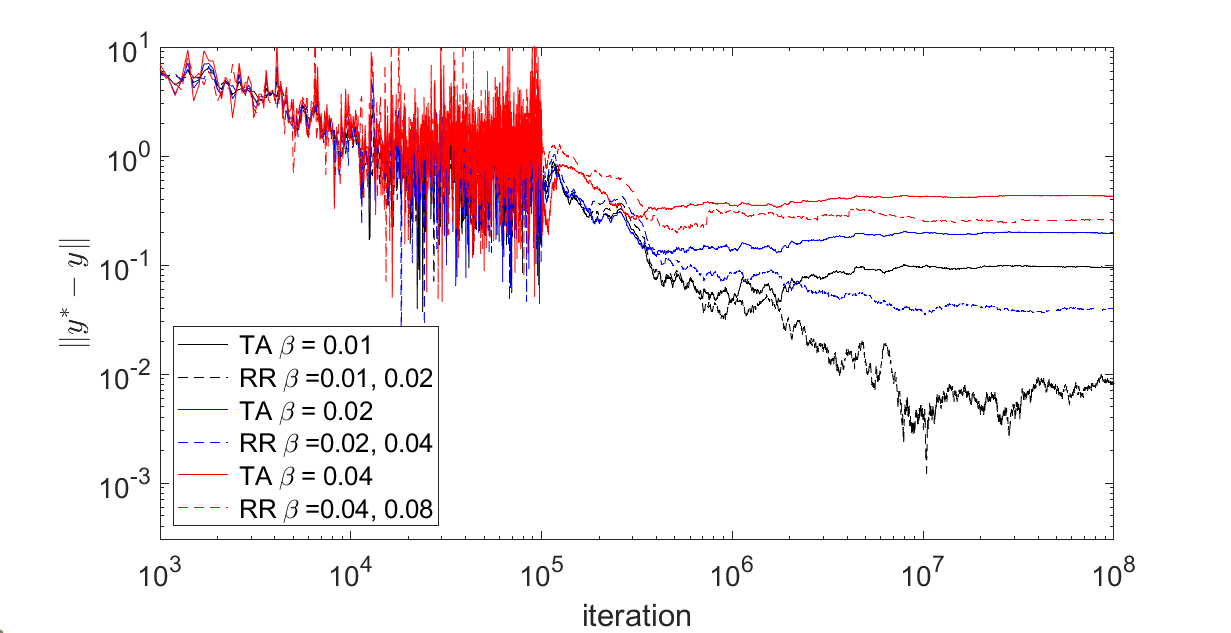}
  \caption{Absolute error in the faster timescale.}
  \label{fig5:sfig2}
\end{subfigure}
\caption{Comparison of Tail-Averaging (TA) and Richard-Romberg (RR) extrapolation in $\beta$.}
\label{fig:fig5}
\end{figure*}

\begin{figure*}
\centering
\begin{subfigure}{.46\textwidth}
  \centering
  \includegraphics[width=\linewidth]{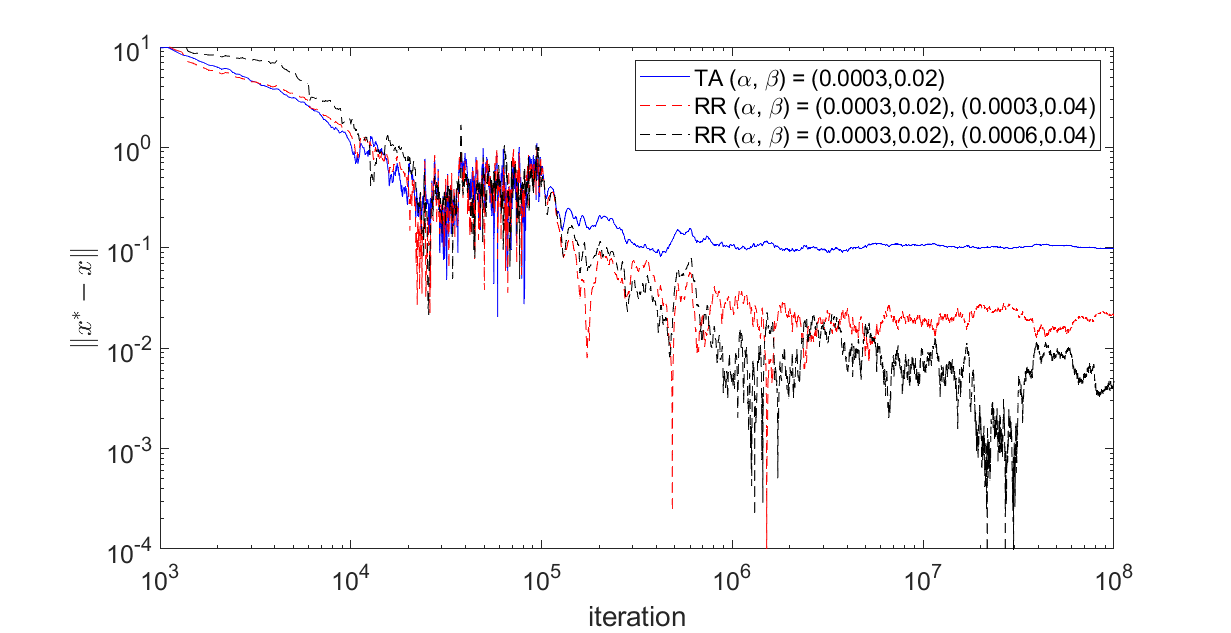}
  \caption{Absolute error in the slower timescale.}
  \label{fig6:sfig1}
\end{subfigure}
\begin{subfigure}{.46\textwidth}
  \centering
  \includegraphics[width=\linewidth]{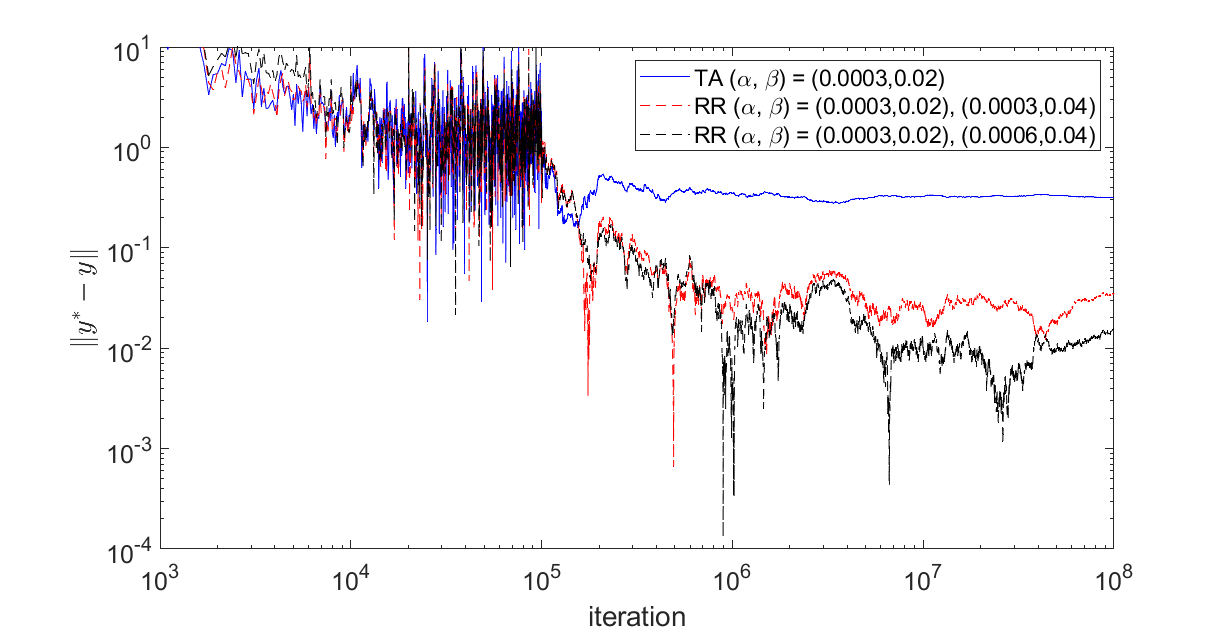}
  \caption{Absolute error in faster timescale.}
  \label{fig6:sfig2}
\end{subfigure}
\caption{Comparison of Tail-Averaging (TA), RR extrapolation in $\beta$, and RR extrapolation in both $\beta$ and $\alpha$.}
\label{fig:fig6}
\end{figure*}

We consider the TTSA iteration \eqref{eq:basic_ttsa_equation} in dimension $d_x = d_y = 2$ driven by a 10-state, irreducible, aperiodic Markov chain. We construct the transition matrix randomly and choose $J_{11}, J_{12}, J_{21}, J_{22}$ such that Assumption \ref{assumption:linear_Hurwitz} hold.

We tested the dependence of the bias and variance of both iterates with respect to $\alpha$ and $\beta$ by varying each individually. After the tail-averaged iterates converged, we calculated the bias as the average distance between the averaged iterate and the true solution, and calculated the variance $\Tr\Var(\cdot)$ as the average square distance from the iterate to the sample mean of the iterates.
For the dependence on $\beta$, we held $\alpha$ constant and varied $\beta$ between $0.03$ and $0.07$.  For the dependence on $\alpha$, we held $\beta$ constant and varied $\alpha$ between $0.0001$ and $0.0005$. 

For the slower iterate $x_t$, Figure~\ref{fig:fig1} shows that the bias scales with both $\beta$ and $\alpha$, while the variance is dependent mostly on $\alpha$ only. For the faster iterate $y_t$, Figure~\ref{fig:fig2}  shows that  the bias depends both on $\beta$ and $\alpha$, and the variance depends on $\beta$. Both results are consistent with our theory.

We also tested the effects of tail-averaging (TA) and Richardson-Romberg (RR) extrapolation with a similar setup. We fixed $\alpha = 0.0003$ and let $\beta = \{ 0.01, 0.02, 0.04, 0.08\}$. In Figure \ref{fig:fig5}, for each $\beta$, we plotted the absolute errors achieved by tail-averaging at stepsize $\beta$ (labeled as ``TA $\beta=$ stepsize''), as well as the errors achieved by RR extrapolation with stepsizes $\beta$ and $2 \beta$ (labeled as ``RR $\beta=$ stepsize, 2$*$stepsize''), which aims to cancel the $\beta$ term in the bias. Compared to the TA iterates (solid line), the corresponding RR extrapolated iterate (the dashed line of the same color) achieved lower errors, corresponding to reduced asymptotic biases.

In addition, we examined the effectiveness of applying RR extrapolation to cancel \emph{both} the $\alpha$ and $\beta$ bias terms. Letting $\alpha = 0.0003, \beta = 0.02$, we compared RR extrapolating on only $\beta$ (using stepsizes $\beta, \alpha$ and $2\beta, \alpha$) with RR extrapolating on both $\beta$ and $\alpha$ (using stepsizes $\beta, \alpha$ and $2\beta, 2\alpha$). In Figure~\ref{fig:fig6}, we see that while the former (red curves) already reduced a large amount of the bias, the latter (black curves) reduced it even further, as predicted by our theoretical results.

\section{Analysis}

We outline the proofs of our main theorems. We focus on  slower iterates; similar ideas apply to faster iterates.

\subsection{Proof Outline of Theorem \ref{theorem:MSE_convergence}}
The first step is to analyze the descent formula for each iterate separately. For the slower iterate, we have
\ifarxiv
\begin{align*}
    \Exs[\|\bar{x}_{t+1}\|^2_{Q_x}] &= \Exs[\|(I-\alpha\Delta)\bar{x}_t\|_{Q_x}^2] + 2\alpha \underbrace{\Exs[\vdot{\bar{x}_t}{w_t^x}_{Q_x}]}_{T_1} + 2\alpha \underbrace{\Exs[\vdot{\bar{x}_t}{J_{12} \bar{y}_t}_{Q_x}]}_{T_2}+ o(\alpha).
\end{align*}
\else
\begin{align*}
    \Exs[\|\bar{x}_{t+1}\|^2_{Q_x}] &= \Exs[\|(I-\alpha\Delta)\bar{x}_t\|_{Q_x}^2] + 2\alpha \underbrace{\Exs[\vdot{\bar{x}_t}{w_t^x}_{Q_x}]}_{T_1}  \\
    &\quad + 2\alpha \underbrace{\Exs[\vdot{\bar{x}_t}{J_{12} \bar{y}_t}_{Q_x}]}_{T_2}+ o(\alpha).
\end{align*}
\fi
The term $T_1$ would have been 0 if the noise sequence were martingale, 
and can be effectively handled with Markovian noises in a standard way by exploiting Assumption \ref{assumption:noise_field}. More pressing issue is handling $T_2$: with naively applying Young's inequality to bound $(ii)$, {\it i.e.,} with $\vdot{\bar{x}_t}{J_{12} \bar{y}_t} \le (c \|\bar{x}_t\|^2 + \frac{J_{\max}}{4c} \|\bar{y}_t\|^2)$, the asymptotic error easily end up being $O(\beta^2/\alpha)$ as in \cite{dalal2018finite, gupta2019finite}, and such an approach can be improved up to at best $O(\beta)$ \citep{doan2022nonlinear}. 

Recent results in \cite{kaledin2020finite, haque2023tight} directly analyzed the descent behavior of $\|T_2\|_{\op}$, 
achieving $O(\alpha)$ asymptotic error for the slower iterate. However, using operator norm often results in extra dependence on dimensions $d_x, d_y$, despite the smoothness condition $J_{\max} = O(1)$ in operator norm. 

Our tweak for this issue is simple: to track the convergence of cross-correlation norm, we employ the Schatten $S^1$-measure for $\|Q_{y}^{1/2} \Exs[\bar{y}_t \bar{x}_t^\top]\|_1$, where $Q_{y}^{1/2}$ term is incorporated to ensure decreasing Lyapunov potential with asymmetric operators. The $S^1$-norm is the best suited for exploiting the smoothness condition without incurring dimension dependence, thanks to the Holder's inequality for matrix Schattern norm:
\begin{align*}
    \|AB\|_1 \le \|A\|_1 \|B\|_{\infty} = \|A\|_1 \|B\|_{\op}.
\end{align*}
Leveraging this property, we can construct the potential function as the sum of three terms (omitting constants):
\begin{align*}
    \Exs[\|\bar{x}_t\|_{Q_x}^2] +\frac{\alpha(\alpha+\beta^2)}{\beta}  \Exs[\|\bar{y}_t\|_{Q_y}^2] + \frac{\alpha}{\beta}\|Q_{y}^{1/2} \Exs[\bar{y}_t \bar{x}_t^\top]\|_1.
\end{align*}
With similar techniques for analyzing the faster iterates and cross-correlation norms, we can obtain a clean $O(\alpha)$ asymptotic error without additional dimension dependence. The full proof is given in Appendix~\ref{appendix:main_theorems}.

\subsection{Proof Outline of Theorem \ref{theorem:distribution_conv}}

Once we have the MSE convergence result, extending the strategies in prior work for the single-timescale SA \citep{dieuleveut2020bridging, huo2023bias}, we first consider two coupling sequences via sharing the common noise sequence $(x_t^1, y_t^1, \xi_t)$ and $(x_t^2, y_t^2, \xi_t)$. The idea is to show that the coupled sequences $\delta_t^x := \bar{x}_t^1 - \bar{x}_t^2$, $\delta_y^y := \bar{y}_t^1 - \bar{y}_t^2$ converge linearly (Lemma \ref{lemma:noise_coupled_convergence}), 
\begin{align*}
    \Exs[\|\delta_t^x\|_{2}^2 ] \lesssim \exp(-c \alpha t) \cdot \Exs\left[\|\delta_0^x\|_{2}^2 + \frac{\alpha}{\beta}\|\delta_0^y\|_{2}^2 \right] .
\end{align*}
Then we can design two sequences coupled in such a way that
$
    (x_t^2, y_t^2, \xi_t) \stackrel{d}{=} (x_{t+1}^1, y_{t+1}^1, \xi_{t+1}). 
$
Combining the two results, the sequence $(x_t^1, y_t^1, \xi_t)$ converges in $L_2$-Wasserstein distribution to a unique stationary point. The remaining details can be found in Appendix \ref{appendix:distributional_convergence}.

\paragraph{Bias and Variance}

Turning to the stationary distributions of the iterates, we observe that $x_\infty$ satisfies 
\begin{align*}
    &\bar{x}_{\infty+1} = (I - \alpha \Delta) \bar{x}_{\infty} - \alpha J_{12} \bar{y}_{\infty} - \alpha w_{\infty}^x, \\
    &\Exs[\bar{x}_{\infty+1}| \xi_{\infty+1} = \xi] = \Exs[\bar{x}_{\infty}| \xi_{\infty} = \xi], \quad \forall \xi \in \Xi.
\end{align*}
Conditioned on the event $\xi_{\infty+1} = \xi$, we have $\xi_{\infty} \sim P^{\dagger}(\cdot | \xi)$, where $\mathcal{P}^{\dagger}$ is the adjoint of the transition kernel $\mathcal{P}$. Using this relation, we can construct a stationary equation for $\Exs[\bar{x}_{\infty}| \xi_{\infty} = \xi]$, and find the explicit expression for biases by integrating the conditional expectation over a stationary distribution $\pi$, {\it i.e.,}
\begin{align*}
    \Exs[\bar{x}_{\infty}] = \int_{\Xi} \Exs[\bar{x}_{\infty} | \xi_{\infty} = \xi] \, \mathrm{d}\pi(\xi) = \alpha \bar{b}_1^x + \beta \bar{b}_2^x + O(\alpha\beta).
\end{align*}

The variance of $\bar{y}_{\infty}$ is relatively simple to bound: 
\begin{align*}
    \Tr(\Var(\bar{y}_{\infty})) \le \Exs[\|\bar{y}_{\infty} - y^*(\bar{x}_{\infty}) \|_2^2] \le O(\beta).
\end{align*}
However, showing the variance upper bound $O(\alpha)$ can not be derived in the same fashion since the MSE bound for $\bar{x}_{\infty}$ is $O(\alpha + \beta^2)$. To derive this, we also construct a stationary equation for the covariance:
\begin{align*}
    \Exs[\bar{x}_{\infty+1}\bar{x}_{\infty+1}^\top| \xi_{\infty+1} = \xi] = \Exs[\bar{x}_{\infty}\bar{x}_{\infty}^\top | \xi_{\infty} = \xi], \ \forall \xi \in \Xi,
\end{align*}
and show that $S^1$-norm of the above is  $O(\alpha)$. Using the inequality $\Tr(A) \le \|A\|_1$ completes the proof.

\bibliographystyle{abbrv}
\bibliography{ref}

\appendix

\begin{appendices}

\section{Technical Lemmas}

\begin{lemma}
    \label{lemma:tr_holder_bound}
    For any two real matrices $A, B$, we have
    \begin{align*}
        \Tr(A^\top B) &\le \|A^\top B\|_1 \le \|A\|_{\infty} \|B\|_{1} = \|A\|_{\op} \|B\|_1. 
    \end{align*}
\end{lemma}
\begin{lemma}
    \label{lemma:operator_to_Q_norm_conversion}
    For a positive definite matrix $Q \succ 0$ and any real matrix $A$, the following holds:
    \begin{align*}
        \|A\|_Q = \|Q^{1/2} A Q^{-1/2}\|_{\op}.
    \end{align*}
\end{lemma}

\begin{lemma}
    \label{lemma:basic_operator_Qnorm_conversion}
    For a positive definite matrix $Q \succ 0$, and for any vectors $x,y$ and a matrix $M$, 
    \begin{align*}
        \vdot{x}{y}_Q &\le \|x\|_Q \|y\|_Q, \ \vdot{Mx}{x}_Q \le \|M\|_{\op} \|x\|_Q^2, \\
        \|Mx\|_Q &\le \|M\|_Q \|x\|_Q \le \sqrt{\kappa(Q)} \|M\|_{\op} \|x\|_Q, 
    \end{align*}
    where $\kappa(Q) = \frac{\sigma_{\max}(Q)}{\sigma_{\min}(Q)}$ is the condition number of $Q$.
\end{lemma}

\begin{lemma}[Lemma C.13 in \cite{haque2023tight}]
    \label{lemma:basic_lyapunov_equation}
    Let $-A$ be a Hurwitz matrix and $Q$ be the solution to 
    \begin{align}
        \label{eq:Lyapunov_equation}
        A^\top Q + QA = I.
    \end{align}
    Then for all $\epsilon \in \left[0, \frac{1}{\|Q\|_{\mathrm{op}} \|A\|_{Q}^2} \right]$, for any matrix $B$, we have
    \begin{align*}
        \|(I-\epsilon A) B\|_{Q} \le (1- \mu \epsilon) \|B\|_{Q},
    \end{align*}
    where $\mu := \frac{1}{2\|Q\|_{\mathrm{op}}}$. In particular, $\|I-\epsilon A\|_Q \le 1 - \mu \epsilon$.
\end{lemma}

\begin{lemma}
    \label{lemma:basic_norm_measure_conversion}
    For any two positive definite matrices $Q_1, Q_2$ and a vector $x$, we have
    \begin{align*}
        \|x\|_{Q_1}^2 \le \frac{\sigma_{\max}(Q_1)}{\sigma_{\min} (Q_2)} \cdot \|x\|_{Q_2}^2.
    \end{align*}
\end{lemma}

\subsection{Auxiliary Lemmas}
We list some useful facts and lemmas here. 

\begin{lemma}
    \label{lemma:ergodic_noise_after_mixing}
    For any $t \ge \tau$, for all $i,j\in \{1,2\}$, we have
    \begin{align*}
        \Exs[\vdot{W_{ij}(\xi_t)}{v_{t-\tau} u_{t-\tau}^\top} | \mathcal{F}_{t-\tau}] &= O(\rho^{\tau} W_{\max} \|v_{t-\tau}\|_2\|u_{t-\tau}\|_2),\\
        \Exs[\vdot{u_i(\xi_t)}{v_{t-\tau}} | \mathcal{F}_{t-\tau}] &= O(\rho^\tau u_{\max} \|v_{t-\tau}\|_2).
    \end{align*}
    where $v_t, u_t$ are any vectors that can be constructed at the $t^{th}$ iteration. 
\end{lemma}

\begin{lemma}
    \label{lemma:wt_reexpress}
    Let two intermediate variables:
    \begin{align*}
        W_{\Delta}^x(\xi) &:= W_{11} (\xi)- W_{12}  (\xi) J_{22}^{-1} J_{21}, \\
        W_{\Delta}^y(\xi) &:= W_{21}(\xi) - W_{22} (\xi) J_{22}^{-1} J_{21}. 
    \end{align*}
    Then, $w_t^x, w_t^y$ can be rewritten as
    \begin{align*}
        w_t^x &= W_{\Delta}^x(\xi_t) \bar{x}_t + W_{12}(\xi_t) \bar{y}_t + W_{\Delta}^x(\xi_t) x^* + u_1(\xi_t), \\
        w_t^y &= W_{\Delta}^y(\xi_t) \bar{x}_t + W_{22}(\xi_t) \bar{y}_t + W_{\Delta}^y(\xi_t) x^* + u_2(\xi_t).
    \end{align*}
\end{lemma}

\begin{lemma}
\label{lemma:sequence_difference_bound}
    For any $t \ge \tau \ge \tau_\alpha$, we have
    \begin{align*}
        \|\bar{x}_t - \bar{x}_{t-\tau}\|_2 &\le 4 \alpha \tau \left(J_{\max} (\kappa_y \|\bar{x}_t\|_2 + \|\bar{y}_t\|_2) + \sigma_x + \beta J_{\max} \sigma_y \right), \\
        \|\bar{y}_t - \bar{y}_{t-\tau}\|_2 &\le 4 \beta \tau \left(J_{\max} (\kappa_y \|\bar{x}_t\|_2 + \|\bar{y}_t\|_2) + \sigma_y \right) + 4 \alpha \kappa_y \tau \sigma_x.
    \end{align*}
\end{lemma}
The following corollary is convenient:
\begin{corollary}
    \label{corollary:sequence_difference_bound}
    If $\alpha \tau \kappa_y \le c_1 \beta\tau \le  c_2 / J_{\max}$ holds with absolute constants $c_1, c_2 > 0$, then for any $t \ge \tau \ge \tau_\alpha$,
    \begin{align*}
        \|\bar{x}_{t-\tau}\|_2 &\le 2 \|\bar{x}_t\|_2 + 8 \alpha \tau J_{\max} \|\bar{y}_t\|_2 + 4 \tau \left( \alpha \sigma_x + \alpha \beta J_{\max} \sigma_y \right), \\
        \|\bar{y}_{t-\tau}\|_2 &\le 4 \beta \tau J_{\max} \kappa_y \|\bar{x}_t\|_2 + 2 \|\bar{y}_t\|_2 + 4 \tau \left( \alpha \kappa_y \sigma_x + \beta \sigma_y \right).
    \end{align*}
\end{corollary}

\begin{lemma}
    \label{lemma:noise_cross_l1_product_y}
    For any $t \ge \tau \ge c \log(\frac{\kappa_y}{\alpha\mu_y})$ with an absolute constant $c > 0$, we have 
    \begin{align*}
        \|\Exs[w_t^x \bar{y}_t^\top ]\|_1 &\le \frac{\mu_y}{8 \kappa_y} \Exs[\|\bar{y}_t\|_2^2] + O(J_{\max}^2 \kappa_y) \beta \tau \Exs[\|\bar{x}_t\|_2^2] +  O(\tau) ((\alpha^2/\beta) \kappa_y \sigma_x^2 +  \beta \sigma_y^2).
    \end{align*}
    Similarly, we can derive the same upper bound for $\|\Exs[w_t^y \bar{y}_t^\top ]\|_1$.  
\end{lemma}

\begin{lemma}
    \label{lemma:noise_cross_l1_product_x}
    For any $t \ge \tau \ge c \log(\frac{\kappa_x}{\alpha\mu_x})$ with an absolute constant $c > 0$, we have 
    \begin{align*}
        \|\Exs[w_t^x \bar{x}_t^\top ]\|_1 &\le \frac{\mu_x}{8 \kappa_x} \Exs[\|\bar{x}_t\|_2^2] + O \left( \alpha\tau J_{\max}^2 + \beta^2 \tau^2 \frac{J_{\max}^4 \kappa_x}{\mu_x} \right) \Exs[\|\bar{y}_t\|_2^2] +  O(\tau) (\alpha \sigma_x^2 + \beta^2 J_{\max} \sigma_y^2).
    \end{align*}
    Similarly, we can derive the same upper bound for $\|\Exs[w_t^y \bar{x}_t^\top ]\|_1$.  
\end{lemma}

\section{Proof of Main Theorems}
\label{appendix:main_theorems}

We recall the definition of $\sigma_x,\sigma_y$ in \eqref{eq:sigma_define}
\begin{align*}
    \sigma_x &:= \max_{\xi \in \Xi} \|u_1(\xi) + W_{\Delta}^x(\xi) x^*\|_2, \nonumber \\ 
    \sigma_y &:= \max_{\xi \in \Xi} \|u_2(\xi) + W_{\Delta}^y(\xi) x^*\|_2.
\end{align*}
Recall that we assume $\beta/\alpha \gg \kappa_y$ in Assumption \ref{assumption:stepsize}, and $\|\Delta\|_{\op} \le J_{\max} \kappa_y$.

\subsection{Proof of Theorem \ref{theorem:MSE_convergence}}
The proof first investigates the convergence of three terms $\Exs[\|\bar{y}_{t}\|_{Q_y}^2]$, $\Exs[\|\bar{x}_{t}\|_{Q_x}^2]$, $\|Q_{y}^{1/2} \Exs[\bar{y}_{t}\bar{x}_t^\top]\|_1$ separately. Then, by constructing the potential function as the following:
\begin{align}
    \label{eq:potential_functions}
    V_t &= \Exs[\|\bar{x}_t\|_{Q_x}^2] + \frac{O(1) J_{\max}^{2} \kappa_y \alpha(\alpha+\beta^2 \tau_\alpha^2 J_{\max} ) }{\mu_x \mu_y \beta} \Exs[\|\bar{y}_t\|_{Q_y}^2] + \frac{O(1) J_{\max}^{1/2} \kappa_y \alpha}{\mu_x \beta} \| Q_y^{1/2} \Exs[ \bar{y}_t \bar{x}_t^\top] \|_{1}, \nonumber \\
    U_{t} &= \Exs[\|\bar{y}_t\|_{Q_y}^2] + \frac{O(1) J_{\max}^{1/2} \kappa_y^{2.5} \alpha}{\mu_y \beta} \|Q_{y}^{1/2} \Exs[\bar{y}_t \bar{x}_t^\top]\|_{1},
\end{align}
and show that they decay in exponential rates. 

\subsubsection{Convergence of $\bar{y}_{t}$} 
We first study the descent behavior of $\bar{y}_t$:
\begin{align*}
    \Exs[\|\bar{y}_{t+1}\|_{Q_y}^2] &\le \Exs \Big[ \|(I - \beta J_{22}) \bar{y}_t \|^2_{Q_y} + \alpha^2 \| J_{22}^{-1} J_{21} (J_{12} \bar{y}_t + \Delta \bar{x}_t + w^x_t) \|^2_{Q_y} + \beta^2 \|w^y_t\|^2_{Q_y} \Big] \\
    &\qquad + 2\alpha \left| \Exs[\vdot{(I-\beta J_{22}) \bar{y}_t}{-J_{22}^{-1}J_{21}(J_{12} \bar{y}_t + \Delta \bar{x}_t + w^x_t)}_{Q_y}] \right| \\
    &\qquad + 2\beta \left| \Exs[\vdot{(I - \beta J_{22})\bar{y}_t}{-w_t^y}_{Q_y} ] \right| + 2\alpha \beta \left| \Exs[\vdot{J_{22}^{-1}J_{21} (J_{12}\bar{y}_t + \Delta\bar{x}_t - w^x_t)}{w^y_t}_{Q_y}] \right|.
\end{align*}
We bound each term:
\begin{enumerate}
    \item Using Lemma \ref{lemma:basic_lyapunov_equation}, we have
    \begin{align*}
        \|(I - \beta J_{22}) \bar{y}_t\|_{Q_y}^2 \le (1 - \mu_y \beta) \|\bar{y}_t\|_{Q_y}^2. 
    \end{align*}
    \item Using the formula in Lemma \ref{lemma:wt_reexpress}, 
    \begin{align*}
        \|w_t^y\|_{Q_y}^2 &\le O(1) \cdot \left(\|Q_y\|_{\op} (W_{\max} \kappa_y)^2 \|\bar{x}_t\|_{2}^2 + \kappa_y W_{\max}^2 \|\bar{y}_t\|_{Q_y}^2 + \|Q_y\|_{\op} \sigma_y^2 \right) \\
        &\le O(1) \cdot \left(J_{\max} \kappa_y^3 \|\bar{x}_t\|_{2}^2 + \kappa_y J_{\max}^2 \|\bar{y}_t\|_{Q_y}^2 + (1/\mu_y) \sigma_y^2 \right),
    \end{align*}
    where we also used Lemma \ref{lemma:basic_operator_Qnorm_conversion} to have $\|W_{22} \bar{y}_t\|_{Q_y}^2 \le \kappa(Q_y) W_{\max}^2 \|\bar{y}_t\|_{Q_y}^2$, and $\kappa(Q_y) = O(\kappa_y)$.
    \item Using Cauchy-Schwarz inequality, we have
    \begin{align*}
        &\| J_{22}^{-1} J_{21} (J_{12} \bar{y}_t + \Delta \bar{x}_t + w^x_t) \|^2_{Q_y} \\
        &\le 3 \|J_{22}^{-1} J_{21} (J_{12} + W_{12}(\xi_t)) \|_{Q_y}^2 \|\bar{y}_t\|_{Q_y}^2  + 3 \|Q_y\|_{\op} \|J_{22}^{-1} J_{21} (\Delta + W_{\Delta}^x(\xi_t)) \|_{\op}^2 \|\bar{x}_t\|_{2}^2\\
        &\qquad + 3 \|J_{22}^{-1} J_{21} (u_1(\xi_t) + W_{\Delta}^x (\xi_t) x^*) \|^2_{Q_y} \\
        &\le O(1) \left(\kappa_y^3 J_{\max}^2 \|\bar{y}_t\|_{Q_y}^2 + J_{\max} \kappa_y^5 \|\bar{x}_t\|_{2}^2 +  (\kappa_y^2/\mu_y) \cdot \sigma_x^2 \right).
    \end{align*}
    \item We separate the cross-product term across $\bar{y}_t$ and $\bar{x}_t$:
    \begin{align*}
        &\left| \Exs[\vdot{(I-\beta J_{22}) \bar{y}_t}{-J_{22}^{-1}J_{21}(J_{12} \bar{y}_t + \Delta \bar{x}_t + w^x_t)}_{Q_y}] \right| \\
        &\le | \underbrace{\Exs[ \vdot{\bar{y}_t}{-J_{22}^{-1}J_{21}(J_{12} \bar{y}_t + \Delta \bar{x}_t )}_{Q_y} ]}_{(i)} |  + \beta | \underbrace{\Exs[\vdot{J_{22} \bar{y}_t}{-J_{22}^{-1}J_{21}(J_{12} \bar{y}_t + \Delta \bar{x}_t )}_{Q_y} ]}_{(ii)} | \\ 
        &\qquad + | \underbrace{\Exs[\vdot{(I-\beta J_{22}) \bar{y}_t}{-J_{22}^{-1}J_{21} w^x_t}_{Q_y}]}_{(iii)} |.
    \end{align*}
    For $(i)$, we can derive that
    \begin{align*}
        -(i) &= \Exs[ \Tr(\bar{y}_t^\top Q_y J_{22}^{-1}J_{21}(J_{12} \bar{y}_t + \Delta \bar{x}_t )) ] \\
        &\le \Tr(\Exs[\bar{y}_t \bar{y}_t^\top Q_y^{1/2}]  Q_y^{1/2} J_{22}^{-1} J_{21} J_{12}) + \Tr(\Exs[\bar{x}_t \bar{y}_t^\top Q_y^{1/2}]  Q_y^{1/2} J_{22}^{-1} J_{21} \Delta) \\
        &\le \Exs[\|\bar{y}_t\|_{Q_y}^2] \cdot \|Q_{y}^{1/2} J_{22}^{-1} J_{21} J_{12} Q_{y}^{-1/2}\|_{\op} + \|Q_y^{1/2} \Exs[\bar{y}_t \bar{x}_t^\top]\|_1 \|Q_y^{1/2} J_{22}^{-1}J_{21}\Delta\|_{\op} \\
        &\le \kappa_y^{3/2} J_{\max} \Exs[\|\bar{y}_t\|_{Q_y}^2] + \kappa_y^{5/2} J_{\max}^{1/2} \cdot \|Q_y^{1/2} \Exs[\bar{y}_t \bar{x}_t^\top]\|_1. 
    \end{align*}
    For (ii), we can simply apply Cauchy-Schwarz inequality with $J_{22}^\top Q_y J_{22}^{-1} = J_{22}^{-1} - Q_y$, to get
    \begin{align*}
        -(ii) &= \Exs[\bar{y}_t^\top J_{22}^\top Q_y J_{22}^{-1}J_{21}(J_{12} \bar{y}_t + \Delta \bar{x}_t) ] \\
        &= \Exs[\bar{y}_t^\top (J_{22}^{-1} - Q_y) J_{21}J_{12} \bar{y}_t  ] + \Exs[\bar{y}_t^\top (J_{22}^{-1} - Q_y) J_{21}J_{12} \bar{y}_t  \Delta \bar{x}_t ] \\
        &\le J_{\max} \kappa_y \Exs[\|\bar{y}_t\|_2^2] + \|\Exs[\bar{x}_t \bar{y}_t^\top Q_y^{1/2}]\|_1 \|Q_y^{-1/2} (J_{22}^{-1} - Q_y) J_{12} J_{21} \Delta\|_{\op} \\
        &\le J_{\max}^2 \kappa_y \Exs[\|\bar{y}_t\|_{Q_y}^2] + (\kappa_y^{2} J_{\max}^{3/2}) \|Q_y^{1/2} \Exs[\bar{y}_t \bar{x}_t^\top]\|_1.
    \end{align*}
    For (iii), we bound the term as
    \begin{align*}
        (iii) &= |\Tr(\Exs[w_t^x \bar{y}_t^\top] (I - \beta J_{22})^\top Q_y J_{22}^{-1} J_{21} ) | \\
        &\le \|Q_y J_{22}^{-1} J_{21} \|_{\op} \cdot  \|\Exs[w_t^x \bar{y}_t^\top]\|_1 \\
        &\le (\kappa_y/\mu_y) \|\Exs[w_t^x \bar{y}_t^\top]\|_1.
    \end{align*}
    Combining (i)-(iii), we get
    \begin{align*}
        &\left| \Exs[\vdot{(I-\beta J_{22}) \bar{y}_t}{-J_{22}^{-1}J_{21}(J_{12} \bar{y}_t + \Delta \bar{x}_t + w^x_t)}_{Q_y}] \right| \\
        &\le \kappa_y^{3/2} J_{\max} \Exs[\|\bar{y}_t\|_{Q_y}^2] + \kappa_y^{5/2} J_{\max}^{1/2} \cdot \|Q_y^{1/2} \Exs[\bar{y}_t \bar{x}_t^\top]\|_1 +(\kappa_y/\mu_y) \|\Exs[w_t^x \bar{y}_t^\top]\|_1. 
    \end{align*}
    
    \item For the cross-product with noise, we get
    \begin{align*}
        |\Exs[\vdot{(I - \beta J_{22}) \bar{y}_t}{w_t^y}_{Q_y}] &\le (1/\mu_y) \|\Exs[w_t^y \bar{y}_t^\top]\|_1. 
    \end{align*}
    \item For the last term, we simply apply Cauchy-Schwartz inequality and use inequalities used before:
    \begin{align*}
        2\alpha\beta |\Exs[\vdot{J_{22}^{-1} J_{21} (J_{12} \bar{y}_t + \Delta \bar{x}_t + w_t^x)}{w_t^y}_{Q_y}]| &\le \alpha^2 \Exs[\|J_{22}^{-1} J_{21} (J_{12} \bar{y}_t + \Delta \bar{x}_t + w_t^x)\|_{Q_y}^2] + \beta^2 \Exs[\|w_t^y\|_{Q_y}^2].
    \end{align*}
\end{enumerate}
Hence to summarize, with $\alpha \ll \beta / \kappa_y^3$ and $\beta \ll 1 / (J_{\max} \kappa_y^2)$, we get
\begin{align*}
    \Exs[\|\bar{y}_{t+1}\|_{Q_y}^2] &\le (1 - 3 \beta \mu_{y} / 4) \Exs[\|\bar{y}_t\|^2_{Q_y}] + O(\kappa_y^3 J_{\max}) \beta^2 \Exs[\|\bar{x}_t\|_{2}^2] \\
    &\qquad + O(1/\mu_y) \beta^2\sigma_y^2 + O(\kappa_y^2/\mu_y) \alpha^2\sigma_x^2 + O(\kappa_y^{5/2} J_{\max}^{1/2} ) \alpha  \|Q_{y}^{1/2} \Exs[\bar{x}_t \bar{y}_t^\top]\|_1 \\
    &\qquad + (2\kappa_y/\mu_y) \alpha \|\Exs[w_t^x \bar{y}_t^\top] \|_1  + (2/\mu_y) \beta \|\Exs[w_t^y \bar{y}_t^\top ]\|_1.
\end{align*}
Then we can invoke Lemma \ref{lemma:noise_cross_l1_product_y} with $\tau = O(\tau_{\alpha})$, and noting that $\beta J_{\max}^2 \kappa_y \ll \mu_y$ to conclude that
\begin{align}
    \label{eq:y_descent_equation}
    \Exs[\|\bar{y}_{t+1}\|_{Q_y}^2] &\le (1 - \beta \mu_{y}/2) \Exs[\|\bar{y}_t\|^2_{Q_y}] + O(\kappa_y^3 J_{\max}) \beta^2 \tau_{\alpha} \Exs[\|\bar{x}_t\|_{2}^2] \nonumber \\
    &\quad + O(\kappa_y^{5/2}J_{\max}^{1/2}) \alpha \|Q_y^{1/2} \Exs[\bar{x}_t \bar{y}_t^\top]\|_1  + O(1/\mu_y) \tau_{\alpha} (\beta^2\sigma_y^2 + \kappa_y^2 \alpha^2\sigma_x^2).
\end{align}

\subsubsection{Convergence of $\bar{x}_t$} 
We start with taking squared-$\|\cdot\|_{Q_x}$ norm for the slower iterates:
\begin{align*}
    \Exs[\|\bar{x}_{t+1}\|^2_{Q_x}] &\le \Exs[\|(I-\alpha\Delta)\bar{x}_t\|_{Q_x}^2 + 2 \alpha^2 \|J_{12} \bar{y}_t\|_{Q_x}^2 + 2 \alpha^2 \|w^x_t\|_{Q_x}^2] \\
    &\quad + 2\alpha |\Exs[\vdot{(I-\alpha \Delta) \bar{x}_t}{J_{12} \bar{y}_t}_{Q_x}]| + 2\alpha |\Exs[ \vdot{(I-\alpha \Delta) \bar{x}_t}{w_t^x}_{Q_x}] | + 2\alpha^2 | \Exs[ \vdot{J_{12}\bar{y}_t}{w^x_t}_{Q_x}]|.
\end{align*}
Following the similar steps for the analysis of $\bar{y}_t$, we show the followings:
\begin{enumerate}
    \item The main drift term satisfies
    \begin{align*}
        \|(I - \alpha \Delta) \bar{x}_t\|_{Q_x}^2 \le (1 - \mu_x \alpha) \|\bar{x}_t\|_{Q_x}^2.
    \end{align*}
    \item For the squared terms, 
    \begin{align*}
        \|J_{12} \bar{y}_t\|_{Q_x}^2 &\le \|Q_x\|_{\op} J_{\max}^2 \|\bar{y}_t\|_2^2, \le (J_{\max}^2 / \mu_x) \|\bar{y}_t\|_2^2, \\
        \|w_t^x\|_{Q_x}^2 &\le 3 \left( \kappa_x (W_{\max} \kappa_y)^2 \|\bar{x}_t\|_{Q_x}^2 + \|Q_x\|_{\op} W_{\max}^2 \|\bar{y}_t\|_2^2 + \|Q_x\|_{\op} \sigma_x^2 \right) \\
        &\le O(1) \cdot \left( \kappa_x \kappa_y^2 J_{\max}^2  \|\bar{x}_t\|_{Q_x}^2 + (J_{\max}^2/\mu_x) \|\bar{y}_t\|_2^2 + (1/\mu_x) \sigma_x^2 \right).
    \end{align*}
    \item For the cross-product term,
    \begin{align*}
        |\Exs &[\vdot{(I-\alpha\Delta) \bar{x}_t}{J_{12} \bar{y}_t}_{Q_x}]| = |\Tr(\Exs[\bar{y}_t \bar{x}_t^\top] (I-\alpha\Delta)^\top Q_x J_{12})| \\
        &\le \|Q_y^{1/2} \Exs[\bar{y}_t \bar{x}_t^\top]  \|_1 \|Q_x^{1/2} Q_{x}^{-1/2} (I-\alpha\Delta)^\top Q_x J_{12} Q_y^{-1/2} \|_{\op} \\
        &\le (J_{\max}^{3/2} / \mu_x) \|Q_y^{1/2} \Exs[\bar{y}_t \bar{x}_t^\top]\|_1. 
    \end{align*}
    \item For the product term with noise, we have
    \begin{align*}
        |\Exs[\vdot{(I-\alpha\Delta) \bar{x}_t}{w_t^x}_{Q_x}]| &\le \|\Exs[ w_t^x \bar{x}_t^\top]\|_1 \|Q_x\|_{\op} \le (1/\mu_x) \|\Exs[ w_t^x \bar{x}_t^\top]\|_1.
    \end{align*}
\end{enumerate}
Writing down the intermediate result, with $\alpha \ll 1/(J_{\max} \kappa_x \kappa_y^3)$, we have
\begin{align*}
    \Exs[\|\bar{x}_{t+1}\|_{Q_x}^2] &\le (1 - \alpha \mu_{x} / 2) \Exs[\|\bar{x}_t\|^2_{Q_x}] + O( J_{\max}^2 / \mu_x) \alpha^2 \Exs[\|\bar{y}_t\|_2^2] + O(1/\mu_x) \alpha^2 \sigma_x^2 \\
    &\qquad + O(J_{\max}^{3/2}/\mu_x) \alpha \|Q_y^{1/2} \Exs[\bar{y}_t\bar{x}_t^\top]\|_1 + (2/\mu_x) \alpha \|\Exs[w_t^x \bar{x}_t^\top]\|_1.
\end{align*}
Invoke Lemma \ref{lemma:noise_cross_l1_product_x} with $\tau = O(\tau_{\alpha})$, and we can conclude that
\begin{align}
    \label{eq:x_descent_equation}
    \Exs[\|\bar{x}_{t+1}\|_{Q_x}^2] &\le (1 - \alpha \mu_{x} / 2) \Exs[\|\bar{x}_t\|^2_{Q_x}] + (J_{\max}^2/\mu_x) (\alpha^2 + \alpha \beta^2 \tau_{\alpha}^2 J_{\max})  \Exs[\|\bar{y}_t\|_{2}^2] \nonumber \\
    &\qquad + \alpha (J_{\max}^{3/2} / \mu_x) \|Q_y^{1/2} \Exs[\bar{y}_t \bar{x}_t^\top]\|_1 + (1/\mu_x) \left(\alpha^2 \tau_{\alpha} \sigma_x^2 + \alpha \beta^2 \tau_{\alpha}^2 J_{\max} \kappa_x^2 \sigma_y^2\right).
\end{align}

\subsubsection{Convergence of Cross-Correlations in $S^1$-Norm} 
We start with unfolding the equation:
\begin{align*}
    \bar{y}_{t+1} \bar{x}_{t+1}^\top &= (I - \beta J_{22}) \bar{y}_t \bar{x}_t^\top (I - \alpha \Delta) - \alpha  (I - \beta J_{22}) \bar{y}_t (J_{12} \bar{y}_t + w^x_t)^\top \\
    &\qquad - \alpha J_{22}^{-1}J_{21}(J_{12}\bar{y}_t + \Delta \bar{x}_t + w_t^x) \bar{x}_t^\top - \beta w_t^y \bar{x}_t^\top \\
    &\qquad + \alpha^2 J_{22}^{-1} J_{21} (J_{12}\bar{y}_t + \Delta \bar{x}_t + w_t^x)(\Delta \bar{x}_t + J_{12} \bar{y}_t + w_t^x)^\top \\
    &\qquad + \alpha \beta \cdot w_t^y (\Delta \bar{x}_t + J_{12} \bar{y}_t + w_t^x)^\top.
\end{align*}
The target norm is $\|\cdot\|_1$ bound on the expectation of the cross-product term. The trick is to multiply $Q_y^{1/2}$ from left on both sides, and use identity $I = Q_y^{-1/2} Q_y^{1/2}$:
\begin{align*}
    \|Q_y^{1/2} \Exs[\bar{y}_{t+1} \bar{x}_{t+1}^\top]\|_1 &\le \|Q_y^{1/2} (I - \beta J_{22}) Q_y^{-1/2} (Q_y^{1/2} \Exs[\bar{y}_t \bar{x}_t^\top]) (I - \alpha \Delta)\|_1 \\
    &\qquad + \alpha  \| Q_y^{1/2} (I - \beta J_{22}) Q_y^{-1/2} (Q_{y}^{1/2} \Exs[\bar{y}_t \bar{y}_t^\top] Q_y^{1/2}) Q_y^{-1/2} J_{12}^\top \|_1 \\
    &\qquad + \alpha  \| Q_y^{1/2} (I - \beta J_{22}) Q_y^{-1/2} Q_{y}^{1/2} \Exs[\bar{w}_t^x \bar{y}_t^\top] \|_1 \\
    &\qquad + \alpha \| Q_y^{1/2} J_{22}^{-1}J_{21}(J_{12} \Exs[\bar{y}_t\bar{x}_t^\top] + \Delta \Exs[\bar{x}_t\bar{x}_t^\top] + \Exs[w_t^x \bar{x}_t^\top])\|_1 + \beta \|Q_y^{1/2} \Exs[w_t^y \bar{x}_t^\top]\|_1 \\
    &\qquad + \alpha^2 \| Q_y^{1/2} J_{22}^{-1} J_{21} \Exs[(J_{12}\bar{y}_t + \Delta \bar{x}_t + w_t^x)(\Delta \bar{x}_t + J_{12} \bar{y}_t + w_t^x)^\top] \|_1 \\
    &\qquad + \alpha \beta \|\Exs[Q_y^{1/2} w_t^y (\Delta \bar{x}_t + J_{12} \bar{y}_t + w_t^x)^\top]\|_1.
\end{align*}
We observe the following:
\begin{enumerate}
    \item $\|Q_y^{1/2} (I-\beta J_{22}) Q_y^{-1/2}\|_{\op} = \|I-\beta J_{22}\|_{Q_y} \le 1-\mu_y \beta$, and therefore 
    \begin{align*}
        \|Q_y^{1/2} (I - \beta J_{22}) Q_y^{-1/2} (Q_y^{1/2} \Exs[\bar{y}_t \bar{x}_t^\top]) (I - \alpha \Delta)\|_1 &\le (1-\mu_y \beta) (1 + \alpha J_{\max} \kappa_y) \|Q_y^{1/2}\Exs[\bar{y}_t \bar{x}_t^\top]\|_1 \\
        &\le (1 - \mu_y \beta/2) \|Q_y^{1/2} \Exs[\bar{y}_t \bar{x}_t^\top]\|_1.
    \end{align*}
    \item In all other terms, we use inequality $\|\Exs[uv^\top]\|_1 \le \frac{1}{2} (\Exs[\|u\|_2^2] + \Exs[\|v\|_2^2])$.  
\end{enumerate}
We omit some algebraic details, and state the desired bounds:
\begin{align*}
    \|Q_y^{1/2} \Exs[\bar{y}_{t+1} \bar{x}_{t+1}^\top]\|_{1} &\le (1-\beta \mu_y / 2) \| Q_y^{1/2} \Exs[\bar{y}_t \bar{x}_t^\top] \|_{1} + \alpha J_{\max} \kappa_y^{3/2}  \|Q_y^{1/2} \Exs[\bar{y}_t \bar{x}_t^\top] \|_1 \\
    &\ + (\alpha J_{\max}^{3/2}) \Exs[\|\bar{y}_t\|_{Q_y}^2] + (\alpha J_{\max}^{1/2} \kappa_y^{5/2}) \Exs[\|\bar{x}_t\|^2_{2}] \\ 
    &\ + (\alpha / \sqrt{\mu_y}) \|\Exs[w_t^x \bar{y}_t^\top]\|_1 + (\alpha\beta J_{\max} / \sqrt{\mu_y}) \|\Exs[w_t^y \bar{y}_t^\top]\|_1 \\ 
    &\ + (\beta / \sqrt{\mu_y}) \|\Exs[w_t^y \bar{x}_t^\top]\|_1 + (\alpha\beta / \sqrt{\mu_y}) (\sigma_x^2 + \sigma_y^2).
\end{align*}
Applying Lemma \ref{lemma:noise_cross_l1_product_x} and  \ref{lemma:noise_cross_l1_product_y}, and using $\alpha \ll \beta / \kappa_y^2$ in Assumption \ref{assumption:stepsize}, we can conclude that
\begin{align}
    \label{eq:xy_descent_equation}
    \|Q_{y}^{1/2} \Exs[\bar{y}_{t+1} \bar{x}_{t+1}^\top]\|_{1} &\le (1-\beta \mu_y / 2) \| Q_y^{1/2} \Exs[\bar{y}_t \bar{x}_t^\top] \|_{1} \nonumber \\
    &\quad + O(\alpha J_{\max}^{3/2} + \beta^3 \tau^2 J_{\max}^{9/2} \kappa_x \kappa_y^{1/2} / \mu_x) \Exs[\|\bar{y}_t\|_{Q_y}^2] + O\left( \alpha J_{\max}^{1/2} \kappa_y^{5/2} \right)  \Exs[\|\bar{x}_t\|_{2}^2] \nonumber \\
    &\quad + O (\alpha\beta \tau_\alpha/ \sqrt{\mu_y}) (\sigma_x^2 +  \sigma_y^2) + (1/\sqrt{\mu_y}) (\beta^3 \tau^2 J_{\max} \kappa_x / \mu_x ) \sigma_y^2. 
\end{align}

\subsubsection{Overall Convergence}
Recall the potential function $V_t$ and $U_t$ in \eqref{eq:potential_functions}:
\begin{align*}
    V_t &= \Exs[\|\bar{x}_t\|_{Q_x}^2] + \frac{O(1) J_{\max}^{2} \kappa_y \alpha(\alpha+\beta^2 \tau_\alpha^2 J_{\max} ) }{\mu_x \mu_y \beta} \Exs[\|\bar{y}_t\|_{Q_y}^2] + \frac{O(1) J_{\max}^{1/2} \kappa_y \alpha}{\mu_x \beta} \| Q_y^{1/2} \Exs[ \bar{y}_t \bar{x}_t^\top] \|_{1}, \nonumber \\
    U_{t} &= \Exs[\|\bar{y}_t\|_{Q_y}^2] + \frac{O(1) J_{\max}^{1/2} \kappa_y^{2.5} \alpha}{\mu_y \beta} \|Q_{y}^{1/2} \Exs[\bar{y}_t \bar{x}_t^\top]\|_{1}.
\end{align*}
We note that $\beta \ll 1 / (\kappa_y^3 \kappa_x J_{\max})$ and $\beta/\alpha \ll 1/ (\kappa_y^3 \kappa_x)$ in Assumption \ref{assumption:stepsize}, and
\begin{align*}
    \|\bar{x}_t\|_2^2 &\le \frac{1}{\sigma_{\min}(Q_x)} \|\bar{x}_t\|_{Q_x}^2 \le J_{\max} \kappa_y \|\bar{x}_t\|_{Q_x}^2, \\
    \|\bar{y}_t\|_2^2 &\le \frac{1}{\sigma_{\min}(Q_y)} \|\bar{y}_t\|_{Q_y}^2 \le J_{\max} \|\bar{y}_t\|_{Q_y}^2.
\end{align*}
Putting altogether, we have
\begin{align}
    V_{t+1} &\le (1 - \alpha \mu_x/2) V_{t} + \frac{\kappa_y^{3/2}}{\mu_x} \left(\alpha^2  \tau_{\alpha} + \frac{J_{\max} \kappa_x \tau_{\alpha}^2}{\mu_x} \alpha \beta^2  \right) \sigma_y^2 + \frac{\tau_\alpha \kappa_y \kappa_x}{\mu_y} \alpha (\alpha + \beta^2 J_{\max} \tau_{\alpha}^2 ) \sigma_y^2 + \frac{\alpha^2 \tau_{\alpha}}{\mu_x} \sigma_x^2. \label{eq:x_mse_descent_final}
\end{align}
Solving this recursion, 
\begin{align*}
    \Exs[\|\bar{x}_t\|_{Q_x}^2] &\le V_{t} \le \exp(-\alpha\mu_x t/ 2 ) V_0 + \frac{\kappa_y^{1/2}}{\mu_x^2} (\alpha \kappa_y \tau_\alpha+ \beta^2 J_{\max} \kappa_x^2 \tau_\alpha^2) \sigma_y^2 + \frac{\tau_\alpha \kappa_y \kappa_x}{\mu_y\mu_x} (\alpha + \beta^2 J_{\max} \tau_\alpha^2) \sigma_y^2  + \frac{\alpha \tau_\alpha}{\mu_x^2} \sigma_x^2,
\end{align*}
for all $t$, hence for all sufficiently large $t \gg \alpha^{-1}$, we have bounds for $\Exs[\|\bar{x}_t\|_{Q_x}^2] \le O_{\texttt{P}} (\alpha)\sigma_x^2 + O_{\texttt{P}}(\alpha + \beta^2) \sigma_y^2$. 

Next, we consider the potential for faster iterates $U_t$. We see that
\begin{align}
    U_{t+1} &\le (1-\beta \mu_{y}/2) U_{t} + \beta^2 \kappa_y^3 J_{\max} \tau_\alpha\Exs[\|\bar{x}_t\|_{2}^2] + (\tau_\alpha/ \mu_y) (\kappa_y^2 \alpha^2 \sigma_x^2 + \beta^2 \sigma_y^2) \nonumber \\
    &\le (1-\beta \mu_{y}/2) U_{t} + \beta^2 \kappa_y^3 J_{\max}^2 \tau_\alpha\exp(-\alpha \mu_x t/2) V_{0} \nonumber \\
    &\quad + O\left(\frac{\kappa_y^2 \tau}{\mu_y} \alpha^2 + \frac{\tau^2 J_{\max}^2 \kappa_y^3}{\mu_x^2} \alpha\beta^2 \right) \sigma_x^2 + O\left( \frac{\tau}{\mu_y} \beta^2 \right) \sigma_y^2, \label{eq:y_mse_descent_final}
\end{align}
which yields 
\begin{align*}
    \Exs[\|y_t\|_{Q_y}^2] &\le U_t \le \exp(-\beta \mu_y t/2) U_0 + \beta \kappa_y^4 \tau_{\alpha} \exp(-\alpha \mu_x t /4) V_0 \\
    &\qquad \quad + O\left(\frac{\kappa_y^2 \tau}{\mu_y^2} \frac{\alpha}{\beta} + \frac{\tau^2 J_{\max} \kappa_y^4}{\mu_x^2} \beta \right) \alpha \sigma_x^2 + 
    O\left( \frac{\tau}{\mu_y^2} \beta \right) \sigma_y^2,     
\end{align*}
assuming $\beta \mu_y \gg \alpha \mu_x$. Thus for sufficiently large $t \gg \alpha^{-1} \log(1/\beta)$, we have $\Exs[\|\bar{y}_t\|_{Q_y}^2] = O_{\texttt{P}}(\alpha^2/\beta + \alpha\beta)\sigma_x^2 + O_{\texttt{P}} (\beta) \sigma_y^2$. This concludes the final error rates as $t \rightarrow \infty$.

\subsection{Proof of Theorem \ref{theorem:distribution_conv}}
\label{appendix:distributional_convergence}

Showing the distributional convergence consists of two steps. First, we setup two sequences $\{(x_t^1, y_t^1, \xi_t)\}_{t\ge0}$, $\{(x_t^2, y_t^2, \xi_t)\}_{t\ge0}$ coupled with the same sequence of Markovian states $\{\xi_t\}_{t\ge0}$. We show that these two sequences will converge in the squared-$L_2$ expectation sense:
\begin{lemma}
    \label{lemma:noise_coupled_convergence}
    Under Assumptions \ref{assumption:linear_Hurwitz}-\ref{assumption:stepsize}, for any two sequences coupled with the same Markovian nosie $(x_t^1, y_t^1, \xi_t)$ and $(x_t^2, y_t^2, \xi_t)$, the following holds:
    \begin{align*}
        \Exs[\|x_t^1-x_t^2\|_{2}^2 ] &\le O_{\texttt{P}}(1) \cdot \Exs\left[\|x_0^1-x_0^2\|_{2}^2 + \frac{\alpha}{\beta}\|\bar{y}_0^1-\bar{y}_0^2\|_{2}^2 \right] \exp(-\alpha \mu_x t / 4), \\
        \Exs [\|\bar{y}_t^1-\bar{y}_t^2\|_{2}^2] &\le O_{\texttt{P}}(1) \cdot \Exs \left[\|x_0^1-x_0^2\|_{2}^2 + \|\bar{y}_0^1-\bar{y}_0^2\|_{2}^2 \right] \exp(-\beta \mu_y t / 4) \\
        &\quad + O_{\texttt{P}}(1) \cdot \Exs\left[\|x_0^1-x_0^2\|_{2}^2 + \frac{\alpha}{\beta} \|\bar{y}_0^1-\bar{y}_0^2\|_{2}^2\right] \beta \exp(-\alpha \mu_x t / 4). 
    \end{align*}
\end{lemma}
We first prove the above lemma and use it to conclude that the distribution of iteration variables converges in Wasserstein distance to a unique stationary distribution.

\subsubsection{Proof of Lemma \ref{lemma:noise_coupled_convergence}} Let us define $\delta^x_t = \bar{x}_t^1 - \bar{x}_t^2, \delta^y_t = \bar{y}_t^1 - \bar{y}_t^2$. Then the stochastic recursion \eqref{eq:basic_ttsa_equation} becomes 
\begin{align*}
    \delta^x_{t+1} = (I - \alpha \Delta) \delta^x_t - \alpha J_{12} \delta_t^y - \alpha \delta^{wx}_{t},
\end{align*}
and for $y$, we have
\begin{align*}
    \delta^y_{t+1} &= (I - \beta J_{22}) \delta^y_t - \alpha J_{22}^{-1} J_{21} (J_{12} \delta_t^y + \Delta \delta_t^x) - (\alpha J_{22}^{-1} J_{21} \delta^{wx}_{t} + \beta \delta^{wy}_{t}),
\end{align*}
where the noise differences are given by:
\begin{align*}
    \delta_{t}^{wx} = W_{\Delta}^x(\xi_t) \delta^x_t + W_{12}(\xi_t) \delta^y_t, \\
    \delta_{t}^{wy} = W_{\Delta}^y(\xi_t) \delta^x_t + W_{22}(\xi_t) \delta^y_t,
\end{align*}
where we used the expression in Lemma \ref{lemma:wt_reexpress}. This can be considered as the same TTSA recursion with $\sigma_x = \sigma_y = 0$. Therefore, the remaining steps are equivalent to the pilot result with $\sigma_x = \sigma_y = 0$, and it leads to 
\begin{align*}
    \Exs[\|\delta_t^x\|_{Q_x}^2] &\le \exp(-\alpha \mu_x t / 2) V_0, \\
    \Exs[\|\delta_t^y\|_{Q_y}^2] &\le \exp(-\beta \mu_y t / 2)U_0 + O_{\texttt{P}} (1) \beta \exp(-\alpha \mu_x t / 4) V_0.
\end{align*}
where we define 
\begin{align*}
    V_t &= \Exs[\|\delta_t^x\|_{Q_x}^2] + \frac{O(1) J_{\max}^{2} \kappa_y \alpha(\alpha+\beta^2 \tau_\alpha^2 J_{\max} ) }{\mu_x \mu_y \beta} \Exs[\|\delta_t^{y}\|_{Q_y}^2] + \frac{O(1) J_{\max}^{1/2} \kappa_y \alpha}{\mu_x \beta} \| Q_y^{1/2} \Exs[ \delta_t^{y} {\delta_t^x}^\top] \|_{1}, \nonumber \\
    U_{t} &= \Exs[\|\delta_t^{y}\|_{Q_y}^2] + \frac{O(1) J_{\max}^{1/2} \kappa_y^{2.5} \alpha}{\mu_y \beta} \|Q_{y}^{1/2} \Exs[\delta_t^{y} {\delta_t^{x}}^\top]\|_{1},
\end{align*}
This shows that $(\bar{x}_t^1,\bar{y}_t^1), (\bar{x}_t^2, \bar{y}_t^2)$ converges exponentially fast with the noise coupling, which in turn means $(x_t^1, y_t^1)$ and $(x_t^2, y_t^2)$ couples exponentially fast since
\begin{align*} 
    \bar{x}_t^1 - \bar{x}_t^2 &= x_t^1 - x_t^2, \\
    \bar{y}_t^1 - \bar{y}_t^2 &= (y_t^1 - y_t^2) - (y^*(x_t^1) - y^*(x_t^2)) = y_t^1 - y_t^2 + O_{\texttt{P}} (\|x_t^1 - x_t^2\|).
\end{align*}

\subsubsection{Distributional Convergence via Coupling}

The steps here mostly follows the proof steps in \cite{huo2023bias}, Appendix A.2.2. We first consider a sequence $(\xi_t^1, x_t^1, y_t^1)_{t\ge 0}$ that starts at $(x_0^1, y_0^1, \xi_0^1) \sim \mu_0$ sampled from some initial distribution $\mu_0$ where $\xi_0^1 \sim \pi$ and $(x_0^1, y_0^1)$ are {\it statistically independent}. Then, we similarly define the initial point distribution of the second sequence $(x^2_{-1}, y_{-1}^2)$ as the same as $(x_0^1, y_0^1)$ and set $(x_0^2, y_0^2)$ be the result of one-step stochastic recursion \eqref{eq:basic_ttsa_equation}, where $\xi_{-1}^2 \sim \mathcal{P}^{\dagger}(\cdot|\xi_0^1)$. Then we couple the Markovian states $\xi_t^1 = \xi_t^2$ for all $t \ge 0$. Now that we have 
\begin{align*}
    (\xi_t^2, x_{t}^2, y_{t}^2) \stackrel{d}{=} (\xi_{t+1}^1, x_{t+1}^1, y_{t+1}^1),
\end{align*}
since $\xi_0^1 \stackrel{d}{=} \xi_1^1$ follws a stationary distribution $\pi$, and $\xi_t^1 = \xi_{t}^2$ is coupled.
Then by definition of Wasserstein distance (with the optimal coupling), using Lemma \ref{lemma:noise_coupled_convergence}, we get
\begin{align*}
    \bar{\mathcal{W}}_2^2 ((x_t^1, y_t^1, \xi_t^1), (x_{t+1}^1, y_{t+1}^1, \xi_{t+1}^1)) \le C\exp(-\alpha \mu_x t/4), \forall t \ge 0,
\end{align*}
and therefore (omitting superscript)
\begin{align*}
    \sum_{t\ge 0} \bar{\mathcal{W}}_2^2 ((x_t, y_t, \xi_t), (x_{t+1}, y_{t+1}, \xi_{t+1})) \le \sum_{t \ge 0} C\exp(-\alpha \mu_x t/4) < \infty,
\end{align*}
with $\alpha > 0$. The probability space over $\Xi \times \mathbb{R}^{d_x} \times \mathbb{R}^{d_y}$ equipped with $\bar{\mathcal{W}}_2$-norm is known to be a Polish space where every Cauchy sequence converges (\cite{villani2009optimal}, Theorem 6.18). Furthermore, convergence in Wasserstein distance implies weak convergence (\cite{villani2009optimal}, Theorem 6.9), hence weak convergence to some distribution $\mu \in \mathcal{P}_2 (\Xi \times \mathbb{R}^{d_x} \times \mathbb{R}^{d_y})$.

\subsubsection{Stationarity of the Limit Distribution}
Next, we show that the sequence converges to a unique stationary distribution $\mu$ regardless of the initial distribution $\mu_0$. To do so, we first show that the sequence has bounded fourth-order moments:
\begin{lemma}
    \label{lemma:fourth_moment_bound}
    Suppose an initial distribution $(\xi_0, x_0, y_0) \sim \mu_0$ that satisfies Assumption \ref{assumption:bounded_fourth_order}. Then for all $t \ge 0$, we have
    \begin{align*}
        \Exs[\|x_t\|_2^4 + \|y_t\|_2^4] < O_{\texttt{P}}(1) \cdot \Exs[\|x_0\|_2^4 + \|y_0\|_2^4] + O_{\texttt{P}}(1) \cdot (\sigma_x^4 + \sigma_y^4). 
    \end{align*}
\end{lemma}
Then we consider two TTSA sequences starting from two arbitrary initial distributions $\mu_0^1, \mu_0^2$. We start with the following lemma that is reminiscent of Lemma A.8 in \cite{huo2023bias} for \eqref{eq:basic_ttsa_equation}:
\begin{lemma}
    \label{lemma:auxiliary_W2_lemma}
    For any two TTSA sequences $(x_t^1, y_t^1, \xi_t^1)\sim \mu_0^1$ and $(x_t^2, y_t^2, \xi_t^2) \sim \mu_0^2$ with bounded fourth-order moments satisfying Assumption \ref{assumption:bounded_fourth_order}, for all $t \ge 0$, we have
    \begin{align}
        \mathcal{W}_2^2 (\mu_t^1(x_t^1), \mu_t^2(x_t^2)) &\le  O_{\texttt{P}} (1) \exp(-\alpha \mu_x t / 8) V_0, \nonumber \\
        \mathcal{W}_2^2 (\mu_t^1(\bar{y}_t^1), \mu_t^2(\bar{y}_{t}^2)) &\le  O_{\texttt{P}} (1) \left(\beta \exp(-\alpha \mu_x t / 8) V_0 + \exp(-\beta \mu_y t / 8) U_0\right), \label{eq:dist_conv_rate_ineq}
    \end{align}
    where 
    \begin{align*}
        V_0 &:= \mathcal{W}_2^2 (\mu_0^1(x_0^1), \mu_0^2(x_0^2)) + \frac{\alpha}{\beta} \mathcal{W}_2^2 (\mu_0^1(\bar{y}_0^1), \mu_0^2(\bar{y}_0^2)) + O_\texttt{P}(\alpha), \\
        U_0 &:=\mathcal{W}_2^2 (\mu_0^1(x_0^1), \mu_0^2(x_0^2)) + \mathcal{W}_2^2 (\mu_0^1(\bar{y}_0^1), \mu_0^2(\bar{y}_0^2)) + O_\texttt{P}(\beta).
    \end{align*}
\end{lemma}
Apply Lemma \ref{lemma:auxiliary_W2_lemma}, we have
\begin{align*}
    \bar{\mathcal{W}}_2 (\mu_t^1, \mu_t^2) < O_{\texttt{P}}(1) \cdot \exp(-\alpha \mu_x t /8) \stackrel{t \rightarrow \infty}{\longrightarrow} 0,
\end{align*}
which in turn implies that all sequences converge to the unique limit distribution $\mu$.

Lastly, we show that $\mu$ is an invariant distribution with $\mu(\xi) = \pi$. By the geometric mixing property of $(\xi_t)_{t\ge 0}$, the limit distribution must satisfy $\mu(\xi) = \pi$ (otherwise, we can derive a contradiction). Thus, for a sequence $(x_t,y_t,\xi_t)$ starting from $\mu$ with marginal $\mu(\xi_0) = \pi$, we know that $\mu_t(\xi_t) = \pi$ for all $t \ge 0$. Thus, using the coupling results, we have
\begin{align*}
    \bar{\mathcal{W}}_2(\mu_1, \mu) &\le \bar{\mathcal{W}}_2(\mu_1, \mu_{t+1}) + \bar{\mathcal{W}}_2(\mu_{t+1}, \mu) \\
    &\le O_\texttt{P}(1) \bar{\mathcal{W}}_2(\mu_0, \mu_{t}) + \bar{\mathcal{W}}_2(\mu_{t+1}, \mu) \stackrel{t \rightarrow \infty}{\longrightarrow} 0,
\end{align*}
where we used $\mu_0 = \mu$.

\subsubsection{Bias Characterization} 
For analyzing the bias if the limite distribution $(x_{\infty}, y_{\infty}, \xi_{\infty}) \sim \mu$, we start from sending $t \rightarrow \infty$ in \eqref{eq:basic_ttsa_equation}
\begin{align*}
    \bar{x}_{t+1} &= (I - \alpha \Delta) \bar{x}_t - \alpha J_{12} \bar{y}_t - \alpha w^x (x_t,y_t;\xi_t), \\
    \bar{y}_{t+1} &= (I - \beta J_{22}) \bar{y}_t - \alpha J_{22}^{-1} J_{21} (J_{12} \bar{y}_t + \Delta \bar{x}_t + J_{21} w^x(x_t,y_t;\xi_t)) + \beta w^y(x_t,y_t;\xi_t)).
\end{align*}
Let 
\begin{align*}
    z^x(\xi) &= \Exs[x_{\infty} - x^*| \xi_{\infty} = \xi], \\
    z^y(\xi) &= \Exs[y_{\infty} - y^*(x_{\infty})| \xi_{\infty} = \xi],
\end{align*} 
and 
\begin{align*}
    w^x(\xi) &= W_{11}(\xi) \Exs[x_{\infty}| \xi_{\infty}=\xi] + W_{12}(\xi) \Exs[y_{\infty}| \xi_{\infty}=\xi] + b_1(\xi), \\
    w^y(\xi) &= W_{21}(\xi) \Exs[x_{\infty}| \xi_{\infty}=\xi] + W_{22}(\xi) \Exs[y_{\infty}| \xi_{\infty}=\xi] + b_2(\xi).
\end{align*} 
We take conditional expectation on $\xi_{\infty+1} = \xi$, we have the backward conditional probability $\xi_{\infty}\sim P^\dagger (\cdot | \xi_{\infty+1}=\xi)$. Let $\mathcal{T}: \Xi \times \mathbb{R}^{*} \rightarrow \Xi \times \mathbb{R}^{*}$ an {\it unnormalized} Markov operator over $\xi$: 
\begin{align*}
    \mathcal{T}\{z\}(\xi') = \int_{\xi \in \Xi} z(\xi) d \mathcal{T}(\xi' | \xi).
\end{align*}
Using the above notation, we can rewrite the recursion as
\begin{align*}
    z^x &= \mathcal{P^\dagger} \{(I - \alpha \Delta) z^x - \alpha J_{12} z^y - \alpha w^x\}, \\
    z^y &= \mathcal{P^\dagger} \{(I - \beta J_{22}) z^y - \alpha J_{22}^{-1} J_{21} (J_{12} z^y + \Delta z^x + w^x) - \beta w^y\}.
\end{align*}
Let $\Pi = 1 \bigotimes \pi$, and note that $\Pi \{W_{ij}\} = 0$ for all $i,j \in \{1,2\}$. We eventually want to characterize $\bar{z}^x := \Exs[x_{\infty} - x^*] = \pi\{z^x\} = \int_{\xi}  z^x(\xi) d\pi(\xi)$ and $\bar{z}^y := \pi\{z^y\}$. Since $\pi P^\dagger = \pi$ by the time-reversing property of the geometrically mixing chain, this implies
\begin{align}
    &\Delta \bar{z}^x + J_{12} \bar{z}^y + \pi \{w^x\} = 0, \nonumber \\
    & J_{22} \bar{z}^y + \pi \{w^y\} = 0. \label{eq:stationary_bias_equation}
\end{align}

To further proceed, let $\delta^x (\xi) = z^x (\xi) - \pi \{z^x\}$, $\delta^y (\xi) = z^y (\xi) - \pi\{z^y\}$, and since $(\mathcal{P^\dagger} - \Pi) \{z^x\} = (\mathcal{P^\dagger} - \Pi) \{\delta^x\}$, we can observe that
\begin{align}
    (I - P^\dagger + \Pi) \{\delta^x\} &= -\alpha (P^\dagger - \Pi) \{\Delta  z^x + J_{12} z^y + w^x\} = - \alpha (P^\dagger - \Pi) \{\Delta  \delta^x + J_{12} \delta^y + w^x\}, \nonumber \\
    (I - P^\dagger + \Pi) \{\delta^y\} &= - \beta (P^\dagger - \Pi) \left\{ J_{22}  z^y + \frac{\alpha}{\beta} J_{22}^{-1} J_{21} (J_{12} z^y + \Delta z^x + w^x) + w^y \right\} \nonumber \\
    &= (I - P^\dagger + \Pi) \{\delta^x\} - \beta (P^\dagger - \Pi) \left\{ J_{22} \delta^y + w^y \right\}. \label{eq:bias_recursion_intermediate} 
\end{align}
Then we note that
\begin{align*}
    w^y(\xi) &= W_{21}(\xi) z^x(\xi) + W_{22}(\xi) z^y(\xi) + u_2(\xi) + W_{21}(\xi) x^* - W_{22}(\xi) J_{22}^{-1} J_{21} (z^x(\xi) + x^*) \\
    &= (W_{21}(\xi) - W_{22}(\xi) J_{22}^{-1} J_{21}) z^x(\xi) + W_{22}(\xi) z^y(\xi) + u_2(\xi) + (W_{21}(\xi) - W_{22}(\xi) J_{22}^{-1} J_{21})x^* \\
    &= W_\Delta^y(\xi) (\delta^x(\xi) + \bar{z}^x) + W_{22}(\xi) (\delta^y(\xi) + \bar{z}^y)  + u_2(\xi) + W_\Delta^y(\xi) x^*, \\
    w^x(\xi) &= (W_{11}(\xi) - W_{12}(\xi) J_{22}^{-1} J_{21}) z^x(\xi) + W_{12}(\xi) z^y(\xi) + u_1(\xi) + (W_{11}(\xi) - W_{12}(\xi) J_{22}^{-1} J_{21}) x^* \\ 
    &= W_\Delta^x(\xi) (\delta^x(\xi) + \bar{z}^x) + W_{12}(\xi) (\delta^y(\xi) + \bar{z}^y) + u_1(\xi) + W_\Delta^x(\xi) x^*.
\end{align*}
Plugging this back into \eqref{eq:stationary_bias_equation} yields
\begin{align*}
    \Delta \bar{z}^x + J_{12} \bar{z}^y + \pi \{ W_{\Delta}^x \circ \delta^x\} + \pi \{W_{12} \circ \delta^y\} = 0, \\
    J_{22} \bar{z}^y + \pi \{W_{\Delta}^y \circ \delta^x\} + \pi \{W_{22} \circ \delta^y\} = 0,
\end{align*}
where we define $(a \circ b)(\xi) = a(\xi) b(\xi)$. In turn, we have
\begin{align}
    \label{eq:stationary_bias}
    \bar{z}^y &= -J_{22}^{-1} (\pi \{W_{\Delta}^y \circ \delta^x\} + \pi \{W_{22} \circ \delta^y\}), \nonumber \\
    \bar{z}^x &= -\Delta^{-1} (\pi\{(W_{\Delta}^x - J_{12} J_{22}^{-1} W_{\Delta}^y)\circ \delta^x\} + \pi\{ (W_{12} - J_{12} J_{22}^{-1} W_{22}) \circ \delta^y\}).
\end{align}
Rearranging \eqref{eq:bias_recursion_intermediate} yields
\begin{align}
    (I - \mathcal{P^\dagger} + \Pi) \{ \delta^x\} &= -\alpha (\mathcal{P^\dagger} - \Pi) \{ (\Delta + W_{\Delta}^x ) \circ \delta^x + (J_{12} + W_{12}) \circ \delta^y + (u_1 + W_{\Delta}^x x^*) \} \nonumber \\
    &\qquad -\alpha (\mathcal{P^\dagger} - \Pi) \{ W_{\Delta}^x \}\bar{z}^x -\alpha (\mathcal{P^\dagger} - \Pi)\{ W_{12} \} \bar{z}^y,  \nonumber\\
    (I - \mathcal{P^{\dagger}} + \Pi) \{\delta^y - \delta^x\} &= - \beta(\mathcal{P^\dagger} - \Pi) \{ W_{\Delta}^y \circ \delta^x + (J_{22} + W_{22}) \circ \delta^y + (u_2 + W_{\Delta}^y x^*) \} \nonumber\\
    &\qquad - \beta(\mathcal{P^\dagger} - \Pi) \{ W_{\Delta}^y \} \bar{z}^x - \beta(\mathcal{P^\dagger} - \Pi) \{W_{22}\} \bar{z}^y . \label{eq:bias_bias_linear_equations}
\end{align}
The operation $(I - \mathcal{P}^\dagger + \Pi)$ is invertible (see Corollary \ref{corollary:inverse_operator_bound}), and thus we can invert the operator $(I - \mathcal{P}^\dagger + \Pi)$. Putting all relationships together leads us to the recursion:
\begin{align}
    \label{eq:delta_final_form}
    \delta^x &= \alpha d^x + O_{\texttt{P}}(\alpha \delta^x + \alpha \delta^y), \nonumber \\
    \delta^y &= \alpha d^x + \beta d^y + O_{\texttt{P}}(\beta \delta^x + \beta \delta^y),
\end{align}
where 
\begin{align}
    \label{eq:d_final_form}
    d^x := -(I - \mathcal{P}^\dagger + \Pi)^{-1} (\mathcal{P}^\dagger - \Pi) \{u_1 + W_{\Delta}^x x^*\}, \nonumber \\
    d^y := -(I - \mathcal{P}^\dagger + \Pi)^{-1} (\mathcal{P}^\dagger - \Pi) \{u_2 + W_{\Delta}^y x^*\},
\end{align}
are independent of the choice of $\alpha,\beta$. Next, we bound the norm of $\delta^x, \delta^y, d^x, d^y$, and thus the norm of $\bar{z}^x, \bar{z}^y$.

\subsubsection{Additional Preliminaries for Bounding Norms} 
Before we proceed, we define the notion of norms that we use in the proof. For vector-valued quantities, let us define $\|v\|_{L^2(\pi)}$ as 
\begin{align*}
    \|v\|_{L^2(\pi)} = \sqrt{\int_{\Xi} \|v\|_2^2 d\pi(\xi)},
\end{align*}
and for the Markov kernel $\mathcal{T}$,
\begin{align*}
    \|\mathcal{T}\|_{L^2(\pi)} := \sup_{\|v\|_{L^2(\pi)} = 1} \|\mathcal{T} \{v\}\|_{L^2(\pi)}.
\end{align*}
For matrices, we use the conjugate norm-pair $\|\cdot\|_1$ and $\|\cdot\|_\infty = \|\cdot\|_{\op}$. Specifically, for matrix-valued quantities, we define $\|A\|_{S^1(\pi)}$ as
\begin{align*}
    \|A\|_{S^1(\pi)} = \int_{\Xi} \|A\|_1 d\pi(\xi),
\end{align*}
and 
\begin{align*}
    \|A\|_{S^\infty(\pi)} = \left(\int_{\Xi} \|A\|_\infty^{\infty} d\pi(\xi)\right)^{1/\infty} = \max_{\xi \in \Xi} \|A(\xi)\|_{\op}.
\end{align*}
The following holder's inequality is crucial to obtain dimension-free bounds on variances:
\begin{lemma}
    \label{lemma:kernel_holder}
    For Markov kernel $\mathcal{T}$ and conditional matrix $A(\xi)$, We have
    \begin{align*}
        \|\mathcal{T} A\|_{S^1(\pi)} &\le \|\mathcal{T}\|_{S^{\infty}(\pi)} \|A\|_{S^1(\pi)}, 
    \end{align*}
    where 
    \begin{align*}
        \|\mathcal{T}\|_{S^{\infty}(\pi)} := \sup_{\|Y\|_{S^\infty(\pi)}\le 1} \|\mathcal{T} Y\|_{S^{\infty}(\pi)}.
    \end{align*}
\end{lemma}
Using the results from Markov chain literature,  we have the following lemma:
\begin{lemma}[Proposition 22.3.5 in \cite{douc2018markov}]
    Let $\mathcal{P}$ be a Markov Kernel on a Boral state-space $\Xi$ with invariant probability $\pi$. Under Assumption \ref{assumption:noise_field}, we have
    \begin{align*}
        \|(\mathcal{P}-\Pi)^k\|_{L^2(\pi)} &\le \sqrt{2c_\rho} \rho^{k/2}, \\
        \|(\mathcal{P}-\Pi)^k\|_{S^{\infty}(\pi)} &\le 2 c_\rho \rho^{k}. 
    \end{align*}
\end{lemma}
The following is the corollary:
\begin{corollary}
    \label{corollary:inverse_operator_bound}
    Under Assumption \ref{assumption:noise_field}, we have
    \begin{align*}
        \max\left( \|(I - \mathcal{P}^{\dagger}+\Pi)^{-1}\|_{L^2(\pi)}, \|(I - \mathcal{P}^{\dagger}+\Pi)^{-1}\|_{S^{\infty} (\pi)} \right) \le 2c_{\rho} / (1-\rho).
    \end{align*}
\end{corollary}

\subsubsection{Norm-Bounds for Stationary Bias}
We show that $\|\delta^x\|_{L_2(\pi)} = O_{\texttt{P}}(\alpha), \|\delta^y\|_{L_2(\pi)} = O_{\texttt{P}}(\beta)$. First, we note that 
\begin{align*}
    \|\delta^x\|_{L^2(\pi)} &\le \alpha( C^x_1 \|\delta^x\|_{L^2(\pi)} + C^x_2  \|\delta^y\|_{L^2(\pi)} + C_{3}^x),
\end{align*}
where
\begin{align*}
    C(\mathcal{P},\pi) &:= \|(I - \mathcal{P}^{\dagger} + \Pi)^{-1}\|_{L^2(\pi)} \|\mathcal{P}^{\dagger} - \Pi\|_{L^2(\pi)} \le \frac{4 c_\rho}{1-\rho} = O(\tau_{\alpha}), \\
    C_1 &:= C(\mathcal{P},\pi) \|\Delta+W_{\Delta}^x\|_{L^2(\pi)} \le C(\mathcal{P},\pi) J_{\max} \kappa_y, \\
    C_2 &:= C(\mathcal{P},\pi) \|J_{12} + W_{12}\|_{L^2(\pi)} \le C(\mathcal{P},\pi) J_{\max}, \\
    C_3 &:= C(\mathcal{P},\pi) (\sigma_x + W_{\max}\kappa_y \|\bar{z}^x\|_{L^2(\pi)} + W_{\max} \|\bar{z}^y\|_{L^2(\pi)}) \\
    &\le C(\mathcal{P},\pi) (\sigma_x + O_{\texttt{P}}(\beta) W_{\max}\kappa_y).
\end{align*}
The last inequality is because
\begin{align*}
    \|\bar{z}^y\|_{L^2(\pi)} &= \int_{\Xi} \|\Exs[\bar{y}_{\infty}|\xi] - \Exs[\bar{y}_{\infty}]\|^2 \pi(d\xi) \le \int_{\Xi} \Exs[\|\bar{y}_{\infty} - \Exs[\bar{y}_{\infty}]\|^2 |\xi] \pi(d\xi) \\
    &= \Exs[\|\bar{y}_{\infty} - \Exs[\bar{y}_{\infty}]\|^2] \le \Exs[\|\bar{y}_{\infty}\|^2] = O_{\texttt{P}}(\beta), 
\end{align*}
by Theorem \ref{theorem:MSE_convergence}, and similarly, we can also show that $\|\bar{z}^x\|_{L^2(\pi)} = O_{\texttt{P}}(\alpha + \beta^2)$. Furthermore,
\begin{align*}
    \|\delta^y\|_{L^2(\pi)} \le \beta( C_1 \|\delta^x\|_{L^2(\pi)} + C_2  \|\delta^y\|_{L^2(\pi)} + C_{3}) + \|\delta^x\|_{L^2(\pi)},
\end{align*}
for the same problem-dependent constants $C_1, C_2, C_3$ as defined above. This concludes that for $\alpha \ll \beta \ll 1/\max(C_1, C_2)$, we have
\begin{align*}
    \|\delta^x\|_{L^2(\pi)} = O_{\texttt{P}}(\alpha), \|\delta^y\|_{L^2(\pi)} = O_{\texttt{P}}(\beta). 
\end{align*}
Similarly, we can show that 
\begin{align*}
    \|d^x\|_{L^2(\pi)} &\le \frac{O(1) \sigma^x}{1-\rho}, \quad
    \|d^y\|_{L^2(\pi)} \le \frac{O(1) \sigma^y}{1-\rho},
\end{align*}
which implies that $\bar{b}_1^y, \bar{b}_2^y = O_{\texttt{P}}(1)$ since 
\begin{align*}
    \|\bar{b}_2^y \|_2 = \left\|J_{22}^{-1} \int_{\Xi} W_{22} d^y d\pi(\xi) \right\|_2 \le \kappa_y \left(\int_{\Xi} \|d^y\|_2 d\pi(\xi)\right) \le \kappa_y \|d^y\|_{L^2(\pi)}. 
\end{align*}
Similarly, we have $\|\bar{b}_1^y\|_2 = O_\texttt{P}(1)$ and $\bar{b}_i^x = O_{\texttt{P}}(1)$ for $i=1,2$. We can plug this result back to \eqref{eq:stationary_bias} to conclude the bias part of Theorem \ref{theorem:distribution_conv}.

\subsubsection{Dimension-Free Bounds for Variances} 

We note that the variance of $x_{\infty}$ is measured by
\begin{align*}
    \|\Var(x_{\infty})\|_1 = \Tr(\Var(x_{\infty})) = \Tr(\Exs[(x_{\infty} - \Exs[x_{\infty}])(x_{\infty} - \Exs[x_{\infty}])^\top]),
\end{align*} 
where the expectation is taken over the stationary distribution $(x_{\infty}, y_{\infty}, \xi_{\infty}) \sim \mu$, and thus we aim bound $\Tr(\Var(x_{\infty}))$. For $y$, it is sufficient to bound $y_{\infty}$ by $O(\beta)$. To see this, note that
\begin{align*}
    \Tr(\Var(y_{\infty})) &= \Exs[\|y_{\infty} - \Exs[y_{\infty}]\|^2] = \Exs[\|y_{\infty} - y^*(x^*)\|^2] + \|\Exs[y_{\infty}] - y^*(x^*) \|^2 \\
    &\le 2\Exs[\|\bar{y}_{\infty}\|^2] + 2\Exs[\|y^*(x_{\infty}) - y^*(x^*)\|^2] + \|\Exs[y_{\infty}(x_{\infty}) - y^*(x^*)] \|^2 \\
    &\le 2\Exs[\|\bar{y}_{\infty}\|^2] + 3\kappa_y \Exs[\|\bar{x}_{\infty}\|^2] = O_{\texttt{P}}(\beta).
\end{align*}
Similarly, we also have that 
\begin{align*}
    \Tr(\Var(x_{\infty})) = O_{\texttt{P}}(\alpha + \beta^2),
\end{align*}
and 
\begin{align*}
    \|\Exs[(x_{\infty} - \Exs[x_{\infty}]) \bar{y}_{\infty}^\top ]\|_1 = O_{\texttt{P}} (\alpha + \beta^2).
\end{align*}

Next, we show that the variance of $x_\infty$ is strictly in order $O(\alpha)$, without $\textrm{poly}(\beta)$ dependence. We first observe that
\begin{align*}
    &(x_{\infty+1}-\Exs[x_{\infty}])(x_{\infty+1}-\Exs[x_{\infty}])^\top \\
    &=(x_{\infty}-\Exs[x_{\infty}])(x_{\infty}-\Exs[x_{\infty}])^\top - \alpha (x_{\infty}-\Exs[x_{\infty}]) (\Delta \bar{x}_{\infty} + J_{12} \bar{y}_{\infty} + w_{\infty}^x)^\top \\
    &\qquad - \alpha (\Delta \bar{x}_{\infty} + J_{12} \bar{y}_{\infty} + w_{\infty}^x)(x_{\infty}-\Exs[x_{\infty}])^\top + \alpha^2 (\Delta \bar{x}_{\infty} + J_{12} \bar{y}_{\infty} + w_{\infty}^x)(\Delta \bar{x}_{\infty} + J_{12} \bar{y}_{\infty} + w_{\infty}^x)^\top.
\end{align*}
Let us define $\Sigma^x(\xi)$ and $\Sigma^{xy}(\xi)$ as the following:
\begin{align*}
    \Sigma^x(\xi) &= \Exs[(x_{\infty} - \Exs[x_{\infty}])(x_{\infty} - \Exs[x_{\infty}])^\top | \xi_{\infty} = \xi], \\
    \Sigma^{xy}(\xi) &= \Exs[(x_{\infty} - \Exs[x_{\infty}])\bar{y}_{\infty}^\top | \xi_{\infty} = \xi] = \Exs[(x_{\infty} - \Exs[x_{\infty}])(y_{\infty} - y^*(x_{\infty}))^\top | \xi_{\infty} = \xi], \\
    \Sigma^{y}(\xi) &= \Exs[\bar{y}_{\infty}\bar{y}_{\infty}^\top | \xi_{\infty} = \xi]. 
\end{align*}
We can then rewrite the recursion compactly:
\begin{align*}
    \Sigma^x = \mathcal{P}^{\dagger} \{\Sigma^x - \alpha (A+A^\top) + \alpha^2 B \},
\end{align*}
where
\begin{align*}
    A(\xi) &= \Sigma^x(\xi) \Delta + \Sigma^{xy}(\xi) J_{12}^\top + \Exs[(x_{\infty} - \Exs[x_{\infty}])^\top w_{\infty}^x | \xi_{\infty} = \xi] \\
    &= \Sigma^x(\xi) (\Delta + W_\Delta^x(\xi))^\top + \Sigma^{xy}(\xi) (J_{12} + W_{12}(\xi))^\top + \delta^x(\xi) ( u_1(\xi) + W_{\Delta}^x (\xi) \Exs[x_{\infty}])^\top, \\
    B(\xi) &= (\Delta + W_{\Delta}^x (\xi)) \Sigma^x(\xi) (\Delta + W_\Delta^x (\xi))^\top + (J_{12} + W_{12}(\xi) ) \Sigma^y(\xi) (J_{12} + W_{12}(\xi))^\top \\
    &\qquad + (\Delta + W_\Delta^x (\xi)) \Sigma^{xy}(\xi) (J_{12} + W_{12}(\xi) )^\top +  (J_{12} + W_{12}(\xi) ) \Sigma^{yx}(\xi) (\Delta + W_\Delta^x (\xi))^\top \\
    &\qquad + (W_{\Delta}^x(\xi) z^x(\xi) + u_1(\xi))(W_{\Delta}^x(\xi) z^x(\xi) + u_1(\xi))^\top + O(\delta^x(\xi) + \delta^y(\xi)).
\end{align*}
Let $\bar{\Sigma}^x = \pi\{\Sigma^x\} = \Exs[\Sigma^x]$, $D^x(\xi) := \Sigma^x(\xi) - \bar{\Sigma}^x$, and similarly define $\bar{\Sigma}^{xy}, D^{xy}$. The steady-state equation is given by
\begin{align}
    \bar{\Sigma}^x \Delta + \Delta \bar{\Sigma}^x + (\bar{\Sigma}^{xy} J_{12}^\top + J_{12} \bar{\Sigma}^{yx}) &=  \alpha \pi\left\{(W_{\Delta}^x(\xi) z^x(\xi) + u_1(\xi))(W_{\Delta}^x(\xi) z^x(\xi) + u_1(\xi))^\top \right\} \nonumber \\
    &\quad + O_{\texttt{P}}(\alpha) (\|\bar{\Sigma}^x\|_{1} + \|\bar{\Sigma}^{xy}\|_1 + \|\bar{\Sigma}^y\|_1 + \|\delta^x\|_{L^2(\pi)} + \|\delta^y\|_{L^2(\pi)} ) \nonumber \\
    &\quad + O_{\texttt{P}}(\|D^x\|_{S^1(\pi)} + \|D^{xy}\|_{S^1(\pi)} + \|\delta^x\|_{L^2(\pi)}). \label{eq:var_steady_state_x}
\end{align}
We also note that $\mathcal{P}^\dagger \{\bar{\Sigma}^x\} = \bar{\Sigma}^x$ and $(\mathcal{P}^\dagger - \Pi) \{\Sigma^x\} = (\mathcal{P}^\dagger - \Pi) \{D^x\}$, and thus similarly to \eqref{eq:bias_recursion_intermediate}, 
\begin{align*}
    D^x &= -\alpha (I - \mathcal{P}^\dagger + \Pi)^{-1} (\mathcal{P}^\dagger - \Pi) \{ O_{\texttt{P}}(D^x + \bar{\Sigma}^x + D^{xy} + \bar{\Sigma}^{xy} + \delta^x) \} \\
    &\qquad + \alpha^2 (I - \mathcal{P}^\dagger + \Pi)^{-1} (\mathcal{P}^\dagger - \Pi) \{O_{\texttt{P}}(\bar{\Sigma}^x + \bar{\Sigma}^{xy} + \bar{\Sigma}^y + \delta^x + \delta^y + 1) \}.
\end{align*}
Taking $\|\cdot\|_{S^1(\pi)}$ of $D^x$, with Lemma \ref{lemma:kernel_holder} and Corollary \ref{corollary:inverse_operator_bound}, we can show that
\begin{align*}
    \|D^x\|_{S^1(\pi)} &\le O_{\texttt{P}} (\alpha) (\|D^x\|_{S^1(\pi)} + \|D^{xy}\|_{S^1(\pi)} + \|\bar{\Sigma}^x\|_{1} + \|\bar{\Sigma}^{xy}\|_{1}) + \alpha^2 O_{\texttt{P}}(1),
\end{align*}
where we also used $\|W_{ij}\|_{S^\infty(\pi)} \le J_{\max}$ for $i,j\in\{1,2\}$ by Assumption \ref{assumption:noise_norm_bound}, and used a Cauchy-Schwarz inequality
\begin{align*}
    \|AB\|_{S^1(\pi)} &\le \int_{\xi} \|A(\xi)\|_1 \|B(\xi)\|_\infty d\pi(\xi) \le \|A\|_{S^1(\pi)} \|B\|_{S^\infty(\pi)}, \\
    \|uv^\top\|_{S^1(\pi)} &\le \int_{\xi} \|u(\xi)\|_2 \|v(\xi)\|_2 d\pi(\xi) \le \|u\|_{L^2(\pi)} \|v\|_{L^2(\pi)},
\end{align*}
with $\|\delta^x\|_{L^2(\pi)} = O_{\texttt{P}}(\alpha), \|\delta^y\|_{L^2(\pi)} = O_{\texttt{P}} (\beta)$. This suggests that as long as $\|D^{xy}\|_{S^1(\pi)}, \|\bar{\Sigma}^x\|_1, \|\bar{\Sigma}^{xy}\|_1 = o_{\texttt{P}}(1)$, we have $\|D^x\|_{S^1(\pi)} = o_{\texttt{P}}(\alpha)$.

To proceed, we also get the expression for $\Sigma^{xy}$:
\begin{align*}
    \Sigma^{xy} = \mathcal{P}^{\dagger} \{\Sigma^{xy} - \beta A' - \alpha B' + O(\alpha\beta) \},
\end{align*}
where
\begin{align*}
    A'(\xi) &= \Sigma^{xy}(\xi) (J_{22}+W_{22}(\xi))^\top + \Sigma^{x}(\xi) W_{21}(\xi) + \delta^x(\xi) (u_2(\xi) + W_{\Delta}^y(\xi) \Exs[x_{\infty}])^\top, \\
    B'(\xi) &= (\Sigma^x(\xi) (\Delta + W_{\Delta}^x(\xi)^\top + \Sigma^{xy} (\xi) (J_{12} + W_{12}(\xi))^\top) (J_{22}^{-1} J_{21})^\top \\
    &\qquad + \Sigma^y(\xi) (J_{12}+W_{12}(\xi))^\top + W_{\Delta}^x(\xi) \Sigma^{yx}(\xi) + \delta^y(\xi) (W_{\Delta}^x(\xi) \Exs[x_{\infty}] + u_1(\xi))^\top.
\end{align*}
and $C'$ is appropriately defined. The steady-state equation is
\begin{align}
    \bar{\Sigma}^{xy} J_{22}^\top + \frac{\alpha}{\beta} \bar{\Sigma}^y J_{12}^\top &= \alpha \pi \left\{ (W_{\Delta}^x(\xi) z^x(\xi) + u_1(\xi))(W_{\Delta}^y(\xi) z^x(\xi) + u_2(\xi))^\top \right\} \nonumber \\
    &\quad + O_{\texttt{P}}(\|D^x\|_{S^1(\pi)} + \|D^{xy}\|_{S^1(\pi)} + \|\delta^x\|_{L^2(\pi)}) + O_{\texttt{P}}(\alpha), \label{eq:var_steady_state_xy}
\end{align}
and the system equation is
\begin{align*}
    D^{xy} &= (I-\mathcal{P}^\dagger + \Pi)^{-1} (\mathcal{P}^\dagger - \Pi) \{\beta O_{\texttt{P}}(D^x + D^{xy} + \bar{\Sigma}^x + \bar{\Sigma}^{xy} + \delta^x) + \alpha O_{\texttt{P}}(D^y + \bar{\Sigma}^y + \delta^y) + O_{\texttt{P}}(\alpha\beta) \}.
\end{align*}
Noting that $\|\delta^x\|_{L^2(\pi)} = O(\alpha), \|\delta^y\|_{L^2(\pi)} = O(\beta)$, we can show that 
\begin{align*}
    \|D^{xy}\|_{S^1(\pi)} \le O_{\texttt{P}}(\beta) (\|D^x\|_{S^1(\pi)} + \|D^{xy}\|_{S^1(\pi)} + \|\bar{\Sigma}^x\|_1 + \|\bar{\Sigma}^{xy}\|_1) + O_{\texttt{P}} (\alpha) (\|D^y\|_{S^1(\pi)} + \|\bar{\Sigma}^{y}\|_{1}) + O_{\texttt{P}}(\alpha\beta).
\end{align*} 
Combining these results, we can conclude that
\begin{align*}
    \|D^{xy}\|_{S^1(\pi)}, \|D^x\|_{S^1(\pi)} = O_{\texttt{P}}(\alpha) (\|\bar{\Sigma}^x\|_1 + \|\bar{\Sigma}^{xy}\|_1) + O_{\texttt{P}}(\alpha\beta).
\end{align*}
Now plugging this back to \eqref{eq:var_steady_state_xy}, we have
\begin{align*}
    \|\bar{\Sigma}^{xy}\|_1 &\le \frac{\kappa_y \alpha}{\beta} \|\bar{\Sigma}^y\|_1 + O_{\texttt{P}} (\|D^x\|_{S^1(\pi)} + \|D^{xy}\|_{S^1(\pi)} + \|\delta^x\|_{L^2(\pi)}) + o_{\texttt{P}}(\alpha),
\end{align*}
yielding $\|\bar{\Sigma}^{xy}\|_{1} = O_{\texttt{P}}(\alpha)$ since $\|\bar{\Sigma}^y\|_1 = O_\texttt{P}(\beta)$. Then using these results, from \eqref{eq:var_steady_state_x}, we can derive that
\begin{align*}
    \Tr(\bar{\Sigma}^x) + &\Tr(\Delta \bar{\Sigma}^x \Delta^{-1}) = 2\Tr(\bar{\Sigma}^x) =  2 \|\bar{\Sigma}^x\|_1 \\
    &\le \|\bar{\Sigma}^{xy} J_{12}^\top\|_1 \|\Delta^{-1}\|_{\op} + O(\alpha) \|\Delta^{-1}\|_{\op} \|W_{\Delta}^x \circ z^x + u_1\|_{L^2(\pi)}^2 + O_{\texttt{P}}(\alpha)  \|\bar{\Sigma}^x\|_1 + o_{\texttt{P}} (\alpha). 
\end{align*}
Therefore, we can conclude that $\|\bar{\Sigma}^x\|_{1} = O_{\texttt{P}}(\alpha)$. Since $\|\bar{\Sigma}^x\|_1 = \Tr(\bar{\Sigma}^x) = \Tr(\Var(x_{\infty}))$, we obtain the last part of the theorem.

\subsubsection{Proof of Corollary \ref{theorem:fine_grained_distributional_convergence_rate}}
This is in fact a corollary of Lemma \ref{lemma:auxiliary_W2_lemma}. To see this, apply Lemma \ref{lemma:auxiliary_W2_lemma} with $\mu_0^2 = \mu$, and then note that under optimal coupling between $\mu_0$ and $\mu$,
\begin{align*}
    \mathcal{W}_2^2(\mu_0(x_0), \mu(x_{\infty})) &\le \Exs[\|x_0 - x_{\infty}\|_2^2] \le 2\Exs[\|x_0 - \Exs[x_{\infty}]\|_2^2] + \Exs[\|x_{\infty} - \Exs[x_{\infty}]\|_2^2] \\
    &= 2\Exs[\|x_0 - \Exs[x_{\infty}]\|_2^2] + 2\Tr(\Var(x_{\infty})). 
\end{align*}
Similarly, 
\begin{align*}
    \mathcal{W}_2^2(\mu_0(\bar{y}_0), \mu(\bar{y}_{\infty})) &\le  2\Exs[\|\bar{y}_0 - \Exs[\bar{y}_{\infty}]\|_2^2] + 2\Tr(\Var(\bar{y}_{\infty})). 
\end{align*}
Then from Theorem \ref{theorem:distribution_conv}, applying $\Tr(\Var(x_{\infty})) = O_{\texttt{P}}(\alpha)$ and $\Tr(\Var(\bar{y}_{\infty})) = O_{\texttt{P}}(\beta)$, we have the lemma.

\subsection{Tail Averaging and Extrapolation}

\subsubsection{Proof of Theorem \ref{theorem:tail_averaging}}
First, let us define $V_k, U_k$ as:
\begin{align*}
    U_k &= \Exs[\|x_k - \Exs[x_{\infty}]\|^2] + \Exs[\|y_k - \Exs[y_{\infty}]\|^2] + O_{\texttt{P}}(\beta), \\
    V_k &= \Exs[\|x_k - \Exs[x_{\infty}]\|^2] + \frac{\alpha}{\beta} \Exs[\|y_k - \Exs[y_{\infty}]\|^2] + O_{\texttt{P}}(\alpha).
\end{align*} 
For all $k \ge t_0 \gg (\alpha\mu_x)^{-1} \log(1/(\alpha\mu_x))$, with Theorem \ref{theorem:fine_grained_distributional_convergence_rate}, we ensure that under an optimal coupling,
\begin{align*}
    \Exs[\|x_k - \Exs[x_{\infty}]\|^2] &\le \Exs[\|x_k - x_{\infty}\|^2] + \Tr(\Var(x_{\infty})) \le O_{\texttt{P}}(\alpha),
\end{align*}
and similarly, $\Exs[\|y_k - \Exs[y_{\infty}]\|^2] \le O_{\texttt{P}}(\beta)$.

\paragraph{Slower Iterate:} We want to analyze 
\begin{align*}
    \Exs[\|\tilde{x}_t - x^*\|_{2}^2] = \Exs[\|\tilde{x}_t - \Exs[x_{\infty}] + (\Exs[x_{\infty}] - x^*)\|_{2}^2] \le 2\Exs[\|\tilde{x}_t - \Exs[x_{\infty}]\|_{2}^2] + 2\Exs[\|\Exs[x_{\infty}] - x^*\|_{2}^2],
\end{align*}
where $x_{\infty} \sim \mu(x)$. To show that this quantity is $O(\alpha)$, it suffices to bound $\Exs[\|\tilde{x}_t - \Exs[x_{\infty}] \|_{Q_x}^2]$ under the optimal coupling. Rewriting this term,
\begin{align*}
    \Exs[\|\tilde{x}_t - \Exs[x_{\infty}] \|_{2}^2] = \frac{1}{(t-t_0)^2} \sum_{k = t_0}^t \Exs[\|x_k - \Exs[x_{\infty}]\|_2^2] + \frac{2}{(t-t_0)^2} \sum_{k = t_0}^t \sum_{l > k }^t \Exs[\vdot{x_k - \Exs[x_{\infty}]}{x_l - \Exs[x_{\infty}]}].
\end{align*}
We first note that by Theorem \ref{theorem:fine_grained_distributional_convergence_rate}, under the optimal coupling between $x_k$ and $x_{\infty}$, we get
\begin{align*}
    \Exs[\|x_k - \Exs[x_{\infty}]\|_{2}^2] &\le 2 \Exs[\|x_k - x_{\infty}\|_{2}^2] + 2 \Exs[\|x_{\infty} - \Exs[x_{\infty}]\|_{2}^2] \\
    &\le \exp(-\alpha \mu_x (k-t_0) / 8) V_{t_0} + \Tr(\Var(x_{\infty})). 
\end{align*}
To proceed, we note that
\begin{align*}
    \Exs[\|\tilde{x}_k - \Exs[x_{\infty}]\|_2^2] &\le \frac{1}{(t-t_0)^2} \sum_{k=0}^{t-t_0} \left( \exp(-\alpha \mu_x k / 8) V_{t_0} + \Tr(\Var(x_{\infty})) \right) \\
    &\qquad + \frac{2}{(t-t_0)^2} \sum_{k=t_0}^{t} \sum_{k' > 0}^{t - k} \Exs[\Exs[\vdot{x_k - \Exs[x_{\infty}]}{x_{k+k'} - \Exs[x_{\infty}]} | \mathcal{F}_k]] \\
    &\le \frac{1}{(t-t_0)^2} \sum_{k=0}^{t-t_0} \left( \exp(-\alpha \mu_x k / 8) V_{t_0} \right) + \frac{\Tr(\Var(x_{\infty}))}{t-t_0} \\
    &\qquad + \frac{2}{(t-t_0)^2} \sum_{k=t_0}^{t} \sum_{k' > 0}^{t - k} \Exs[\|x_k - \Exs[x_{\infty}]\|_2 \cdot \|\Exs[x_{k+k'} - x_{\infty} | \mathcal{F}_k]\|_2 ].
\end{align*}
To bound the second term, we first note that for any $k' > 0$, we use an optimal coupling between $x_{k+k'} | \mathcal{F}_k$ and $x_{\infty}$, and again apply Theorem \ref{theorem:fine_grained_distributional_convergence_rate}:
\begin{align*}
    \Exs[\| \Exs[x_{k+k'} | \mathcal{F}_k] - \Exs[x_{\infty}]\|_{2}^2 ] &\le \Exs[\Exs[ \|x_{k+k'} - x_{\infty} \|_2^2 | \mathcal{F}_k]] \le \exp(-\alpha \mu_x k' / 8) V_k.
\end{align*}
Using Cauchy-Schwarz inequality, we have
\begin{align*}
    \sum_{k=t_0}^t \sum_{k' > 0}^{t - k} &\Exs[\|x_k - \Exs[x_{\infty}]\|_2 \cdot \|\Exs[x_{k+k'} - x_{\infty} | \mathcal{F}_k]\|_2 ] \\
    &\le \sum_{k=t_0}^t \sqrt{\Exs[\|x_k - \Exs[x_{\infty}]\|_2^2]} \cdot  \left(\sum_{k' > 0}^{t - k} \sqrt{\Exs[\Exs[\|x_{k+k'} - x_{\infty}\|_2^2 | \mathcal{F}_k] ]} \right) \\
    &\le \sum_{k=t_0}^t \sqrt{V_k} \cdot \frac{\sqrt{V_k}}{\alpha \mu_x} \le \sum_{k=t_0}^t \frac{1}{\mu_x} = O_{\texttt{P}}(t-t_0),
\end{align*}
where we used that $V_k = O_{\texttt{P}}(\alpha)$ for all $k \ge t_0$. 
Plugging this, we can conclude that
\begin{align*}
    \Exs[\|\tilde{x}_k - \Exs[x_{\infty}]\|_2^2] &\le \frac{O_{\texttt{P}}(1)}{t-t_0}.
\end{align*}

\paragraph{Faster Iterate:} In this case, we first note that 
\begin{align*}
    \|\tilde{y}_t - y^*\|_2^2 &\le 2\|\tilde{y}_t - y^*(\tilde{x}_t)\|_2^2 + 2\|y^*(\tilde{x}_t) -  y^*(x^*) \|_2^2 \\
    &\le 4\|\tilde{\bar{y}}_t - \Exs[\bar{y}_{\infty}]\|_2^2 +  4\|\Exs[\bar{y}_{\infty}]\|_2^2 + 2\kappa_y^2 \|\tilde{x}_t -  x^* \|_2^2,
\end{align*}
where $\tilde{\bar{y}}_k := \frac{1}{t-t_0} \sum_{t'=t_0}^t \bar{y}_t$. The second term is squared-bias in order $O_{\texttt{P}}(\beta^2)$, and the third term inherits the error analysis from slower iterates. Thus, we focus on bounding the first term.

Following the same process for slower iterates, we first note that
\begin{align*}
    \Exs[\|\tilde{\bar{y}}_t - \Exs[\bar{y}_{\infty}] \|_{2}^2] &= \frac{1}{(t-t_0)^2} \sum_{k = t_0}^t \Exs[\|\bar{y}_k - \Exs[\bar{y}_{\infty}]\|_2^2] + \frac{2}{(t-t_0)^2} \sum_{k = t_0}^t \sum_{l > k }^t \Exs[\vdot{\bar{y}_k - \Exs[\bar{y}_{\infty}]}{\bar{y}_l - \Exs[\bar{y}_{\infty}]}] \\
    &\le \frac{1}{(t-t_0)^2} \sum_{k = t_0}^t \Exs[\|\bar{y}_k - \Exs[\bar{y}_{\infty}]\|_2^2] \\
    &\qquad + \frac{2}{(t-t_0)^2} \sum_{k = t_0}^t \sum_{l > k }^t \Exs[\|\bar{y}_k - \Exs[\bar{y}_{\infty}]\|_2 \cdot \|\Exs[\bar{y}_{k+k'} - \bar{y}_{\infty} | \mathcal{F}_k]\|_2 ].
\end{align*}
For the first term, we invoke Corollary \ref{theorem:fine_grained_distributional_convergence_rate}, under optimal coupling, we have
\begin{align*}
    \Exs[\|\bar{y}_k - \Exs[\bar{y}_{\infty}]\|_2^2] &\le 2\Exs[\|\bar{y}_k - \bar{y}_{\infty}\|_2^2] + 2 \Exs[\|\bar{y}_{\infty} - \Exs[\bar{y}_{\infty}]\|_2^2] \\
    &\le \beta \exp(-\alpha \mu_x (k-t_0) / 8) V_{t_0} + \exp(-\beta \mu_y (k-t_0) / 8) U_{t_0} + \Tr(\Var(\bar{y}_{\infty})).
\end{align*}
For the second term, with Corollary \ref{theorem:fine_grained_distributional_convergence_rate}, we have
\begin{align*}
    \Exs[\| \Exs[\bar{y}_{k+k'} | \mathcal{F}_k] - \Exs[\bar{y}_{\infty}]\|_{2}^2 ] &\le \Exs[\Exs[ \|\bar{y}_{k+k'} - \bar{y}_{\infty} \|_2^2 | \mathcal{F}_k]] \le \beta\exp(-\alpha \mu_x k' / 8) V_k + \exp(-\beta\mu_y k' / 8) U_k,
\end{align*}
and again using Cauchy-Schwarz inequality, we can show that
\begin{align*}
    \sum_{k=t_0}^t \sum_{k' > 0}^{t - k} &\Exs[\|\bar{y}_k - \Exs[\bar{y}_{\infty}]\|_2 \cdot \|\Exs[\bar{y}_{k+k'} - \bar{y}_{\infty} | \mathcal{F}_k]\|_2 ] \\
    &\le \sum_{k=t_0}^t \sqrt{\Exs[\|\bar{y}_k - \Exs[\bar{y}_{\infty}]\|_2^2]} \cdot  \left(\sum_{k' > 0}^{t - k} \sqrt{\Exs[\Exs[\|\bar{y}_{k+k'} - \bar{y}_{\infty}\|_2^2 | \mathcal{F}_k] ]} \right) \\
    &\le \sum_{k=t_0}^t \sqrt{U_k} \cdot \left(\frac{\sqrt{\beta V_k}}{\alpha \mu_x} + \frac{\sqrt{U_k}}{\beta\mu_y} \right) \le O_{\texttt{P}} (t-t_0) \left(\sqrt{\frac{\beta^2}{\alpha}} \frac{1}{\mu_x} + \frac{1}{\mu_y} \right).
\end{align*}
On the other hand, we can apply Cauchy-Schwarz inequality in different ways:
\begin{align*}
    \sum_{k=t_0}^t \sum_{k' > 0}^{t - k} &\Exs[\|\bar{y}_k - \Exs[\bar{y}_{\infty}]\|_2 \cdot \|\Exs[\bar{y}_{k+k'} - \bar{y}_{\infty} | \mathcal{F}_k]\|_2 ] \\
    &\le \sum_{k=t_0}^t \sqrt{\Exs[\|\bar{y}_k - \Exs[\bar{y}_{\infty}]\|_2^2]} \cdot  \left(\sum_{k' > 0}^{t - k} \sqrt{\Exs[\Exs[\|\bar{y}_{k+k'} - \bar{y}_{\infty}\|_2^2 | \mathcal{F}_k] ]} \right) \\
    &\le \sqrt{t-t_0} \sum_{k=t_0}^t \sqrt{U_k} \cdot \left(\sqrt{\frac{\beta V_k}{\alpha \mu_x}} + \sqrt{\frac{U_k}{\beta\mu_y}} \right) \le O_{\texttt{P}} (t-t_0)^{3/2} \left(\frac{\beta}{\sqrt{\mu_x}} + \sqrt{\frac{\beta}{\mu_y}} \right).
\end{align*}
Summarizing the results, we can conclude that
\begin{align*}
    \frac{1}{(t-t_0)^2} \Exs[\|(\tilde{y}_k - y^*(\tilde{x}_k)) - \Exs[\bar{y}_{\infty}]\|_2^2] &\le \frac{O_{\texttt{P}}(1)}{t-t_0} + \min\left( \sqrt{\frac{\beta^2}{\alpha}} \frac{ O_{\texttt{P}}(1) }{t-t_0}, \sqrt{\frac{ O_{\texttt{P}}(\beta) }{t-t_0}}\right).
\end{align*}
This concludes Theorem \ref{theorem:tail_averaging}.

\subsubsection{Proof of Corollary \ref{corollary:extrapolation}}
We note that
\begin{align}
    \Exs[\|\zeta_t^x - x^*\|_{2}^2] &\le 2\Exs[\|\zeta_t^x - \Exs[(2x_{\infty}^{2\alpha,2\beta} - x_{\infty}^{\alpha,\beta})]\|_2^2] + 2\|\Exs[(2x_{\infty}^{2\alpha,2\beta} - x_{\infty}^{\alpha,\beta})] - x^*\|_{2}^2 \nonumber \\
    &\le 16\Exs[\|\tilde{x}_t^{2\alpha,2\beta} - \Exs[x_{\infty}^{2\alpha,2\beta}]\|_2^2] + 4\Exs[\|\tilde{x}_t^{\alpha,\beta} - \Exs[x_{\infty}^{\alpha,\beta}]\|_2^2] + 2\|\Exs[(2x_{\infty}^{2\alpha,2\beta} - x_{\infty}^{\alpha,\beta})] - x^*\|_{2}^2. \label{eq:extraploation_bound}
\end{align}
Note that from Theorem \ref{theorem:distribution_conv}, 
\begin{align*}
    2x_{\infty}^{\alpha,\beta} - x_{\infty}^{2\alpha,2\beta} - x^* &= 2(x_{\infty}^{\alpha,\beta} - x^*) - (x_{\infty}^{2\alpha,2\beta} - x^*) = O_{\texttt{P}}(\beta^2).
\end{align*}
The first and second terms in \eqref{eq:extraploation_bound} can be bounded by $O_{\texttt{P}}(1)/(t-t_0)$, following exactly same steps in the proof of Theorem \ref{theorem:tail_averaging}. The result for faster iterates can also be derived similarly.

\section{Deferred Proofs}

\subsection{Proof of Lemma \ref{lemma:TTSA_diff_rewrite}}

    The stochastic approximation equation becomes
    \begin{align*}
        (x_{t+1}-x^*) &= (x_t-x^*) - \alpha F(x_t,y^*(x_t)) + \alpha (F(x_t,y^*(x_t)) - F(x_t, y_t)) - \alpha w^x(x_t,y_t; \xi_t) \\
        &= (x_t-x^*) - \alpha H(x_t) - \alpha J_{12} (y_t - y^*(x_t)) - \alpha w_t^x(x_t,y_t; \xi_t), \\
        (y_{t+1} - y^*(x_{t+1})) &= (y_t - y^*(x_{t})) + (y^*(x_t) - y^*(x_{t+1})) - \beta G(x_t, y_t) - \beta w^y(x_t,y_t; \xi_t) \\
        &= (y_t - y^*(x_{t})) - J_{22}^{-1} J_{21} (x_t - x_{t+1}) - \beta G(x_t, y_t) - \beta w^y(x_t,y_t; \xi_t) \\
        &= (y_t - y^*(x_{t})) - \alpha J_{22}^{-1} J_{21} (F(x_t, y_t) + w^f(x_t, y_t;\xi_t)) - \beta G(x_t, y_t) - \beta w^y(x_t,y_t; \xi_t) \\
        &= (y_t - y^*(x_{t})) - \alpha J_{22}^{-1} J_{21} (F(x_t, y_t) - H(x_t) + H(x_t)) - \beta G(x_t, y_t) \\
        &\qquad - \alpha J_{22}^{-1} J_{21} w^x(x_t, y_t;\xi_t) - \beta w^y(x_t,y_t; \xi_t).
    \end{align*}
    Using $H(x^*) = 0$, $G(x, y^*(x))=0$, we can rewrite the recursion as \eqref{eq:new_recursion}.

\subsection{Proof of Lemma \ref{lemma:fourth_moment_bound}}
We can start with a coarse bound on $\|\bar{y}_{t+1}\|_{Q_y}^2$:
\begin{align*}
    \|\bar{y}_{t+1}\|_{Q_y}^2 &\le \|(I - \beta J_{22}) \bar{y}_t \|^2_{Q_y} + \alpha^2 \| J_{22}^{-1} J_{21} (J_{12} \bar{y}_t + \Delta \bar{x}_t + w^x_t) \|^2_{Q_y} + \beta^2 \|w^y_t\|^2_{Q_y} \\
    &\qquad + 2\alpha \left| \vdot{(I-\beta J_{22}) \bar{y}_t}{-J_{22}^{-1}J_{21}(J_{12} \bar{y}_t + \Delta \bar{x}_t + w^x_t)}_{Q_y} \right| \\
    &\qquad + 2\beta \left| \vdot{(I - \beta J_{22})\bar{y}_t}{-w_t^y}_{Q_y}  \right| + 2\alpha \beta \left| \vdot{J_{22}^{-1}J_{21} (J_{12}\bar{y}_t + \Delta\bar{x}_t - w^x_t)}{w^y_t}_{Q_y} \right| \\
    &\le (1-\beta \mu_y/2) \|\bar{y}_t\|_{Q_y}^2 + O_{\texttt{P}} (\beta^2) \|\bar{x}_t\|_2^2 + O_{\texttt{P}} (\alpha) \sigma_x^2 + O_{\texttt{P}} (\beta) \sigma_y^2. 
\end{align*}
Thus, taking the square on both sides, we get
\begin{align*}
    \|\bar{y}_{t+1}\|_{Q_y}^4 &\le (1-\beta \mu_y/2)^2 \|\bar{y}_t\|_{Q_y}^4 + 2 (1-\beta \mu_y/2) \|\bar{y}_t\|_{Q_y}^2 \left( O_{\texttt{P}} (\beta^2) \|\bar{x}_t\|_2^2 + O_{\texttt{P}} (\alpha) \sigma_x^2 + O_{\texttt{P}} (\beta) \sigma_y^2 \right) \\
    &\quad + \left( O_{\texttt{P}} (\beta^2) \|\bar{x}_t\|_2^2 + O_{\texttt{P}} (\alpha) \sigma_x^2 + O_{\texttt{P}} (\beta) \sigma_y^2 \right)^2 \\
    &\le (1-\beta \mu_y/4) \|\bar{y}_t\|_{Q_y}^4 + \left( O_{\texttt{P}} (\beta^2) \|\bar{x}_t\|_2^4 + O_{\texttt{P}} (\alpha) \sigma_x^4 + O_{\texttt{P}} (\beta) \sigma_y^4 \right). 
\end{align*}
Similarly for $\bar{x}_t$, we have
\begin{align*}
\|\bar{x}_{t+1}\|_{Q_x}^2 &\le \|(I-\alpha\Delta)\bar{x}_t\|_{Q_x}^2 + 2 \alpha^2 \|J_{12} \bar{y}_t\|_{Q_x}^2 + 2 \alpha^2 \|w^x_t\|_{Q_x}^2 \\
&\qquad + 2\alpha |\vdot{(I-\alpha \Delta) \bar{x}_t}{J_{12} \bar{y}_t}_{Q_x}]| + 2\alpha |\vdot{(I-\alpha \Delta) \bar{x}_t}{w_t^x}_{Q_x} | + 2\alpha^2 | \vdot{J_{12}\bar{y}_t}{w^x_t}_{Q_x} | \\
&\le (1-\alpha\mu_x/2) \|\bar{x}_t\|_{Q_x}^2 + O_{\texttt{P}} (\alpha) \|\bar{y}_t\|_{2}^2 + O_{\texttt{P}} (\alpha) \sigma_x^2, 
\end{align*}
and thus,
\begin{align*}
\|\bar{x}_{t+1}\|_{Q_x}^4 &\le (1-\alpha\mu_x/4) \|\bar{x}_t\|_{Q_x}^4 + O_{\texttt{P}} (\alpha) \|\bar{y}_t\|_{2}^4 + O_{\texttt{P}} (\alpha) \sigma_x^4.
\end{align*}
Taking potential $V_t = \Exs[\|\bar{x}_{t}\|_{Q_x}^4 + \frac{O_{\texttt{P}}(1) \alpha}{\beta} \|\bar{y}_{t}\|_{Q_y}^4]$, we have
\begin{align*}
    V_{t+1} \le (1-\alpha \mu_x / 4) V_t + O_{\texttt{P}}(\alpha) (\sigma_x^4 + \sigma_y^4),
\end{align*}
which leads to 
\begin{align*}
    \Exs[\|\bar{x}_t\|_{Q_x}^4] \le V_{t} \le \exp(-\alpha \mu_x t / 4) V_0 + O_{\texttt{P}}(\sigma_x^4 + \sigma_y^4). 
\end{align*}
Plugging this back to the recursion for $y$, we also have
\begin{align*}
    \Exs[\|\bar{y}_t\|_{Q_y}^4] \le \exp(-\beta \mu_x t / 4) \Exs[\|\bar{y}_0\|_{Q_y}^4] + O_{\texttt{P}}(\sigma_x^4 + \sigma_y^4).
\end{align*}
Converting $\|\cdot\|_{Q_x}$ and $\|\cdot\|_{Q_y}$ to $\|\cdot\|_2$ norm concludes the lemma.

\subsection{Proof of Lemma \ref{lemma:auxiliary_W2_lemma}}

The proof strategy is to consider three cases separately when $t$ is small and large. Let $c_1 > 0$ be some sufficiently large absolute constant.

\paragraph{Case (i) $t \le c_1 \cdot \tau_\alpha$:} In this case, consider optimal coupling between $\mu_0^1, \mu_0^2$, and apply Lemma \ref{lemma:sequence_difference_bound}:
\begin{align*}
    \Exs[\|x_t^1 - x_t^2\|_2^2] &\le 3 \Exs[\|x_0^1 - x_0^2\|_2^2] + 3 \Exs[\|x_t^1 - x_0^1\|_2^2] + 3 \Exs[\|x_t^1 - x_0^1\|_2^2] \\
    &\le 3 \mathcal{W}_2^2 (\mu_0^1(x_0^1), \mu_0^2(x_0^2)) + O_{\texttt{P}} (\alpha^2 \tau^2). 
\end{align*}
Since in this case $\exp(-\alpha \mu_x t / 4) > 1/2$ for $t < O(1) \tau_{\alpha} \ll 1/(\alpha \mu_x)$, the $x$ part in the inequality \eqref{eq:dist_conv_rate_ineq} holds. The $y$ part of \eqref{eq:dist_conv_rate_ineq} can be shown similarly.

\paragraph{Case (ii) $c_1 \cdot \tau_{\alpha} < t$:}
We consider a coupling on $\xi_\tau^1$ and $\xi_\tau^2$ first. Let $\nu^1, \nu^2$ be probability distributions over $\Xi \times \Xi$ such that
\begin{align*}
    &\nu^1(\xi^1, \xi^2) \propto \indic{\xi^1 = \xi^2} \cdot \min(\mu_\tau^1(\xi^1), \mu_\tau^2(\xi^2)), \\
    &\nu^2(\xi^1, \xi^2) \propto \max(0, \mu_\tau^1(\xi^1) - \mu_\tau^2(\xi^1)) \times \max(0, \mu_\tau^2(\xi^2) - \mu_\tau^1(\xi^2)).
\end{align*}
The coupling distribution decides $\nu^1$ with probability $1 - \texttt{TV}(\mu_\tau^1(\xi_\tau^1), \mu_\tau^2(\xi_\tau^2))$, and $\nu^2$ with probability $\texttt{TV}(\mu_\tau^1(\xi_\tau^1), \mu_\tau^2(\xi_\tau^2))$ to sample $(\xi_\tau^1, \xi_\tau^2)$. Then it samples $(x_\tau^1, y_\tau^1) \sim \mu_\tau^1(\cdot | \xi_\tau^1)$ and $(x_\tau^2, y_\tau^2) \sim \mu_\tau^2(\cdot | \xi_\tau^2)$. When $\nu^1$ is selected, we couple two sequences by setting $\xi_t^1 = \xi_t^2$ for all $t \ge 0$, and invoke Lemma \ref{lemma:noise_coupled_convergence} to show that
\begin{align*}
    \Exs[\|x_t^1-x_t^2\|_{2}^2 \mid \nu^1] &\le O_{\texttt{P}} (1) \cdot \Exs\left[\|x_\tau^1-x_\tau^2\|_{2}^2 + \frac{\alpha}{\beta} \|y_\tau^1-y_\tau^2\|_{2}^2 \mid \nu^1 \right] \exp(-\alpha \mu_x (t-\tau) / 4),
\end{align*}
where we also took expectation over the optimal coupling for $(x_\tau^1,y_\tau^1)|\xi_0^1$ and $(x_\tau^2,y_\tau^2)|\xi_\tau^2$. When $\nu^2$ is selected, we let the two sequences independently evolve, and using Lemma \ref{lemma:fourth_moment_bound} to show that
\begin{align*}
    \Exs&[\|x_t^1-x_t^2\|_{2}^2  \cdot \indic{\nu^2} ] \le O_{\texttt{P}} (1) \sqrt{\Exs[\|x_t^1 -x_t^2\|_{2}^4 + \|\bar{y}_t^1 - \bar{y}_t^2\|_{2}^4 ]} \cdot \sqrt{\texttt{TV}(\mu_0^1(\xi_0^1), \mu_0^2(\xi_0^2)} \\
    &\le O_{\texttt{P}} (1) \sqrt{\Exs[\|x_t^1\|_2^4 + \|x_t^2\|_{2}^4 + \|\bar{y}_t^1\|_2^4 + \|\bar{y}_t^2\|_{2}^4 ]} \cdot \sqrt{\texttt{TV}(\mu_\tau^1(\xi_\tau^1), \mu_\tau^2(\xi_\tau^2)} \\
    &\le \sqrt{O_{\texttt{P}}(1) \TV(\mu_\tau^1(\xi_\tau^1), \mu_\tau^2(\xi_\tau^2))}.
\end{align*}

Given the above results, let $c_2 > 0$ be another sufficiently large absolute constant. Now if $t \le c_2 \log(\beta/\alpha) / (\alpha \mu_x)$, then we set $\tau = c_1 \tau_{\alpha} / 4$ such that $t > 4 \tau$. With $\TV(\mu_\tau^1(\xi_\tau^1), \mu_\tau^2(\xi_\tau^2)) < \rho^\tau \ll (\mu_x \alpha)^{O(1)} \le \exp(-\alpha \mu_x t / 4)$, we have
\begin{align*}
    \Exs[\|x_t^1-x_t^2\|_{2}^2] &\le O_{\texttt{P}} (1) \cdot \Exs\left[\|x_\tau^1-x_\tau^2\|_{2}^2 + \frac{\alpha}{\beta} \|y_\tau^1-y_\tau^2\|_{2}^2 + O_{\texttt{P}}(\alpha) \right] \exp(-\alpha \mu_x t / 8) \\
    &\le O_{\texttt{P}} (1) \cdot V_0 \exp(-\alpha \mu_x t / 8).
\end{align*}
On the other hand, if $t > c_2 \log(\beta/\alpha) / (\alpha \mu_x)$, then we take $\tau = t/8$. In this case, we instead invoke the MSE result in Theorem \ref{theorem:MSE_convergence}, which gives
\begin{align*}
    \Exs[\|x_\tau^1 - x_\tau^2\|_2^2] &\le 2 \Exs[\|\bar{x}_{\tau}^1\|_2^2] + 2 \Exs[\|\bar{x}_{\tau}^2\|_2^2] \le O_{\texttt{P}}(1) \cdot \exp(-\alpha \mu_x \tau / 4) \ll O_{\texttt{P}}(\alpha), \\
    \Exs[\|\bar{y}_\tau^1 - \bar{y}_\tau^2\|_2^2] &\le 2 \Exs[\|\bar{y}_{\tau}^1\|_2^2] + 2 \Exs[\|\bar{y}_{\tau}^2\|_2^2] \\
    &\le O_{\texttt{P}}(1) \cdot \left(\exp(-\beta \mu_y \tau / 4)+ \beta \exp(-\alpha \mu_x \tau / 4)\right) \ll O_{\texttt{P}}(\alpha). 
\end{align*}
Together with $\TV(\mu_\tau^1(\xi_\tau^1), \mu_\tau^2(\xi_\tau^2)) < \rho^{t/8} \le \exp(-\alpha \mu_x t / 4)$, we get the same conclusion that
\begin{align*}
    \Exs[\|x_t^1-x_t^2\|_{2}^2] &\le O_{\texttt{P}} (1) \cdot \Exs\left[\|x_\tau^1-x_\tau^2\|_{2}^2 + \frac{\alpha}{\beta} \|y_\tau^1-y_\tau^2\|_{2}^2 + O_{\texttt{P}}(\alpha) \right] \exp(-\alpha \mu_x t / 8) \\
    &\le O_{\texttt{P}} (1) \cdot V_0 \exp(-\alpha \mu_x t / 8).
\end{align*}
The inequality for $y$ in \eqref{eq:dist_conv_rate_ineq} can also be similarly proven.

\section{Proof of Technical Lemmas}

\subsection{Proof of Lemma \ref{lemma:tr_holder_bound}}
This result follows immediately from the fact that $\Tr(AB) \le \|AB\|_1$, and Hölder's inequality applied to matrix p-Schattern norm.

\subsection{Proof of Lemma \ref{lemma:operator_to_Q_norm_conversion}}
By definition of $\|A\|_Q$, we start from
\begin{align*}
    \|A\|_Q^2 &= \max_{\|x\|_Q \le 1} \|Ax\|_Q^2 = \max_{\|x\|_Q \le 1} (x^\top A^\top Q A x) = \max_{\|z\|_2 \le 1} (z^\top Q^{-1/2} A^\top Q A Q^{-1/2} z) \\
    &= \max_{\|z\|_2 \le 1} \|Q^{1/2} A Q^{-1/2} z\|_2^2 = \|Q^{1/2}AQ^{-1/2}\|_{\op}^2.
\end{align*}

\subsection{Proof of Lemma \ref{lemma:basic_operator_Qnorm_conversion}}
By definition for $\vdot{\cdot}{\cdot}_Q$, 
\begin{align*}
    \vdot{x}{y}_Q = x^\top Q y \le \|x^\top Q^{1/2}\|_2 \|Q^{1/2} y\|_2 = \|x\|_Q \|y\|_Q.
\end{align*}
Next, we observe that
\begin{align*}
    \vdot{Mx}{x}_Q = x^\top M^\top Q x = \Tr(xx^\top M^\top Q) \le \|Qxx^\top\|_1 \|M\|_{\op}.
\end{align*}
Then since $Qxx^\top$ is a rank-1 matrix, we have
\begin{align*}
    \|Qxx^\top\|_1 = \|Qxx^\top\|_{2} = \sqrt{x^\top Q x} = \|x\|_Q.
\end{align*}
Finally, by definition of $\|\cdot\|_Q$,
\begin{align*}
    \|Mx\|_Q \le \|M\|_Q \|x\|_Q. 
\end{align*}
Then, note that
\begin{align*}
    \|M\|_Q^2 = \|Q^{1/2}MQ^{-1/2}\|_{\op}^2 \le \kappa(Q) \|M\|_{\op}^2, 
\end{align*}
yielding the proof.

\subsection{Proof of Lemma \ref{lemma:basic_norm_measure_conversion}}
This can be shown from the definition of Q-norm:
\begin{align*}
    \|x\|_{Q_1}^2 = x^\top Q_1 x \le \sigma_{\max}(Q_1) \|x\|_2^2 \le \frac{\sigma_{\max}(Q_1)}{\sigma_{\min}(Q_2)} \|x\|_{Q_2}^2.
\end{align*}

\subsection{Proof of Lemma \ref{lemma:ergodic_noise_after_mixing}}
Let $\pi_{\tau} = \PP_{\xi_t}(\cdot | \mathcal{F}_{t-\tau})$ be a distribution over $\Xi$. By Assumption \ref{assumption:noise_field}, 
\begin{align*}
    \|\pi_{\tau} - \pi\|_1 \le c_{\rho} \rho^\tau. 
\end{align*}
Furthermore, we know that
\begin{align*}
    \int_{\Xi} W_{ij}(\xi) d\pi(\xi) = 0.
\end{align*}
Thus, 
\begin{align*}
    \Exs[\vdot{W_{ij}(\xi)}{u_{t-\tau} v_{t-\tau}^\top} | \mathcal{F}_{t-\tau}] &\le \Exs_{\xi \sim \pi} [\vdot{W_{ij}(\xi)}{u_{t-\tau} v_{t-\tau}^\top}] + W_{\max} \|u_{t-\tau} v_{t-\tau}^\top\|_{1} \|\pi_{\tau}-\pi\|_1 \\
    &\le W_{\max} \|u_{t-\tau}\|_2 \|v_{t-\tau}\|_2 \cdot c_{\rho} \rho^{\tau}.
\end{align*}
The second inequality also follows similarly.

\subsection{Proof of Lemma \ref{lemma:wt_reexpress}}
    Note that $x_t = \bar{x}_t + x^*$ and $y_t = \bar{y}_t - J_{22}^{-1} J_{21} \bar{x}_t + J_{22}^{-1} J_{21} x^*$. Plugging these to $w_t^x = W_{11}(\xi_t) x_t + W_{12}(\xi_t) y_t + u_1(\xi_t)$ and similarly to $w_t^y$ yields the expressions.

\subsection{Proof of Lemma \ref{lemma:sequence_difference_bound}}
By the recursion in \eqref{eq:new_recursion}, 
\begin{align*}
    \|\bar{x}_{t+1}\|_2 &\le (1 + \alpha \|\Delta + W_{\Delta}^x(\xi_t)\|_{\op}) \|\bar{x}_t\|_2 + \alpha (\|J_{12} + W_{12}(\xi_t)\|_{\op} \|\bar{y}_t\|_2 + \|W_{\Delta}^x(\xi_t) x^* + u_1(\xi_t) \|_2) \\
    &\le (1 + \alpha \kappa_y J_{\max}) \|\bar{x}_t\|_2 + \alpha (J_{\max} \|\bar{y}_t\|_2 + \sigma_x), \\
    \|\bar{x}_{t+1} - \bar{x}_t\|_2 &\le \alpha (\kappa_y J_{\max} \|\bar{x}_t\|_2 + J_{\max}\|\bar{y}_t\|_2 + \sigma_x).
\end{align*}
Similarly, we have
\begin{align*}
    \|\bar{y}_{t+1}\|_2 &\le (1 + \beta J_{\max})\|\bar{y}_t\|_2 + \beta (J_{\max} \kappa_y \|\bar{x}_t\|_2 + \sigma_y) + \kappa_y \alpha (\kappa_y J_{\max} \|\bar{x}_t\|_2 + J_{\max} \|\bar{y}_t\|_2 + \sigma_x), \\
    \|\bar{y}_{t+1}-\bar{y}_t\|_2 &\le \beta (J_{\max} \|\bar{y}_t\|_2 + J_{\max} \kappa_y \|\bar{x}_t\|_2 + \sigma_y) + \kappa_y \alpha (\kappa_y J_{\max} \|\bar{x}_t\|_2 + J_{\max} \|\bar{y}_t\|_2 + \sigma_x).
\end{align*}
Adding two equations, 
\begin{align*}
    \kappa_y \|\bar{x}_{t+1}\|_2 + \|\bar{y}_{t+1}\|_2 &\le (1 + 2\beta J_{\max})  (\kappa_y \|\bar{x}_{t}\|_2 + \|\bar{y}_{t}\|_2) + \beta \sigma_y + \alpha \kappa_y\sigma_x. 
\end{align*}
Solving this recursively, we get
\begin{align*}
    \kappa_y \|\bar{x}_{t}\|_2 + \|\bar{y}_{t}\|_2 &\le (1 + 2\beta \tau J_{\max})  (\kappa_y \|\bar{x}_{t-\tau}\|_2 + \|\bar{y}_{t-\tau}\|_2) + \tau(\beta \sigma_y + \alpha \kappa_y\sigma_x). 
\end{align*}
Using this result, we have
\begin{align*}
    \|\bar{x}_{t} - \bar{x}_{t-\tau}\|_2 &\le \sum_{i=t-\tau+1}^{t} \|\bar{x}_{i} - \bar{x}_{i-1}\|_2 \le \alpha J_{\max} \sum_{i=t-\tau+1}^{t} \left(\kappa_y \|\bar{x}_i\|_2 + \|\bar{y}_i\|_2 \right) + \alpha\tau \sigma_x \\
    &\le 2\alpha\tau J_{\max} \left(\kappa_y \|\bar{x}_{t-\tau}\|_2 + \|\bar{y}_{t-\tau}\|_2 \right) + 2\alpha\tau \sigma_x + \alpha \beta J_{\max} \tau \sigma_y.
\end{align*}
Similarly, 
\begin{align*}
    \|\bar{y}_{t} - \bar{y}_{t-\tau}\|_2 &\le \sum_{i=t-\tau+1}^{t} \|\bar{y}_{i} - \bar{y}_{i-1}\|_2 \le (\beta + \alpha\kappa_y) J_{\max} \sum_{i=t-\tau+1}^{t} \left(\kappa_y \|\bar{x}_i\|_2 + \|\bar{y}_i\|_2 \right) + \tau(\beta \sigma_y + \alpha\kappa_y \sigma_x) \\
    &\le 2\tau J_{\max} \beta \left(\kappa_y \|\bar{x}_{t-\tau}\|_2 + \|\bar{y}_{t-\tau}\|_2 \right) + 2\tau (\alpha \kappa_y \sigma_x + \beta \sigma_y).
\end{align*}
Finally, from these two equations, note that 
\begin{align*}
    \kappa_y \|\bar{x}_{t} - \bar{x}_{t-\tau}\|_2 + \|\bar{y}_{t} - \bar{y}_{t-\tau}\|_2 &\le 8 \beta J_{\max} \tau (\kappa_y \|\bar{x}_t\|_2 + \|\bar{y}_t\|_2) + 8 (\alpha \kappa_y \sigma_x + \beta \sigma_y).
\end{align*}
Plugging this back with $\|\bar{x}_{t-\tau}\|_2 \le \|\bar{x}_t\|_2 + \|\bar{x}_t - \bar{x}_{t-\tau}\|_2$ and $\|\bar{y}_{t-\tau}\|_2 \le \|\bar{y}_t\|_2 + \|\bar{y}_t - \bar{y}_{t-\tau}\|_2$, we get the lemma.

\subsection{Proof of Lemma \ref{lemma:noise_cross_l1_product_y}}

    To begin with, we start with unfolding the expression as
    \begin{align*}
        \Exs[w_t^x \bar{y}_t^\top ] &= \Exs[W_{11}(\xi_t) \bar{x}_t \bar{y}_t^\top ] + \Exs[W_{12}(\xi_t) \bar{y}_t \bar{y}_t^\top ] + \Exs[u_1(\xi_t) \bar{y}_t^\top ]. 
    \end{align*}
    To proceed, we first note that
    \begin{align*}
        \Exs[W_{11}(\xi_t) \bar{x}_t \bar{y}_t^\top ] &= \Exs[W_{11}(\xi_t) \bar{x}_{t-\tau} \bar{y}_{t-\tau}^\top ] + \Exs[W_{11}(\xi_t) (\bar{x}_t - \bar{x}_{t - \tau}) \bar{y}_{t-\tau}^\top ] + \Exs[W_{11}(\xi_t) \bar{x}_t(\bar{y}_t - \bar{y}_{t-\tau})^\top ].
    \end{align*}
    For each term in the above, we have the following inequalities:
    \begin{enumerate}
        \item Using the mixing-time assumption, we can show that
        \begin{align*}
            \|\Exs[W_{11}(\xi_t) \bar{x}_{t-\tau} \bar{y}_{t-\tau}^\top ]\|_1 &\le \rho^\tau \cdot \Exs[ \Exs[ \max_{\xi_t \in \Xi} \| W_{11}(\xi_t) \bar{x}_{t-\tau} \bar{y}_{t-\tau}^\top \|_1 | \mathcal{F}_{t-\tau}]] \\
            &\le \rho^\tau W_{\max} \Exs[\|\bar{x}_{t-\tau} \bar{y}_{t-\tau}^\top\|_1] = \rho^\tau W_{\max} \Exs[\|\bar{x}_{t-\tau}\|_2 \|\bar{y}_{t-\tau}\|_2] \\
            &\le O(1) \rho^\tau W_{\max} \Exs[ \|\bar{x}_t\|_2^2 + \|\bar{y}_t\|_2^2 + \alpha^2 \kappa_y^2 \tau^2 \sigma_x^2 + \beta^2 \tau^2 \sigma_y^2 ] \\
            &\le \frac{\mu_y}{32 \kappa_y} \Exs[\|\bar{y}_{t}\|^2_2] + \rho^{\tau} W_{\max} \cdot O\left( \frac{\alpha^2 \mu_y^2}{\kappa_y^2} \Exs[\|\bar{x}_{t}\|_2^2] + \tau^2 (\alpha^2 \kappa_y^2 \sigma_x^2 + \beta^2 \sigma_y^2) \right).
        \end{align*}
        \item For the next term, we apply Lemma \ref{lemma:sequence_difference_bound} and Corollary \ref{corollary:sequence_difference_bound}:
        \begin{align*}
            \|\Exs[W_{11}(\xi_t) &(\bar{x}_t - \bar{x}_{t-\tau}) \bar{y}_{t-\tau}^\top ]\|_1 \le W_{\max} \Exs[\|(\bar{x}_t - \bar{x}_{t-\tau}) \bar{y}_{t-\tau}^\top\|_1] \\
            &\le \alpha \tau J_{\max} W_{\max} \cdot \Exs[(\kappa_y \|\bar{x}_t\|_2 + \|\bar{y}_t\|_2) \|\bar{y}_{t-\tau}\|_2] + \alpha\tau W_{\max} (\sigma_x + \beta J_{\max} \sigma_y) \Exs[\|\bar{y}_{t-\tau}\|_2] \\
            &\le O(1) \cdot \alpha\tau J_{\max}^2 \left(\Exs[\beta J_{\max} \kappa_y^2 \|\bar{x}_t\|_2^2 + (\kappa_y \|\bar{x}_t\|_2 \|\bar{y}_t\|_2 + \|\bar{y}_t\|_2^2) + (\alpha^2 \kappa_y^2 \sigma_x^2 + \beta^2 \sigma_y^2)] \right) \\
            &\qquad + O(1) \cdot \alpha\tau J_{\max} (\sigma_x + \beta J_{\max} \sigma_y) (J_{\max}\beta\kappa_y\tau \|\bar{x}_t\| + \|\bar{y}_t\| + \tau (\alpha \kappa_y \sigma_x + \beta \sigma_y)) \\
            &\le \frac{\mu_y}{32\kappa_y} \Exs[\|\bar{y}_t\|_2^2] + O(1) J_{\max}^2 \kappa_y^2  \tau^2 \left( \beta^2 J_{\max} + \frac{\alpha^2 \kappa_y J_{\max}^2 }{\mu_y} \right) \Exs[\|\bar{x}_t\|_2^2] \\
            &\qquad + O\left(\frac{J_{\max}^2 \kappa_y}{\mu_y}\right) \left(\alpha^2 \tau^2 \sigma_x^2+ J_{\max}^2 \alpha \beta^2 \tau^2 \sigma_y^2 \right).
        \end{align*}
        We can simplify it further later, using the condition that 
        $\alpha \ll \beta / \kappa_y$ and $W_{\max} \le J_{\max}$.
        
        \item For the last term, similarly, 
        \begin{align*}
            \|\Exs[W_{11}(\xi_t) &\bar{x}_t (\bar{y}_t - \bar{y}_{t-\tau})^\top ]\|_1 \le W_{\max} \Exs[\|\bar{x}_t\|_2 \|\bar{y}_t - \bar{y}_{t-\tau}\|_2] \\
            &\le O(1) \beta \tau J_{\max}^2 \cdot \Exs[(\kappa_y \|\bar{x}_t\|_2 + \|\bar{y}_t\|_2) \|\bar{x}_{t}\|_2] + O(1) \tau J_{\max} \Exs[\|\bar{x}_{t}\|_2] (\alpha \kappa_y \sigma_x + \beta \sigma_y) \\
            &\le O(1) \beta\tau J_{\max}^2 \Exs[\kappa_y \|\bar{x}_t\|_2^2 + \|\bar{y}_t\|_2^2] + O(1) \tau ((\alpha^2/\beta) \kappa_y \sigma_x^2 + \beta\sigma_y^2). 
        \end{align*}
    \end{enumerate}
    Combining these inequalities, and given that $\beta\tau \ll \kappa_y^{-2} / J_{\max}$ in Assumption \ref{assumption:stepsize}, we can conclude that
    \begin{align*}
         \|\Exs[W_{11}(\xi_t) \bar{x}_t \bar{y}_t^\top]\|_1 &\le \frac{\mu_y }{16 \kappa_y} \Exs[\|\bar{y}_t\|_2^2] + O(1) \beta \tau J_{\max}^2 \kappa_y \Exs[\|\bar{x}_t\|_2^2] + O(1) \tau \left((\alpha^2/\beta) \kappa_y \sigma_x^2 + \beta \sigma_y^2 \right). 
    \end{align*}
    Similarly, we can also show that
    \begin{align*}
        \|\Exs[W_{12}(\xi_t) \bar{y}_t \bar{y}_t^\top]\|_1 &\le \frac{\mu_y}{16\kappa_y} \Exs[\|\bar{y}_t\|_2^2] + O(1) \beta\tau J_{\max}^2 \kappa_y \Exs[\|\bar{x}_t\|_2^2] + O(1) \tau ((\alpha^2/\beta) \kappa_y \sigma_x^2 + \beta \sigma_y^2).
    \end{align*}
    
    For the last one, we proceed as 
    \begin{align*}
        \|\Exs[u_1(\xi_t) \bar{y}_t^\top ]\|_1 &\le  \|\Exs[u_1(\xi_t) \bar{y}_{t-\tau}^\top ]\|_1 + \|\Exs[u_1(\xi_t) (\bar{y}_t - \bar{y}_{t-\tau})^\top ]\|_1 \\
        &\le \rho^\tau u_{\max} \Exs[\|\bar{y}_{t-\tau}\|_2] + u_{\max} \Exs[\|\bar{y}_t - \bar{y}_{t-\tau}\|_2]\\
        &\le O(1) (\rho^\tau + \beta \tau J_{\max}) u_{\max} \Exs[\kappa_y \|\bar{x}_t\|_2 + \|\bar{y}_t\|_2] + \tau  u_{\max} (\alpha \kappa_y \sigma_x + \beta\sigma_y) \\
        &\le O(1) \beta \tau J_{\max}^2 \Exs[\kappa_y \|\bar{x}_t\|_2^2 + \|\bar{y}_t\|_2^2] + O(1) \tau ((\alpha^2/\beta) \kappa_y \sigma_x^2 + \beta \sigma_y^2), 
    \end{align*}
    where in the last inequality, we use $u_{\max} \le \sigma_y$.
    Combining all the above inequalities yields the lemma.

\subsection{Proof of Lemma \ref{lemma:noise_cross_l1_product_x}}

    To begin with, we start with unfolding the expression as
    \begin{align*}
        \Exs[w_t^x \bar{x}_t^\top ] &= \Exs[W_{11}(\xi_t) \bar{x}_t \bar{x}_t^\top ] + \Exs[W_{12}(\xi_t) \bar{y}_t \bar{x}_t^\top ] + \Exs[u_1(\xi_t) \bar{x}_t^\top ]. 
    \end{align*}
    To proceed, we first note that
    \begin{align*}
        \Exs[W_{11}(\xi_t) \bar{x}_t \bar{x}_t^\top ] &= \Exs[W_{11}(\xi_t) \bar{x}_{t-\tau} \bar{x}_{t-\tau}^\top ] + \Exs[W_{11}(\xi_t) (\bar{x}_t - \bar{x}_{t - \tau}) \bar{x}_{t-\tau}^\top ] + \Exs[W_{11}(\xi_t) \bar{x}_t(\bar{x}_t - \bar{x}_{t-\tau})^\top ].
    \end{align*}
    For each term in the above, we have the following inequalities:
    \begin{enumerate}
        \item Using the mixing-time assumption, we can show that
        \begin{align*}
            \|\Exs[W_{11}(\xi_t) \bar{x}_{t-\tau} \bar{x}_{t-\tau}^\top ]\|_1 &\le \rho^\tau \cdot \Exs[ \Exs[ \max_{\xi_t \in \Xi} \| W_{11}(\xi_t) \bar{x}_{t-\tau} \bar{x}_{t-\tau}^\top \|_1 | \mathcal{F}_{t-\tau}]] \\
            &\le \rho^\tau W_{\max} \Exs[\|\bar{x}_{t-\tau} \bar{x}_{t-\tau}^\top\|_1] = \rho^\tau W_{\max} \Exs[\|\bar{x}_{t-\tau}\|_2^2] \\
            &\le O(1) \rho^\tau W_{\max} \Exs[ \|\bar{x}_t\|_2^2 + \alpha^2 J_{\max}^2 \|\bar{y}_t\|_2^2 + \alpha^2 \tau^2 \sigma_x^2 ] \\
            &\le \frac{\mu_x}{32 \kappa_x} \Exs[\|\bar{x}_{t}\|^2_2] + \rho^{\tau} J_{\max} O(\alpha^2 \mu_x^2 \Exs[\|\bar{y}_{t}\|_2^2] + \tau^2 \alpha^2 \sigma_x^2).
        \end{align*}
        
        \item For the next term, we apply Lemma \ref{lemma:sequence_difference_bound} and Corollary \ref{corollary:sequence_difference_bound}:
        \begin{align*}
            \|\Exs[W_{11}(\xi_t) &(\bar{x}_t - \bar{x}_{t-\tau}) \bar{x}_{t-\tau}^\top ]\|_1 \le W_{\max} \Exs[\|(\bar{x}_t - \bar{x}_{t-\tau}) \bar{x}_{t-\tau}^\top\|_1] \\
            &\le O(1) \alpha \tau J_{\max}  W_{\max} \cdot \Exs[(\kappa_y \|\bar{x}_t\|_2 + \|\bar{y}_t\|_2) \|\bar{x}_{t-\tau}\|_2] + O(1) \alpha\tau W_{\max} (\sigma_x + \beta J_{\max} \sigma_y) \Exs[\|\bar{x}_{t-\tau}\|_2] \\
            &\le O(1) \cdot \alpha\tau J_{\max}^2  \left(\Exs[ \kappa_y \|\bar{x}_t\|_2^2 + \|\bar{y}_t\|_2^2 + \tau^2\alpha^2 ( \sigma_x^2 + \beta^2 J_{\max}^2 \sigma_y^2) \right) \\
            &\le \frac{\mu_x}{32\kappa_x} \Exs[\|\bar{x}_t\|_2^2] + O(1) \alpha\tau J_{\max}^2  \Exs[\|\bar{y}_t\|_2^2] + (\alpha\tau)^3 (\sigma_x^2 + \beta^2 J_{\max}^2 \sigma_y^2),
        \end{align*}
        where we use the condition that 
        $\alpha \tau \ll 1 / (J_{\max} \kappa_x^2)$ and $W_{\max} \le J_{\max}$.
    \end{enumerate}
    Combining these inequalities, and given that $\beta\tau \ll \kappa_y^{-2} / J_{\max}$ in Assumption \ref{assumption:stepsize}, we can conclude that
    \begin{align*}
         \|\Exs[W_{11}(\xi_t) \bar{x}_t \bar{x}_t^\top]\|_1 &\le \frac{\mu_x }{16 \kappa_x} \Exs[\|\bar{x}_t\|_2^2] + \alpha \tau J_{\max}^2  \Exs[\|\bar{y}_t\|_2^2] + \alpha \tau \sigma_x^2. 
    \end{align*}

    We also need to check the cross term: 
    \begin{align*}
        \Exs[W_{12}(\xi_t) \bar{y}_t \bar{x}_t^\top] &= \Exs[W_{12}(\xi_t) \bar{x}_{t-\tau} \bar{y}_{t-\tau}^\top ] + \Exs[W_{12}(\xi_t) (\bar{y}_t - \bar{y}_{t - \tau}) \bar{x}_{t-\tau}^\top ] + \Exs[W_{12}(\xi_t) \bar{y}_t(\bar{x}_t - \bar{x}_{t-\tau})^\top ].
    \end{align*}
    First term can be bounded similarly using the geometric mixing assumption. For the second term,
    \begin{align*}
        &\|\Exs[W_{12}(\xi_t)(\bar{y}_t - \bar{y}_{t-\tau}) \bar{x}_{t-\tau}^\top ]\|_1 \le W_{\max} \Exs[\|(\bar{y}_t - \bar{y}_{t-\tau}) \bar{x}_{t-\tau}^\top\|_1] \\
        &\le O(1) \beta \tau J_{\max}^2 \cdot \Exs[(\kappa_y \|\bar{x}_t\|_2 + \|\bar{y}_t\|_2) \|\bar{x}_{t-\tau}\|_2] + O(1) \tau J_{\max} (\alpha \kappa_y\sigma_x + \beta \sigma_y) \Exs[\|\bar{x}_{t-\tau}\|_2] \\
        &\le \frac{\mu_x}{32\kappa_x} \Exs[\|\bar{x}_t\|_2^2] + O(1) \frac{\beta^2 \tau^2  J_{\max}^4 \kappa_x }{\mu_x}  \Exs[\|\bar{y}_t\|_2^2] + O \left(\frac{\alpha^2 \tau^2 J_{\max}^2 \kappa_y^2 \kappa_x}{\mu_x} \sigma_x^2 + \frac{\kappa_x \beta^2 \tau^2 J_{\max}^2}{\mu_x} \sigma_y^2\right).
    \end{align*}
    For the third term, similarly, we have
    \begin{align*}
        &\|\Exs[W_{12}(\xi_t) \bar{y}_{t} (\bar{x}_t - \bar{x}_{t-\tau})^\top ]\|_1 \le W_{\max} \Exs[\|\bar{y}_t (\bar{x}_t - \bar{x}_{t-\tau})^\top \|_1] \\
        &\le O(1) \alpha \tau J_{\max}^2 \cdot \Exs[(\kappa_y \|\bar{x}_t\|_2 + \|\bar{y}_t\|_2) \|\bar{y}_{t}\|_2] + O(1) \tau J_{\max} (\alpha \sigma_x + \beta^2 J_{\max} \sigma_y) \Exs[\|\bar{y}_{t}\|_2] \\
        &\le \frac{\mu_x}{32\kappa_x} \Exs[\|\bar{x}_t\|_2^2] + O(1) \tau J_{\max}^2 (\alpha + \beta^2 J_{\max}) \Exs[\|\bar{y}_t\|_2^2] + O \left(\alpha \tau \sigma_x^2 + \beta^2 \tau J_{\max} \sigma_y^2\right).
    \end{align*}

    For the last one, we proceed as 
    \begin{align*}
        \|\Exs[u_1(\xi_t) \bar{x}_t^\top ]\|_1 &\le  \|\Exs[u_1(\xi_t) \bar{x}_{t-\tau}^\top ]\|_1 + \|\Exs[u_1(\xi_t) (\bar{x}_t - \bar{x}_{t-\tau})^\top ]\|_1 \\
        &\le \rho^\tau u_{\max} \Exs[\|\bar{x}_{t-\tau}\|_2] + u_{\max} \Exs[\|\bar{x}_t - \bar{x}_{t-\tau}\|_2]\\
        &\le (\rho^\tau + \alpha \tau J_{\max}) u_{\max} \Exs[\kappa_y \|\bar{x}_t\|_2 + \|\bar{y}_t\|_2] + \tau  u_{\max} \alpha\sigma_x  \\
        &\le \alpha \tau J_{\max}^2 \Exs[\kappa_y \|\bar{x}_t\|_2^2 + \|\bar{y}_t\|_2^2] + \tau \alpha\sigma_x^2, 
    \end{align*}
    where in the last inequality, we use $u_{\max} \le \sigma_x$.
    Combining all the above inequalities yields the lemma.

\end{appendices}

\end{document}